\documentclass{aa}
\usepackage{epsfig}
\usepackage{graphicx}
\usepackage{txfonts}
\begin{document}

\title{Metallicity distribution of red giants in the Inner Galaxy from Near Infrared spectra}

\author{C. Gonz\'alez-Fern\'andez\inst{1}, A. Cabrera-Lavers\inst{1,2}, P.L. Hammersley\inst{1}, \and F. Garz\'on\inst{1,3}}
   
\offprints{carlos.gonzalez@iac.es}

\institute{Instituto de Astrof\'{\i}sica de Canarias, E-38205 La Laguna, Tenerife, Spain\\ \and GTC Project Office, E-38205 La Laguna, Tenerife, Spain\\ \and Departamento de Astrof\'{\i}sica, Universidad de La Laguna, E-38205 La Laguna, Tenerife, Spain\\ }

\date{Received XX; accepted XX}


\abstract
{The existence in the Milky Way of either a long thin bar with a half length of 4.5 kpc and a position angle of around 45$^\circ$ with respect to the Sun--Galactic centre line or a bulge+bar complex, thicker and shorter, with a smaller tilt respect to the Sun-GC line, has been a matter of discussion in recent decades.}
{In this paper, we present low resolution (R=500) near-infrared spectra for selected and serendipitous sources in six inner in-plane Galactic fields at $l$=7$^\circ$, 12$^\circ$, 15$^\circ$, 20$^\circ$, 26$^\circ$ and 27$^\circ$, with the aim of analysing the stellar content present in those fields.}
{From the equivalent widths of the main features of the K band spectra (the NaI, CaI and CO bandheads) we have derived the metallicities of the sources by means of the empirical scale obtained by Ram\'{\i}rez et al. (2000) and Frogel et al. (2001) for luminous red giants.}
{Our results show how the mean metallicity of the sample varies with Galactic longitude. We find two groups of stars, one whose [Fe/H] is similar to the values obtained for the bulge in other studies (Molla et al. 2000; Schultehis et al 2003), and a second one with a metallicity similar to that of the inner parts of the disc (Rocha-Pinto et al. 2006). The relative density of both groups of stars in our sample varies in a continuous way from the bulge to the disc. This could hint at the existence of a single component apart from the disc and bulge, running from $l$=7$^\circ$ to $l$=27$^\circ$ and able to transport disc stars inwards and bulge stars outwards, which could be the Galactic bar that has been detected in previous works as an overdensity of stars located at those galactic coordinates (Hammersley et al. 1994, 2000; Picaud et al. 2003).}
{}

\keywords{Galaxy: abundances --- Galaxy: general --- Galaxy: stellar content --- Galaxy: structure --- Stars: late-type}

\authorrunning{C. Gonz\'alez-Fern\'andez et al.}
\titlerunning{Metallicity distribution of inner MW red giants}

\maketitle

%

\section{Introduction}

	There is  substantial consensus for the presence of a bar in the inner Galaxy. It was suggested for the first time by de Vaucouleurs (1964) and the first evidence was derived from the asymmetries in the infrared (IR) light distribution (e.g., Blitz\& Spergel 1991; Dwek et al. 1995) and in the source counts  (Weinberg 1992; Hammersley et al. 1994; Stanek et al. 1994), which show a strong increase towards  positive longitudes in the Galactic Plane (GP). However, the definition of the exact nature and parameters  of this structure are still controversial. While some authors refer to a small  bar with a moderated position angle of 15--30 degrees with respect to the Sun--Galactic Centre direction (Dwek et al. 1995; Nikolaev \& Weinberg 1997; Stanek et al. 1997; Binney et al. 1997; Freudenreich 1998; Sevenster et al. 1999; Bissantz \& Gerhard 2002; Babusiaux \& Gilmore 2005), other researchers point to a larger bar with a half length of 4 kpc and a position angle around 45 degrees. 
	
	Hammersley et al. (2000) observed a strong overdensity of K2--3III type stars on the GP at \mbox{$l=27^\circ$} that could also be detected at lower galactic longitudes, but more reddened in accordance with a longer distance to the observer. This overdensity dissapears once one looks either 
	above or below the GP at the same galactic longitudes. This was interpreted as a strong evidence for the existence of an in-plane bar with  a half-length of $\sim$4~kpc and a position angle of 43$^\circ$, clearly distinguishable from  the triaxial bulge but running into it near $l$=12$^\circ$. This overdensity of stars was also analysed by comparing near-infrared (NIR) counts with predictions of the {Besan\c{c}on Galactic model} (Robin et al. 2003) with similar conclusions (Picaud et al. 2003). Very recently, this result  has also been confirmed by observations in the mid-infrared with GLIMPSE data (Benjamin et al. 2005), a range where effects due to extinction are even lower than in the NIR, hence the penetration in the innermost parts of the Galaxy is higher (for a compilation of evidence in favor to the existence of this long bar in the Milky Way see L\'opez-Corredoira et al. 2006). The predominant observed populations differ from the NIR, where old populations dominate, to MIR, where a mixture of both old and young stars prevails. But such studies as that of Wozniak (2007) demonstrate that bars mix  their stellar content very efficiently (with a diffusion timescale of around 100 Myr), so even if stellar formation is concentrated in certain areas of the structure, there is no expected difference in the stellar population mix, and thus both in the NIR and MIR the same structures should be observed.
	
	All the above suggests that, rather than a short-scale bar, there is an altogether different structure in the innermost Galaxy ($|l|<5^\circ$). This inner structure could therefore be a triaxial body with axial ratios of 1:0.49:0.37 (L\'opez-Corredoira et al. 2005), but such a flat feature cannot be responsible by itself for the observed star counts in the Galactic plane for $l>0^\circ$ (L\'opez-Corredoira et al. 1999), as also noted by Nishiyama et al. 2005, with NIR photometry of red clump stars. The effect of the bulge in the red clump counts is predominant at higher latitudes, while the effect of the long thin bar is more constrained to the Galactic Plane at larger longitudes (15$^\circ <$l$<$27$^\circ$), thus there are two very different large-scale triaxial structures coexisting in the inner Galaxy (Cabrera-Lavers et al. 2007). Furthermore,  the possibility of smaller nonaxisymetric structures such as a secondary bar in The Milky Way cannot be discarded at all (Unavane \& Gilmore 1998; Alard 2001; Nishiyama et al. 2005), thus more observations are needed to be conclusive.
	
	A spectral analysis of selected sources around this overdensity would yield evidence for one of the two possible configurations of the central part of the Galaxy. Either the disc by itself is responsible for the majority of the observed counts, except for those in the bulge, or there is another structural component that also contributes to the stellar content. The majority of these objects lie very close to the GP, and the high value of extinction means that they cannot be observed at visible wavelengths, NIR observations being necessary for this purpose. In the H and K bands, there is a series of molecular lines, OH, H$_2$O and CO (or more complex carbon molecules in carbon stars), as well as a number of metal lines (Si, Na, Ca, Fe, etc.; see, for example, Kleinmann \& Hall 1986; Origlia et al. 1993; Wallace \& Hinkle 1997; Ram\'{\i}rez et al. 1997; Lan\c{c}on\& Wood 2000). The relative strength of these lines will allow the spectral type to be accurately determined, as well as other physical information, such as the metallicity (Ram\'{\i}rez et al. 1997; Frogel et al. 2001; Schultheis et al. 2003). 
	
	In this paper, we describe low resolution spectroscopic observations of a sample of inner Galactic stars in order to derive  their spectral type and  metallicities. Sources have been selected in different lines of sight towards the inner Galaxy, so they are expected to correspond with the supposed bar, bulge and disc locations, in order to compare the results for each component looking for  differences between them.

\section{Sample selection}
\begin{figure}[!h]
		\centering
		\resizebox*{4.4cm}{4.4cm}{\epsfig{file=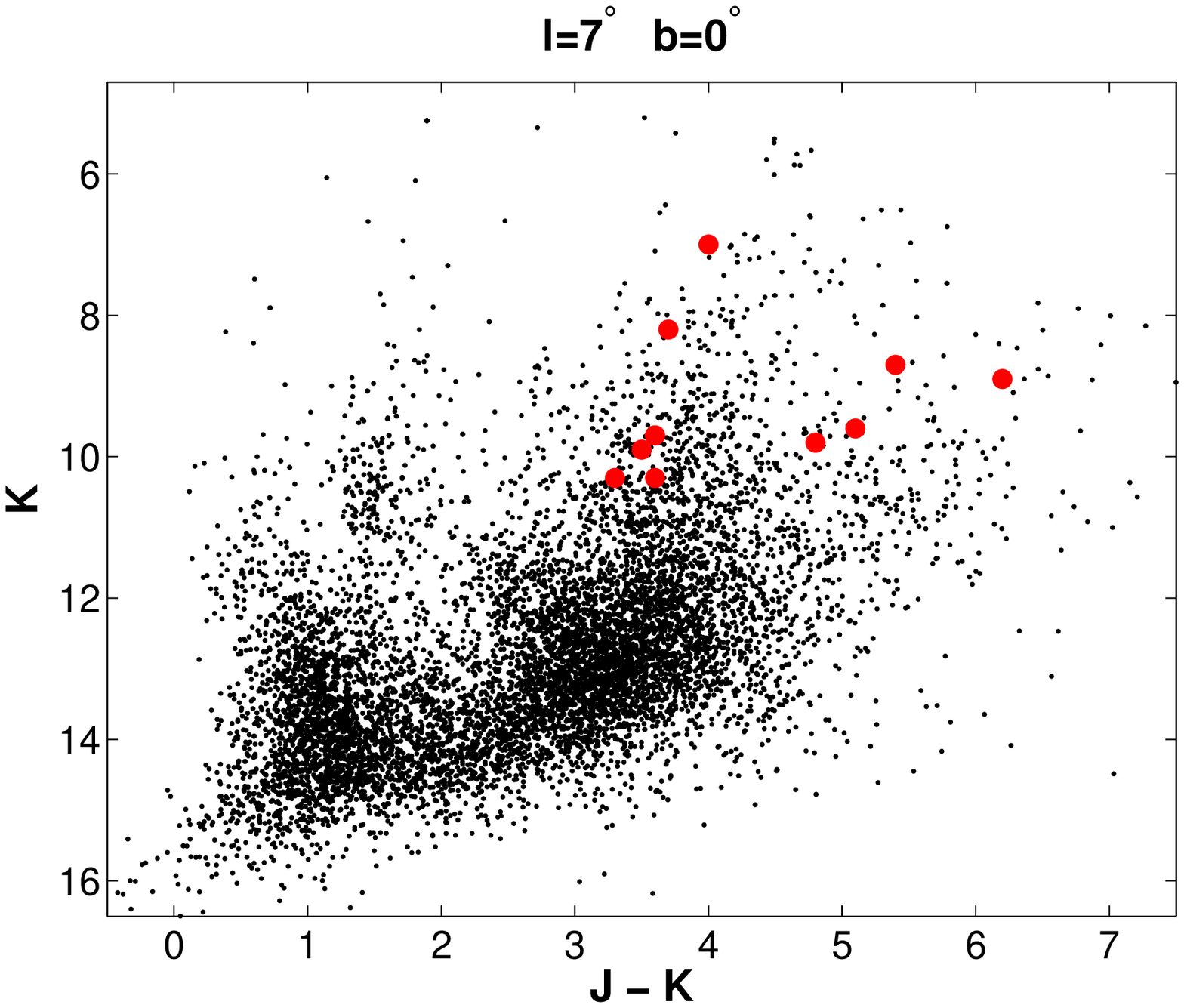,height=4.4cm}}\resizebox*{4.4cm}{4.4cm}{\epsfig{file=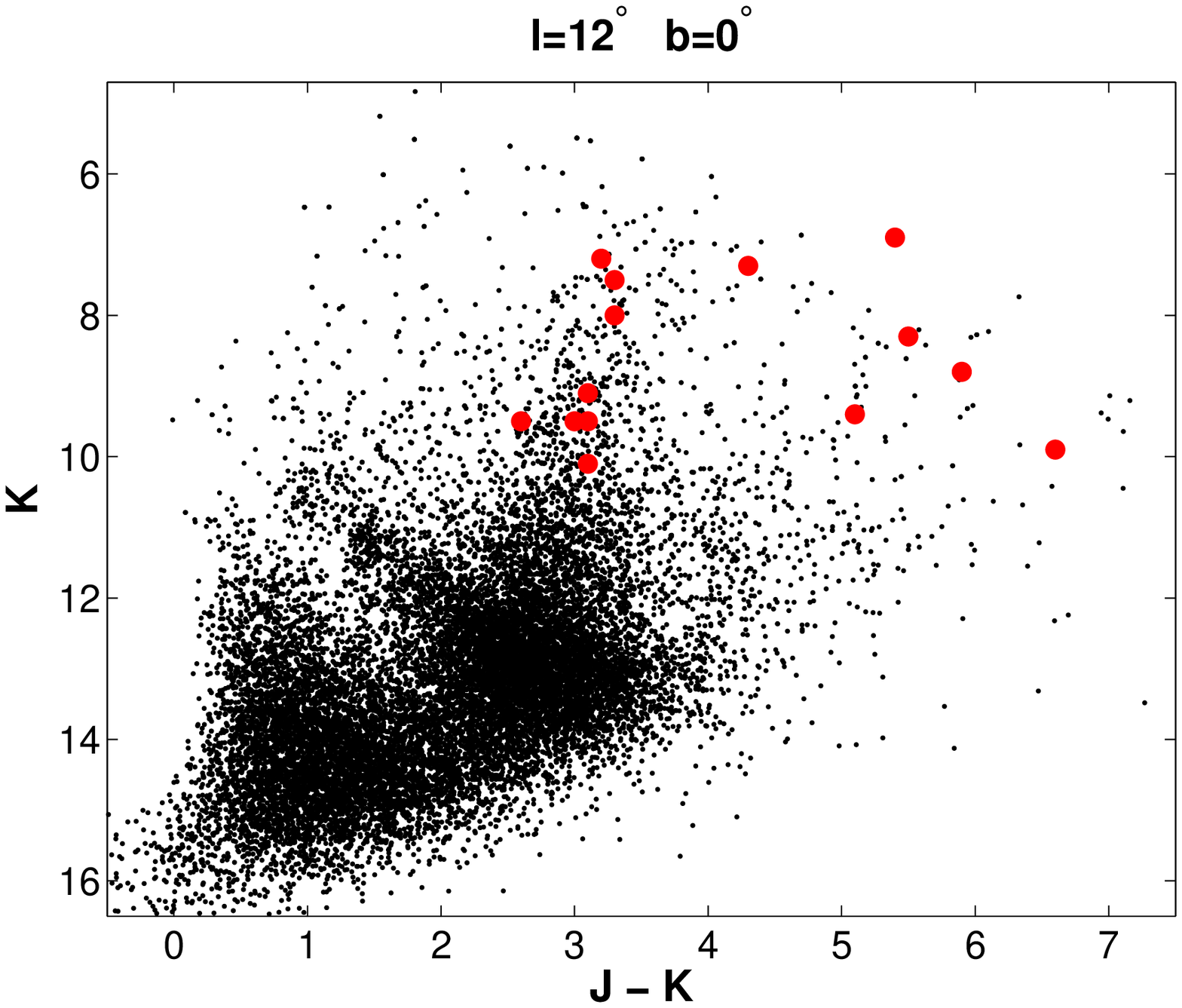,height=4.4cm}}
		\resizebox*{4.4cm}{4.4cm}{\epsfig{file=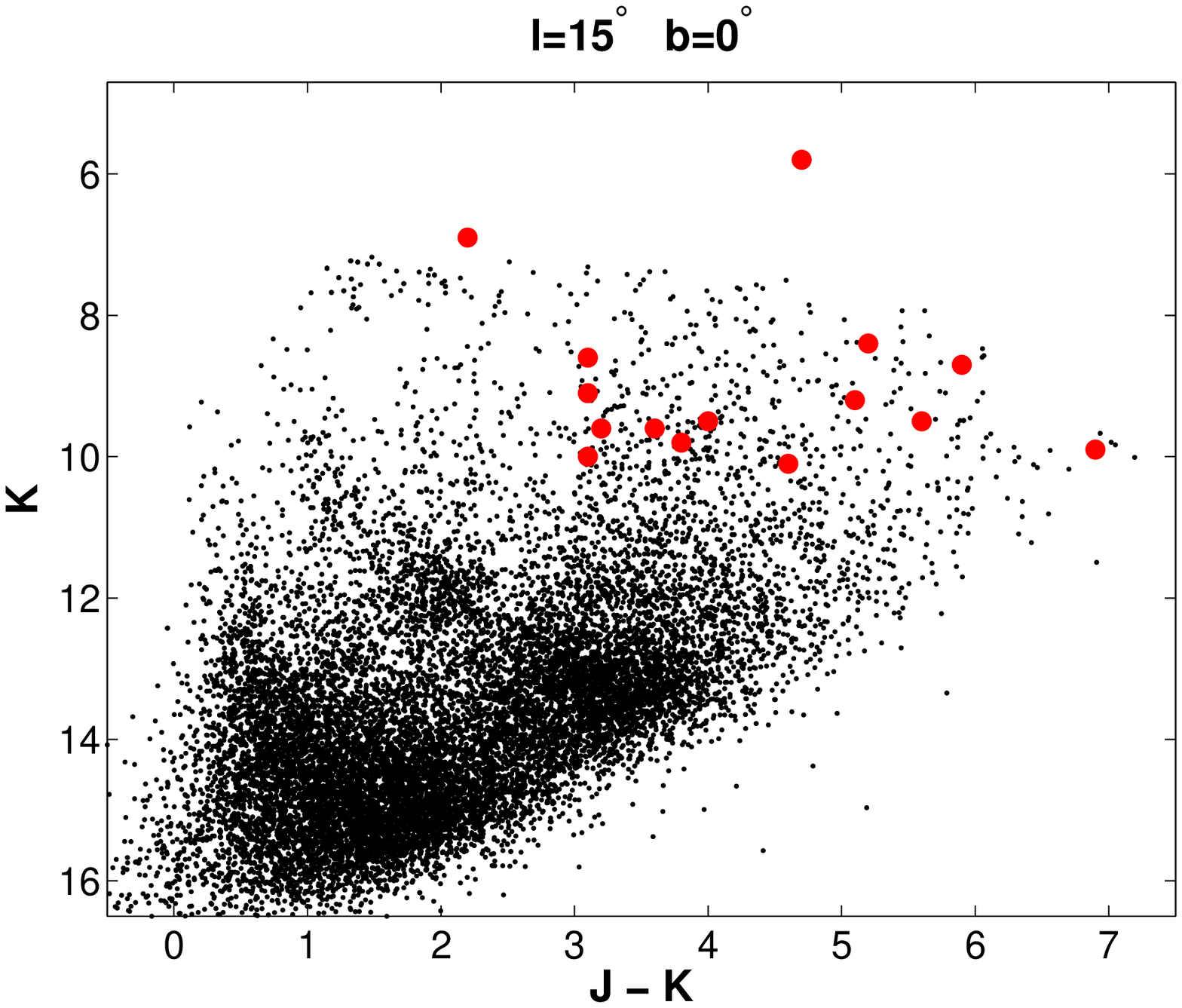,height=4.4cm}}\resizebox*{4.4cm}{4.4cm}{\epsfig{file=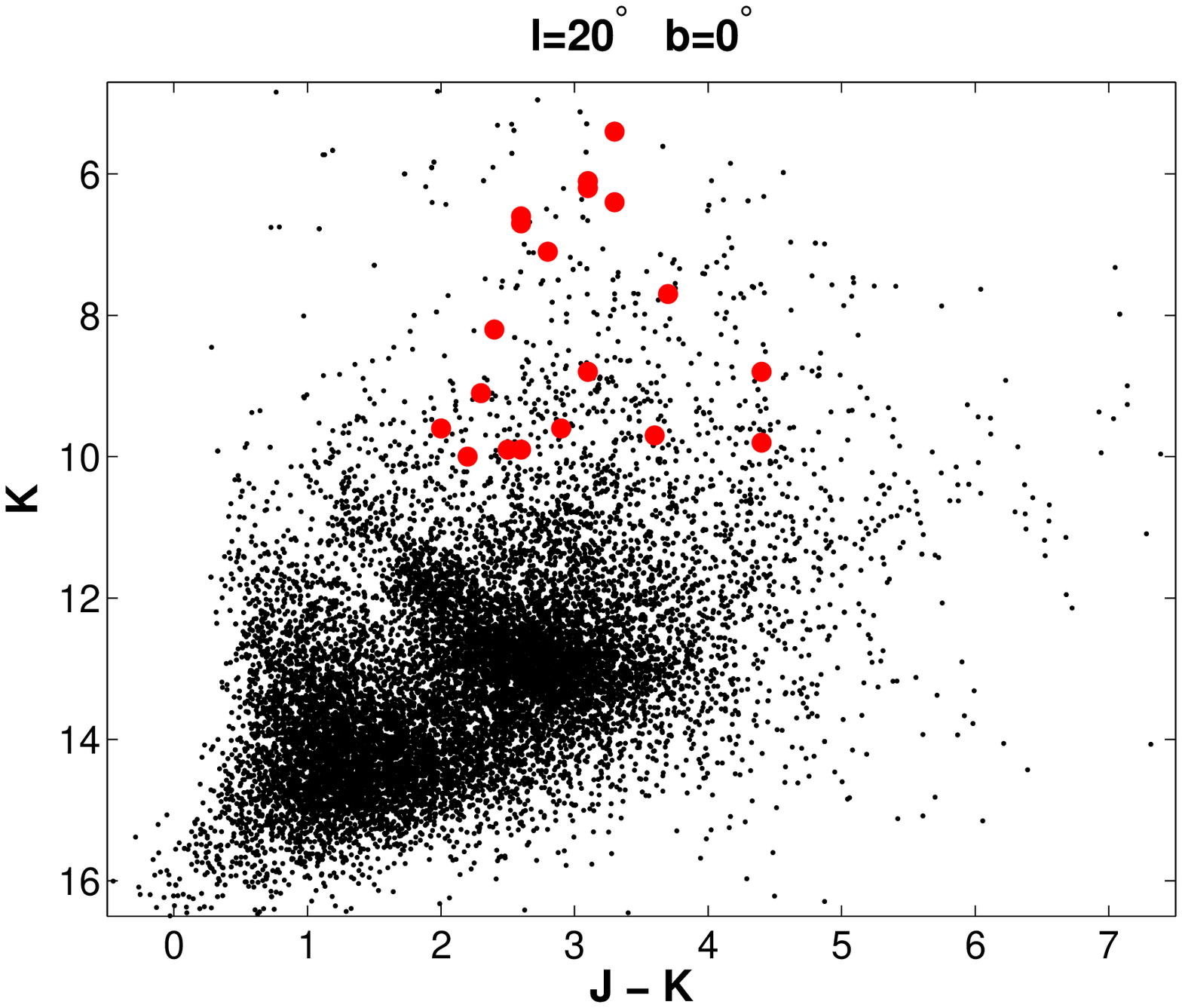,height=4.4cm}}
		\resizebox*{4.4cm}{4.4cm}{\epsfig{file=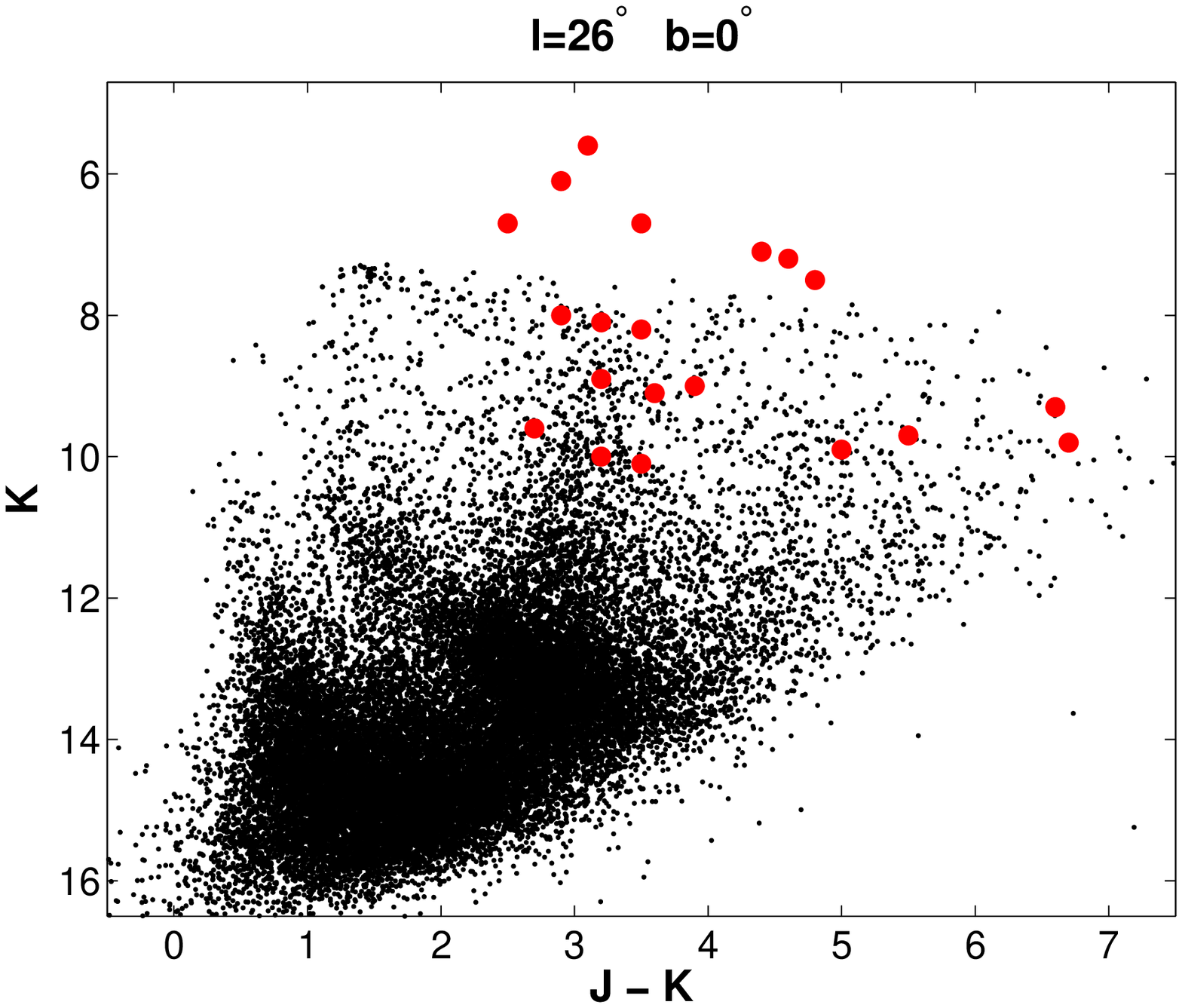,height=4.4cm}}\resizebox*{4.4cm}{4.4cm}{\epsfig{file=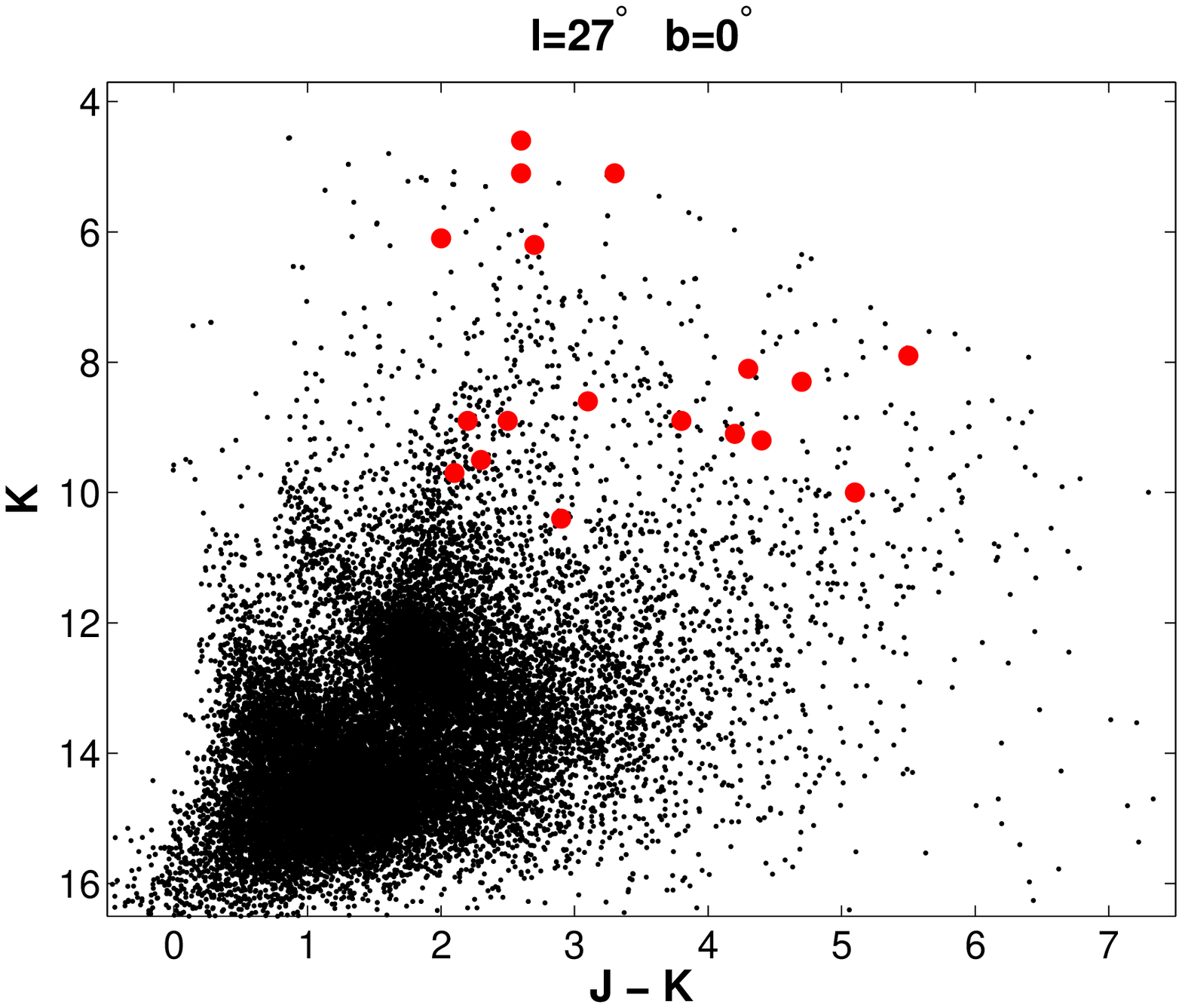,height=4.4cm}}
		\caption{TCS-CAIN NIR CMDs of the fields from which we selected the target stars (l=7$^\circ$, 10$^\circ$, 15$^\circ$, 20$^\circ$, 26$^\circ$, 27$^\circ$, b=0$^\circ$), marked as big dots. Observe these stars exhibit large $(J-K)$ colours, being located in the CMD region corresponding to the K and M giants.}
		\label{CMDs}
	\end{figure}

\begin{figure}[!h]
		\centering
		\epsfig{file=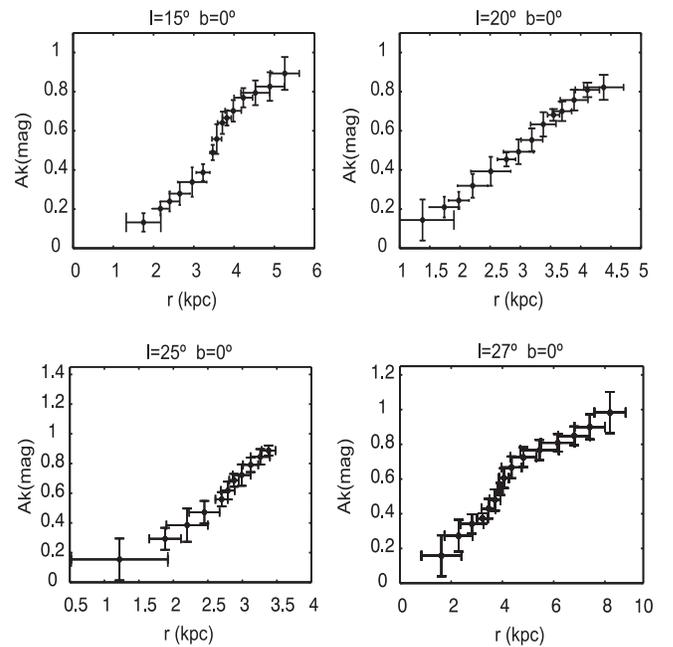,height=8.5cm,width=8.5cm}
		\caption{Example of the extinction measurements derived from TCS-CAIN NIR CMD's along l=10$^\circ$, 20$^\circ$, 25$^\circ$, 27$^\circ$, b=0$^\circ$.}
		\label{extinciones}
	\end{figure}

	The sources were chosen according to their locations in NIR colour--magnitude diagrams (CMDs), since it is possible to estimate the total extinction from these diagrams (see Figs \ref{CMDs} and \ref{extinciones}, and  L{\'o}pez-Corredoira et al. 2001 for details), from the position of a star over the CMD we can derive the Galactic component that it belongs to. Also, since the aim of our study is galactic giants, we can select the object stars in a fashion that minimizes possible contamination from other populations in the sample. However, the infrared HR diagram is still not well calibrated, and the relation of the spectral type with the position over the CMD is not totally understood. Currently, we can only be certain for a few spectral types (e.g. K2III) and even then  there are problems. Even though  the more extreme IR sources (such as late M giants and carbon stars) dominate the  brighter IR magnitudes, their luminosity function is not well known. So to ensure the cleanliness of our set of stars, further filtering must be performed, as will be described in section \ref{samfil}.

	NIR CMDs are obtained from the TCS-CAIN catalogue (Cabrera-Lavers et al. 2006). This catalogue contains roughly 500 near-plane Galactic fields covering $\sim$42 deg$^2$ of the sky in the J, H and K$_s$ bands with a position accuracy of 0.2" and a photometric accuracy of $\sim$0.1 mag in the three filters. The limiting magnitudes of this survey (defined in terms of completeness of the counts) are 1--1.5 mag fainter than those of the 2MASS or DENIS NIR surveys, thus it constitutes a very valuable tool for analysing the innermost Galaxy. The sample stars were chosen from  selected fields where the limiting magnitude of the catalogue in K$_s$ lies between 14 and 14.5 mag. These stars are always brighter than K=10.5, so there are no completeness effects in this analysis.

	Due to the high stellar density, we often find serendipitous detections of stars that fall along the slit when observing a target source. These spectra have also been used for the analysis. The initial sample of target stars contains 212 inner Galactic sources (to which we will add in later steps all the serendipitous sources), corresponding to six different in-plane fields:

	\begin{table}[!h]
		\caption[]{Initial sample of stars.}
		\label{tabstar}
		\begin{center}
		\begin{tabular}{cccc}
		\hline
		Gal. lon. & Object stars & Field stars & Total \\
		\hline
		\hline
		7 & 13 &  7 & 20 \\
		12 & 16 & 21 & 37  \\
		15 & 16 & 19 & 35  \\
		20 & 20 & 19 & 39  \\
		26 & 24 & 15 & 39  \\
		27 & 24 & 18 & 42  \\
		\hline
		\end{tabular}
		\end{center}
	\end{table}

\section{Observation and data reduction}
	For all these sources we obtain low resolution spectra using the 3.6 m Telescopio Nazionale Galileo (TNG) in two different observing runs. The first one comprised three days (2004 June $25$, $26$, 27) and the second two more nights (2006 July 5, and $6$). Spectra were obtained using NICS (Near-Infrared Camera and Spectrometer), a FOSC-type cryogenic focal reducer, equipped with two interchangeable cameras feeding a Rockwell Hawaii 1024$\times$1024 array. The camera used for the spectroscopic observations has a focal ratio of f/4.3, corresponding to a plate scale of 0.25 arcsec pix$^{-1}$ (Oliva \& Gennari 1995; Baffa et al. 2000). We used a low resolution spectroscopic mode with a NICS HK grism, spanning a complete 1.40--2.5 $\mu$m spectral range. A 0.75 arcsec width slit was used, yielding a spectral resolving power of R$\sim$500. Since the main NIR features (such as the molecular lines of OH, H$_2$O and CO, or the metal lines of Na, Ca, Fe, Si, Mg, etc.) are well spread out between 1.5 and 2.4 $\mu$m, high resolution is not required for this kind of study. 

	The K magnitudes of the target stars run from about 5 to 10.5, thus typical acquisition consisted of a series of images in one position (position \emph{A}) of the slit and then offsetting the telescope by $\sim$35" along the slit (to position \emph{B}), repeating this process in an \emph{ABBA}  cycle. Integration times varied accordingly to the magnitude of the stars, from 30 s for the brighter standards to 180 s for the faintest giants. In some cases, whenever real-time inspection of the data made it necessary, the cycle was repeated twice in order to increase the signal-to-noise ratio of the spectra, a parameter that is well above S/N=20. For the identification of target stars, a single 5 s K image is obtained to locate the source accurately. The positional accuracy of TCS-CAIN is about 0.2", which together with the pointing accuracy of TNG made it unnecessary to change the instrumental configuration of NICS, so that, in all the cases,  the target sources were well within the slit. To optimize the observation procedure, we selected slit angles that, where possible, produced spectra of more than one object. Depending of the field, these secondary objects could be object stars selected from TCS-CAIN or serendipitous sources.

	The reduction of the spectra was performed following the standard path, combining the several sky flat-fields obtained every night, prior to the observations. Assuming that atmosphere changes on a timescale larger than our integration times, which is true for the shorter exposures, substraction of two consecutive images, each one in positions A and B, allows us to clean the spectra of telluric emission, both in strong OH lines and diffuse continuous background. 

	For the longer exposures, which for dimmer objects can reach $\sim$200s, sky signal varies noticeably between exposures, so an extra cleaning pass is needed to subtract sky emission from the object spectra. But since from the differentiation of two consecutive pointings A and B we obtain a positive and a negative spectral trace, we can again combine them, thus supressing residual effects present in both features.

	Since our object fields often have a high stellar density, serendipitous objects often fall on the slit, so we have to extract spectra even in extreme regions of the detector, where the spectral trace is curved in the spatial direction. To take into account and correct this effect and others such as possible tilts of the spectra, before extracting them a Gaussian is fitted to the spectrum along the spatial direction for every column. The maxima of the series of Gaussian is fitted to a $3^{rd}$ degree polynomial, giving us a estimate of the deformation of our spectrum. Besides this correction, performing a Gaussian fit allows us to determine accurately  the number of significant pixels to be extracted, $3\sigma$ above and below the Gaussian maximum, thus minimizing the noise introduced in the final spectrum. 

	Wavelength calibration was performed using an argon lamp, with very accurately measured lines. Following staff indications, a 2D fit was performed, choosing a fourth-degree polynomial in the dispersion direction and a third-degree one in the spatial direction.

	Apart from the programme targets, some standard stars from the Hipparcos catalogue were observed (4--6 objects per night). Those stars have well-known spectral types (from K0 III to M7 III, along with some A0V and O5V for calibration purposes) and will be used for comparison with the spectra of the target stars to help in their spectral classification.

	To correct our spectra for telluric absortion, we consider the variations in atmospheric absorption to be linear with the airmass, so comparing the spectra of standard star (of types A0V and O5V) obtained at different heights with a spectrum of the same type from the library of Pickles (1998) we can measure the behaviour of atmospheric extinction with $\lambda$ and airmass, and so remove it from our spectra.
 
\section{Equivalent widths}
\begin{figure}[!h]
		\centering
		\resizebox*{8cm}{6cm}{\epsfig{file=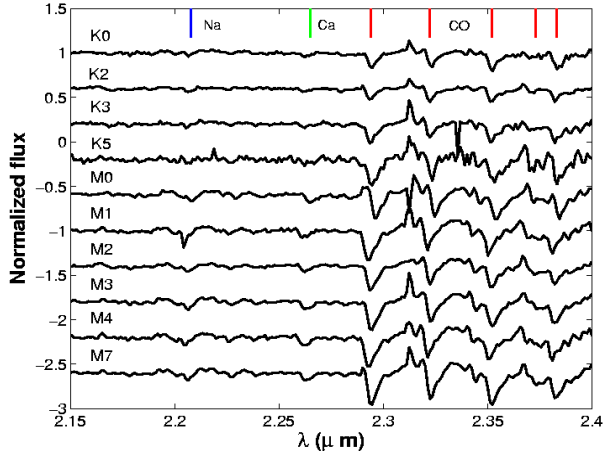,height=6cm}}
		\caption{Sample of K-band spectra for stars from K to M spectral types, showing the mean features observed in this wavelength range.}
		\label{ejemplo}
	\end{figure}

	From the many spectral features present in the spectra, we have selected some of the most prominent ones, such as NaI, CaI, and the (2,0) band of $^{12}$CO in the K-band (see Table \ref{tabla3} for the wavelengths), which are in common with those selected in Ram\'{\i}rez et al. (1997), Frogel et al. (2001) or Schultheis et al. (2003). In the H-band, the strongest feature is the line of SiI (1.59 $\mu$m) although  in very cool stars (beyond $\sim$M2) the major contributor to this feature is OH, so a resolution of at least R$\sim$2000 is necessary to distinguish the two components (Origlia et al. 1993). This fact, together with the observed constant behaviour of EW [Si(1.59$\mu$m)] for spectral types beyond M2 (see Fig. 5a in Origlia et al. 1993), render this feature useless in our study, which focuses on the equivalent widths of the K-band features in our spectra (Fig. \ref{ejemplo}).

	In our analysis the equivalent widths of the selected features (NaI, CaI and CO(2,0)) were measured with respect to a continuum level defined as the best first-order fit of bands near the features. These bandpasses adopted for the continuum and for the features themselves are listed in Table \ref{tabla2} (with the references where those values were taken) and the features themselves are shown in Figure \ref{ejemplo}.

	In order to estimate the accuracy of the equivalent width measurements, we have used the different spectra obtained for the standard stars on the three nights and compare the values obtained for the equivalent widths of the selected features for the several detections of the same star. Since we have several object and standard stars measured with more than one spectrum, we can obtain a standard deviation for the equivalent width of each feature and each star. These deviations are very stable, and do not show any dependence on the equivalent width when comparisons are made between different stars, so we use them as estimates of the uncertainties in the equivalent widths of $\sigma EW(Ca)$=0.61 \AA, $\sigma EW(Na)$=0.46 \AA\ and \mbox{$\sigma EW(CO)$=0.81 \AA}, respectively.

	\begin{table}
		\caption[]{Definition of band edges for continuum and features.}
		\label{tabla3}
		\begin{center}
		\begin{tabular}{ccc}
		\hline
		Feature Name & Band passes ($\mu$m) & Reference \\ 
		\hline
		\hline
		NaI feature & 2.204 - 2.211 &  R97\\
		NaI continuum \#1 & 2.191 - 2.197 &R97\\
		NaI continuum \#2 & 2.213 - 2.217 &R97\\
		CaI feature & 2.258 - 2.269 & R97\\
		CaI continuum \#1 & 2.245 - 2.256 &R97\\
		CaI continuum \#2 & 2.270 - 2.272 &R97\\
		$^{12}$CO feature & 2.291 - 2.302& F01\\
		$^{12}$CO continuum \#1 & 2.230 - 2.237 &F01\\
		$^{12}$CO continuum \#2 & 2.242 - 2.258 & F01\\
		$^{12}$CO continuum \#3 & 2.268 - 2.279 & F01\\
		$^{12}$CO continuum \#4 & 2.284 - 2.291 & F01\\
		\hline
		\end{tabular}
		\end{center}
		References - R97: Ram\'{\i}rez et al. (1997) and F01: Frogel \mbox{et al. (2001)}
	\end{table}

	The estimation of the metallicity from Frogel et al. (2001) is based on spectra with resolution of R$\sim$1500, while that of our data is R=500. There is a dependence of the equivalent width on resolution, and so to use Frogel's calculations, we must ensure that our measurements are compatible with those at higher resolutions.

	To do this, we use synthetic spectra from the NextGen grid (Hauschildt et al. 1999). We can degrade the models to the two resolutions to see how the equivalent widths of the features of interested compare.

	\begin{figure}[!h]
		\centering
		\resizebox*{8cm}{7cm}{\epsfig{file=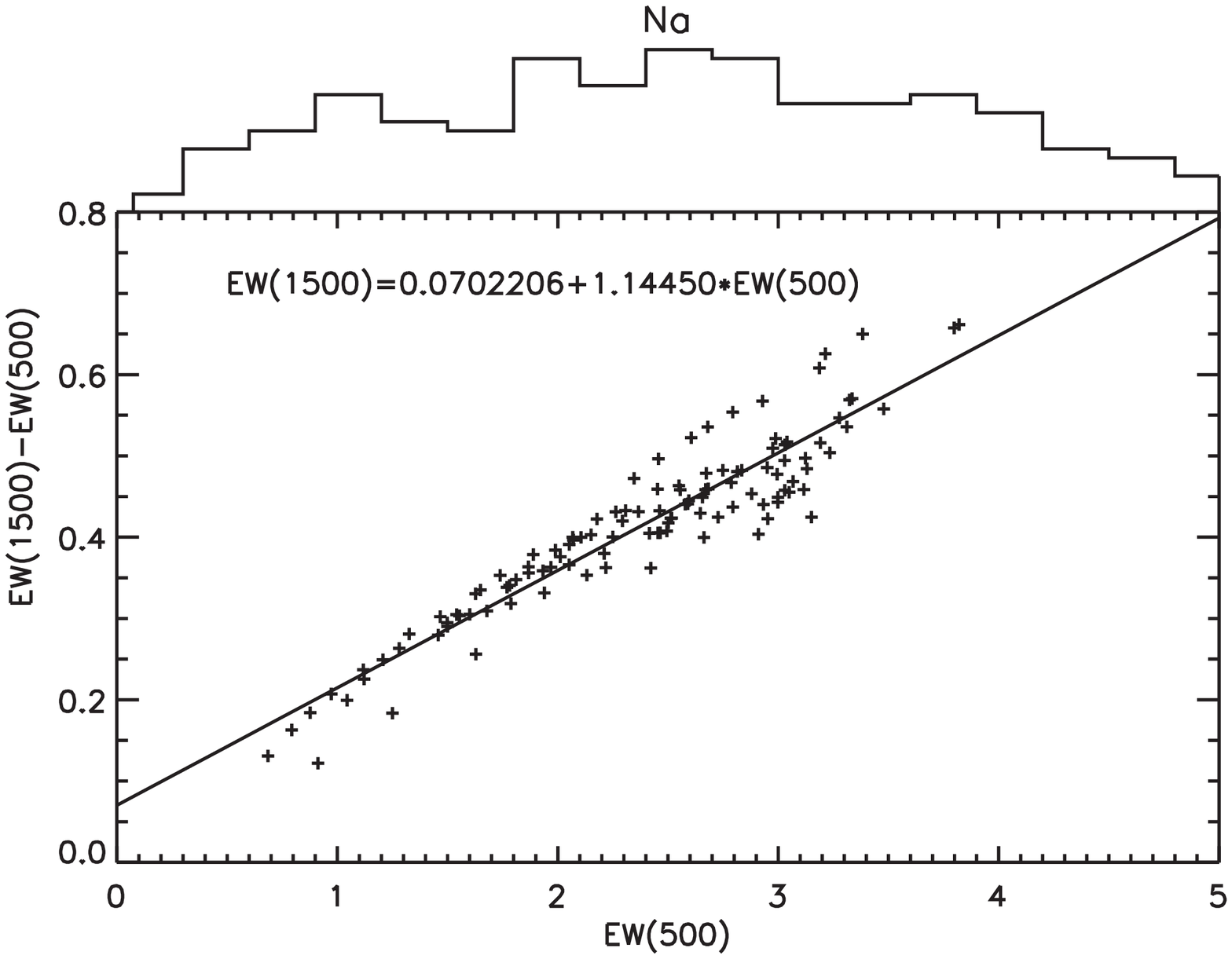}}
		\resizebox*{8cm}{7cm}{\epsfig{file=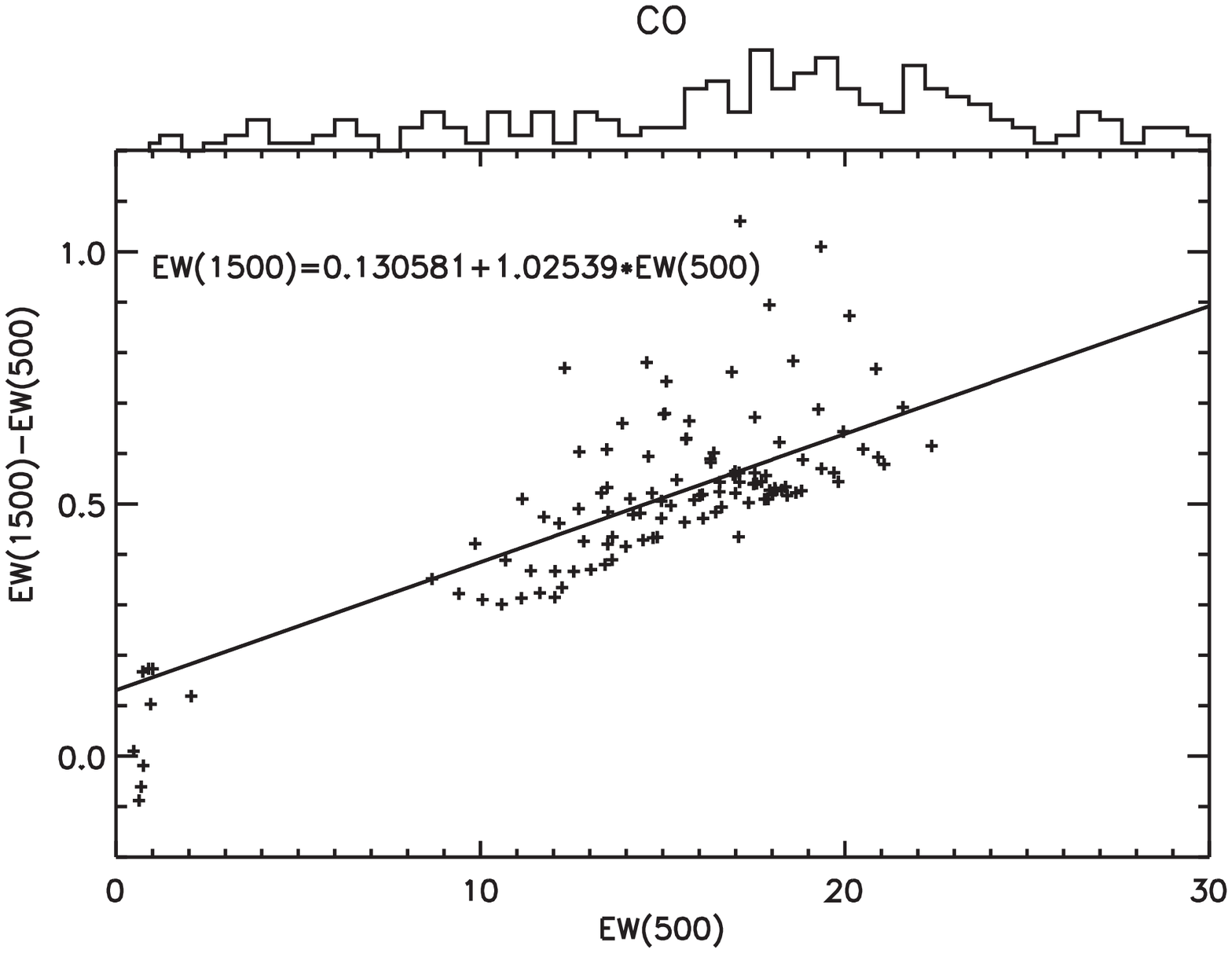}}
		\caption{Relation between the equivalent widths obtained for R=1500 and R=500. Above each plot, an histogram with the distribution of all our measurements, serendipitous and target stars, is plotted.}
		\label{cal_anch}
	\end{figure}

	As can be seen in Fig. \ref{cal_anch}, both the CO and Na features measured at R=500 seem to be linearly related to those derived at R=1500, at least for our range of values, and so we use a linear transformation prior to the application of the Frogel et al. (2001) expressions. Unfortunately, although the same kind of relation should be expected, nothing can be said about the behaviour of the Ca feature. The ranges used to derive this equivalent width are influenced by more metallic lines and molecular bands (for example Ni, Sc and the CN band, see Vanhollebeke et al. 2006 for a detailed list and discussion) than the ones for the other two features, thus making the modelling and extrapolation between resolutions harder. As a result, over the range of metallicities computed for the NextGen library (a grid at [M/H]=0.0, -0.3, -0.5, -0.7), the equivalent widths derived are restricted to EW(Ca)$<$3, while our sample distribution peaks at EW(Ca)$\sim$5. Taking this into account, we have chosen not to perform any transformation over this feature.

	Even so, as Frogel et al. (2001) discuss in their paper, the Na feature presents the heavier dependence with metallicity. Since when measuring equivalent widths with lower resolutions the presence of unresolved metallic lines that affect both the continuum and the object line itself seems to be the major effect to take into account, we expect that the correction for the Ca feature would be of the order of, if not smaller than, that for Na. These values, according to the calculations of the aforementioned paper, would translate into an uncertainty of around 10\% in the estimation of the metallicity.

\section{Data analysis}
\subsection{Sample filtering}
\label{samfil}

	In our sample of stars, all the targets were supposed to be K and M giants according to their location in the CMDs (Fig. \ref{CMDs}), but nothing can be said a priori about the nature of the serendipitous detections that we also plan to use. Comparing the relation between equivalent widths and temperature for giant and dwarf stars, in Ram\'{\i}rez et al. (1997) the authors define a parameter, which we will refer as $lg$ in this work, that seems to be a powerful luminosity indicator:

	\begin{equation}
		lg=\log\left(\frac{EW(CO)}{EW(Na)+EW(Ca)}\right)
	\end{equation} 

	With this parameter, it is possible to distinguish between giant and dwarfs over the temperature range between 3400 and 4600 K (see Fig. 11 in Ram\'{\i}rez et al. (1997)). However, it cannot be used to separate other populations, such as  supergiants and red giants (Schultheis et al. 2003). To avoid interference from such sources in our study, we use Blum et al.'s (2003) classification, which, based on two indexes that measure the strength of the CO and H$_{2}$O features, can disentangle giants from supergiants and long period variables (such as Miras). For a certain feature M, the strength is obtained with the formula:

	\begin{equation}
		M=\left(1-\frac{F_{M}}{F_{cont}}\right)\cdot100
	\end{equation} 

	Where $F_{M}$ and $F_{cont}$ are the fluxes of the feature and the continuum. The CO feature is a 0.015 $\mu$m band centred on 2.302 $\mu$m, while its continuum is derived at an identical interval with its centre in 2.284 $\mu$m. The calculation of the H$_{2}$O index requires a bit more complicated procedure: a continuum is obtained with a quadratic fit to two intervals, one of them in the H-band of the spectra, at (1.68,1.72) and (2.20,2.29) $\mu$m and then this fit is compared to a band 0.015 $\mu$m wide at 2.0675 $\mu$m. It has to be noted that these indexes are not equivalent widths, and thus they are expressed in percentile units.

\begin{figure}[!h]
		\centering
		\resizebox*{8cm}{6cm}{\epsfig{file=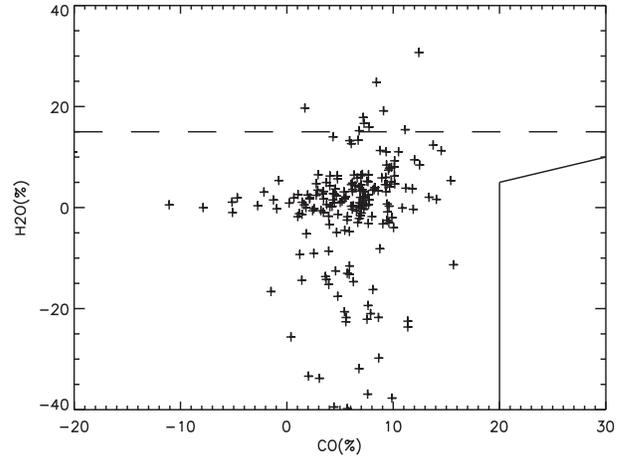,height=6cm}}
		\caption{Distribution of H$_{2}$O and CO for our sample of stars. Solid lines mark the supergiant region, and the dashed one the threshold for LPV stars. We have found 10 (4.7\% of the total) of these in our sample.}
		\label{blum}
	\end{figure}

	When comparing the strengths obtained at R=500 with those at R=1500, differences lie below 5\% for our range of values. In their study, Blum and collaborators find that all the LPV (long period variable) stars on their sample have large values of water absorption, while supergiants have stronger CO lines than giant stars, and verify the relation H$_{2}$O$<$-5+0.5$\times$CO. As Fig. \ref{blum} shows, there are no supergiants in our sample, which was to be spected since we have selected our stars from CMDs, and given the intrinsic magnitude of these population, any supergiant lying between us and the central 4 kpc of the Galaxy would appear conspicuously bright.

	Ram\'{\i}rez et al. (1997) establish a set of thresholds for their calculated parameter, $0.3 < lg < 0.7$; within these boundaries, stars are considered to be giants. Proper allowance should be made to account for observational errors, which yield an uncertainty for $lg$ of $\pm0.05$, derived from the dispersions of the equivalent widths. In order to do so, we could end upwith a cut for our giant region of $0.35 < lg < 0.65$, a more conservative range that allows us to ensure cleanliness in the filtered sample, but that could exclude stars near the Ram\'{\i}rez's boundary. In contrast, if we want to be sure to include these stars, we may use a wider range, $0.25 < lg < 0.75$, but again we would be contaminating our sample with stars that genuinely lie outside the range of applicability of Ram\'{\i}rez et al.'s (1997) study.

	To solve this problem, we do a clustering analysis over the ([Fe/H],lg) space (for details on [Fe/H] calculations, see Sect. 5.3). Since the method for calculating the metallicity is only applicable to giant stars, those that are not are more likely to have extreme values of [Fe/H]. Thus we have a way to point at those stars, at least a fraction of them, that our conservative threshold in lg has put again into the sample.

	An easy way to perform this study is to do a hierarchical clustering analysis over our sample. We choose to use Ward's clustering method (Ward 1963), an iterative procedure that, for a sample of n datapoints, starts with a partition of the data $P_{n}$, consisting of n clusters of one datapoint each.  In each successive step, two of the remaining clusters merge, until a partition $P_{1}$ is reached, in which all data points fall within one cluster.

	This merging is performed minimising a merit function, the total square sum of errors. This function (called SSE) is calculated over each cluster as:

	\begin{equation}
		SSE\left( C_{n} \right) = \sum_{i=1}^{n}\left(x_{i}-\frac{1}{n}\sum_{j=1}^{n}x_{j}\right)^2
	\end{equation} 
	
	For a cluster with n points. Ward's method searches for the new partition $P_{i}$ which has  the minimum sum of SSE of all the possible partitions containing i clusters. The sum of SSE is a measure of the loss of information resulting from swapping the original $P_{n}$, in which each cluster is one datapoint and that has a SSE equal to zero, with a new distribution of $i$ points, each one of them being the mean of the original datapoints assigned to the $ith$ cluster. 

	\begin{figure}[!h]
		\centering
		\resizebox*{8cm}{6cm}{\epsfig{file=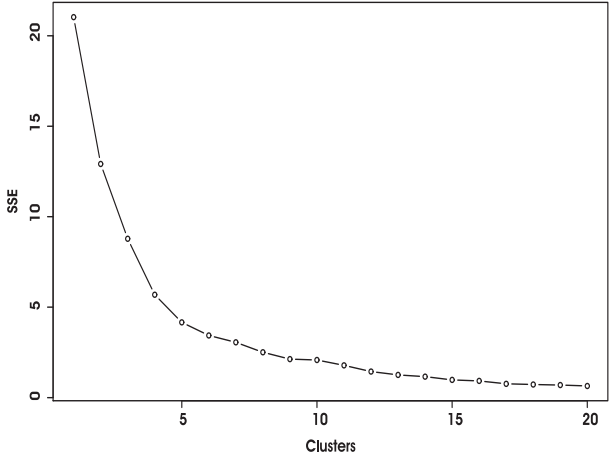,height=6cm}}
		\caption{SSE variation with the number of clusters for our dataset.}
		\label{sse}
	\end{figure}

	With this in mind, we can set up a method for choosing the optimal partition for our dataset. Since we are working with a more or less continuous distribution, the SSE will decrease monotonically with the number of clusters, as can be seen in Figure \ref{sse}. Too large a number of clusters would mean that we are partitioning our date according to small scale differences, probably within observational or statistical errors, and too small a number of clusters would not show the inherent structure of the dataset, if there is one. 

\begin{figure}[!h]
		\centering
		\resizebox*{8cm}{6cm}{\epsfig{file=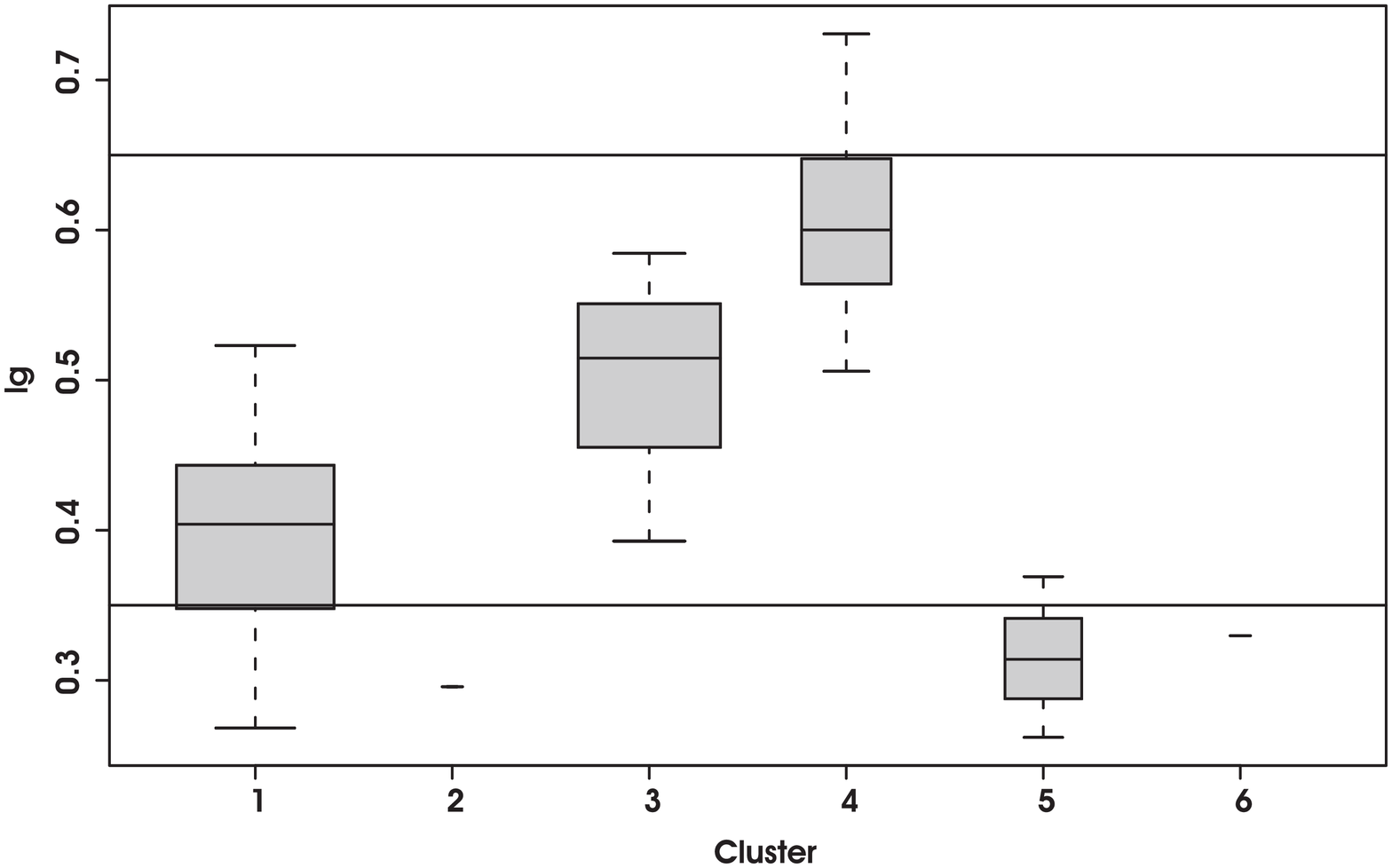,height=6cm}}
		\caption{Distribution of lg for the resultant clusters. In the box plot each box's sides represent the first and third quartiles, its width varies proportionally to the number of sources within the considered longitude, the central line marks the median and the whiskers account for the maximum and minimum, in absence of outliers. The horizontal lines mark Ram\'{\i}rez's lg limits.}
		\label{boxlg}
	\end{figure}

	As a rule of thumb, the optimal partitioning corresponds to the elbow of the SSE function, which in our case corresponds to a partition with six clusters.\footnote{Actually, there is no practical difference, for our purposes, between between choosing five or six clusters, since this step only breaks cluster 1 in Fig. \ref{boxlg} into two groups well within our limits.} Since it is also a measure of the mean interdistance between group members, we can see that with $i>5$ the difference in SSE between $P_{i}$ and $P_{i-1}$ with each new partition is smaller (except for some fluctuations), since we are building up clusters that lie next to one another, whereas with $i<5$, the variations in the SSE with i are bigger, since we are building clusters that really represent separate structures.

	\begin{table}[!h]
		\caption[]{Number of elements and mean and dispersion of the metallicity and lg for each one of the clusters.}
		\label{tabclus}
		\begin{center}
		\begin{tabular}{cccccc}
		\hline
		Cluster& N & $<[Fe/H]>$ & $\sigma_{[Fe/H]}$ &$<lg>$ &$\sigma_{lg}$\\
		\hline
		\hline
		1 & 59 & -0.16 & 0.16 &  0.40 & 0.06 \\
		2 & 1 & 1.61 & -- &  0.30 & -- \\
		3 & 48 & -0.46 & 0.16 &  0.50 & 0.05 \\
		4 & 19 & -0.87 & 0.23 &  0.61 & 0.06 \\
		5 & 14 & -0.76 & 0.28 &  0.31 & 0.03 \\
		6 & 1 & -3.12 & -- &  0.33 & -- \\
		\hline
		\end{tabular}
		\end{center}
	\end{table}

	As can be seen in Table \ref{tabclus} and Fig. \ref{boxlg}, three of the six clusters fall inside the boundaries in lg, and thus we use clusters 1, 3 and 4 for our analysis.\footnote{Although Table \ref{tabclus} contains the final sample of stars, both targets and serendipitous detections, the total number of stars in it   is significantly lower than in Table 1. This is so because since the latter contains all the target stars, while the former details only those spectra with derivable metallicity, hence the fainter objects, those with unmeasurable lines, etc., have already been taken out; this loss of objects is greater than the added number of serendipitous spectra, since there is only a small fraction of random detections that meets all the requirements (S/N ratio, lg parameter, etc.) of the filtering process.}

	Using only a sample of late-type giants serves a dual purpose: first, this enables us to use Ram\'{\i}rez's relations between line strengths to obtain metallicities; second, working with stars that are at a similar evolutionary stage ensures that any trend or relation that we can derive from them in fact reflects the underlying behaviour of the structures they belong to, and not the different chemophysical characteristics of different evolutionary stages of a single stellar population.

\subsection{Identifying spectral types}

	In Ram\'{\i}rez et al. (1997) we can find an exhaustive discussion of the dependence of our selected spectral features (CO, NaI and CaI) with temperature and luminosity (hence, the spectral type) of the stars. They modelled the relationship between the effective temperature and MK spectral type by a second-order polynomial fit of the form:

	\begin{equation}
		T_{eff}= 4565.3 - 142.8 \times G + 3.8 \times G^2
		\label{tCO}
	\end{equation}

	Where G is an integer assigned to each spectral type, in a way that  G=0 corresponds to a K0 star and G=13 to a M7 giant star (so K5=5, M0=6, etc.). This relationship, valid in the range \mbox{3360 K $<$ T$_{eff}$ $<$ 4870 K}, was obtained after making an extense compilation of T$_{eff}$ measurements from different researchers (see details in \S3 of Ram\'{\i}rez et al. 1997).

	Converting the spectral types of their sample of 43 K0III to M6III stars into $T_{eff}$ by means of this relationship, they obtained a single linear fit to the well-defined dependence on effective temperature of EW(CO) for giant stars that allows the $T_{eff}$ to be calculated with an uncertainty of $\pm$150 K. 

\begin{figure}[!h]
		\centering
		\resizebox*{8cm}{6cm}{\epsfig{file=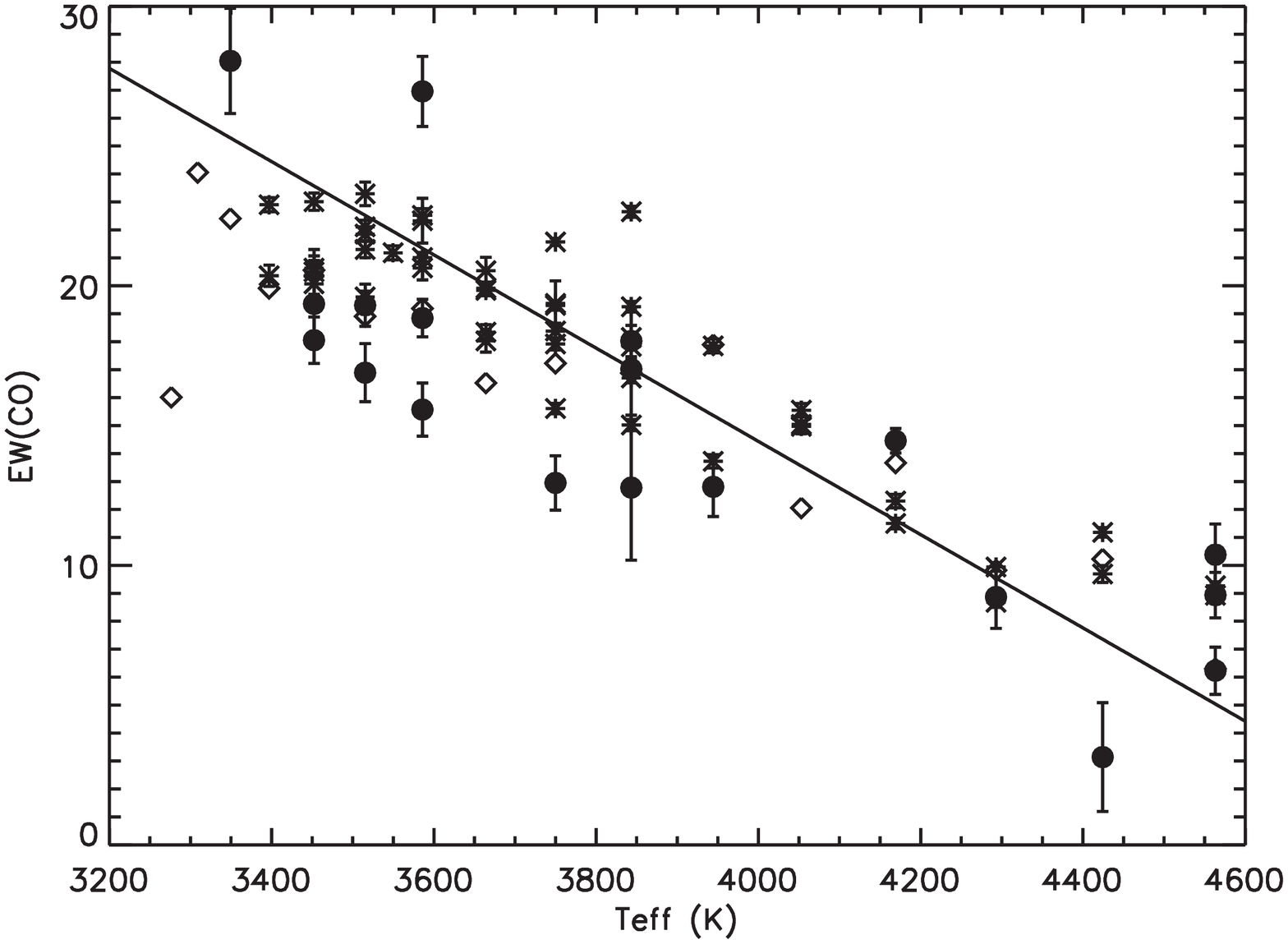,height=6cm}}
		\caption{Relation between effective temperature and equivalent width of the CO features for the standard stars of our sample (filled circles), the data from  Ram\'{\i}rez et al. (1997) (asterisks) and the calculations for Pickles' library stars (diamonds). The solid line represents the linear fit for the whole of the data.}
		\label{COTeff}
	\end{figure}

	We have repeated this analysis with our control sample of standard stars, with a well-known spectral type, and the derived relationship between EW(CO) and T$_{eff}$ for giant stars has been fitted by a linear fit,

	\begin{equation}
		T_{eff}=(4895\pm130)-(62\pm7) \times EW(CO)
		\label{gCO}
	\end{equation}

	which is consistent within the uncertainties with the relationship obtained by Ram\'{\i}rez et al. (1997). The fitted polynomial and the CO equivalent widths of the standard stars are shown in Figure \ref{COTeff}. The Ram\'{\i}rez et al. sample is also plotted for comparison.

\begin{figure}[!h]
		\centering
		\resizebox*{8cm}{6cm}{\epsfig{file=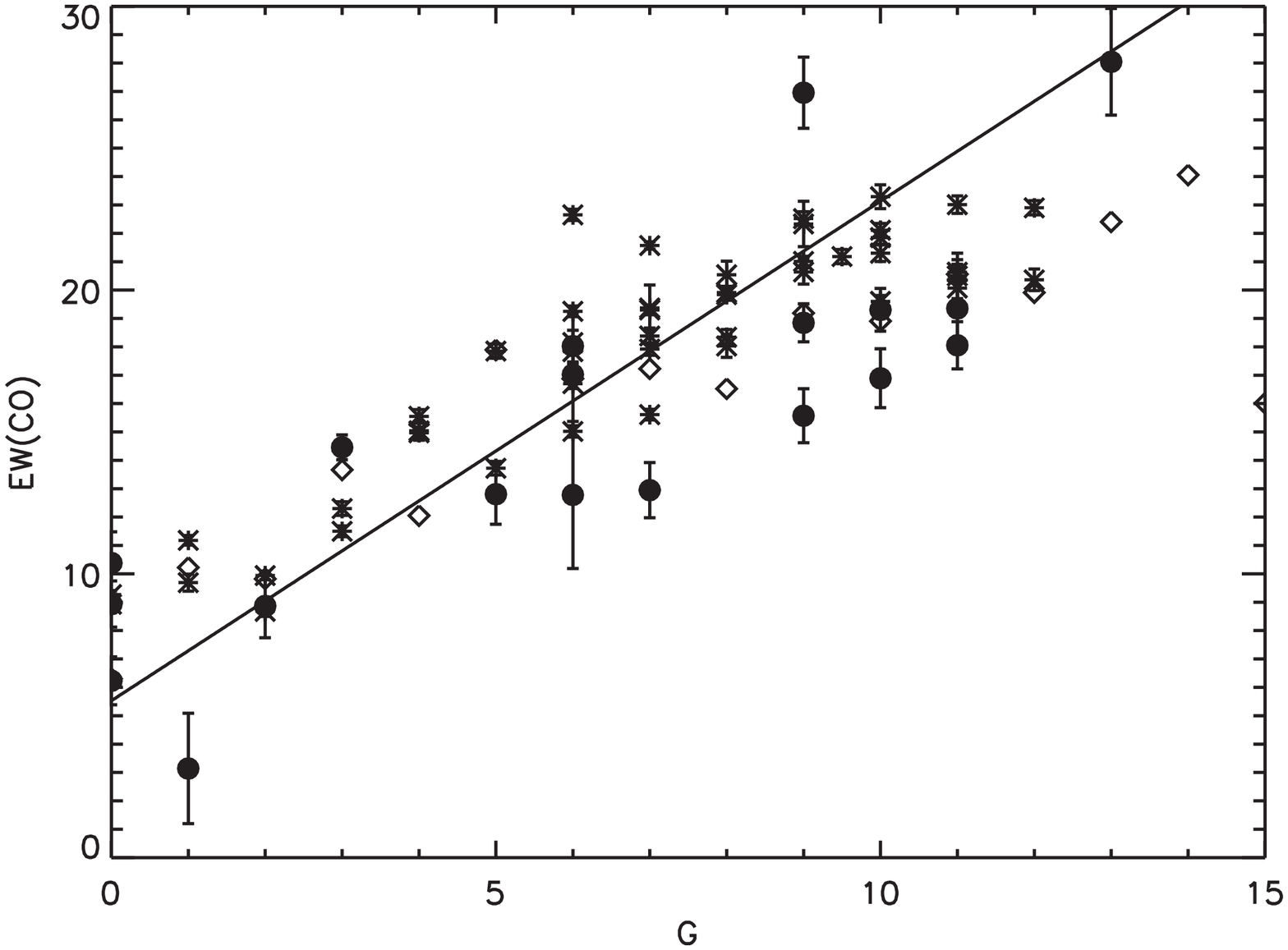,height=6cm}}
		\caption{Relation between the equivalent with of the CO bandhead and the spectral type (defined by the integer number G) for the standard stars of our sample (filled circles), the data from  Ram\'{\i}rez et al. (1997) (asterisks) and the calculations for Pickles' library stars (diamonds). The solid line represents the linear fit for the whole of the data.}
		\label{comparison}
	\end{figure}

	Once it can be assumed that EW(CO) and T$_{eff}$ are related, we  propose to invert the process and obtain the spectral types of the stars (G) by means of the CO equivalent width. The relationship between both quantities is shown in Fig. \ref{comparison} and can be expressed by a linear fit of the form
	
	\begin{equation}
		G=(0.56\pm0.04)\times EW(CO) - (3.0\pm0.5)
		\label{eqCO}
	\end{equation}
	
\begin{figure}[!h]
		\centering\resizebox*{8cm}{6cm}{\epsfig{file=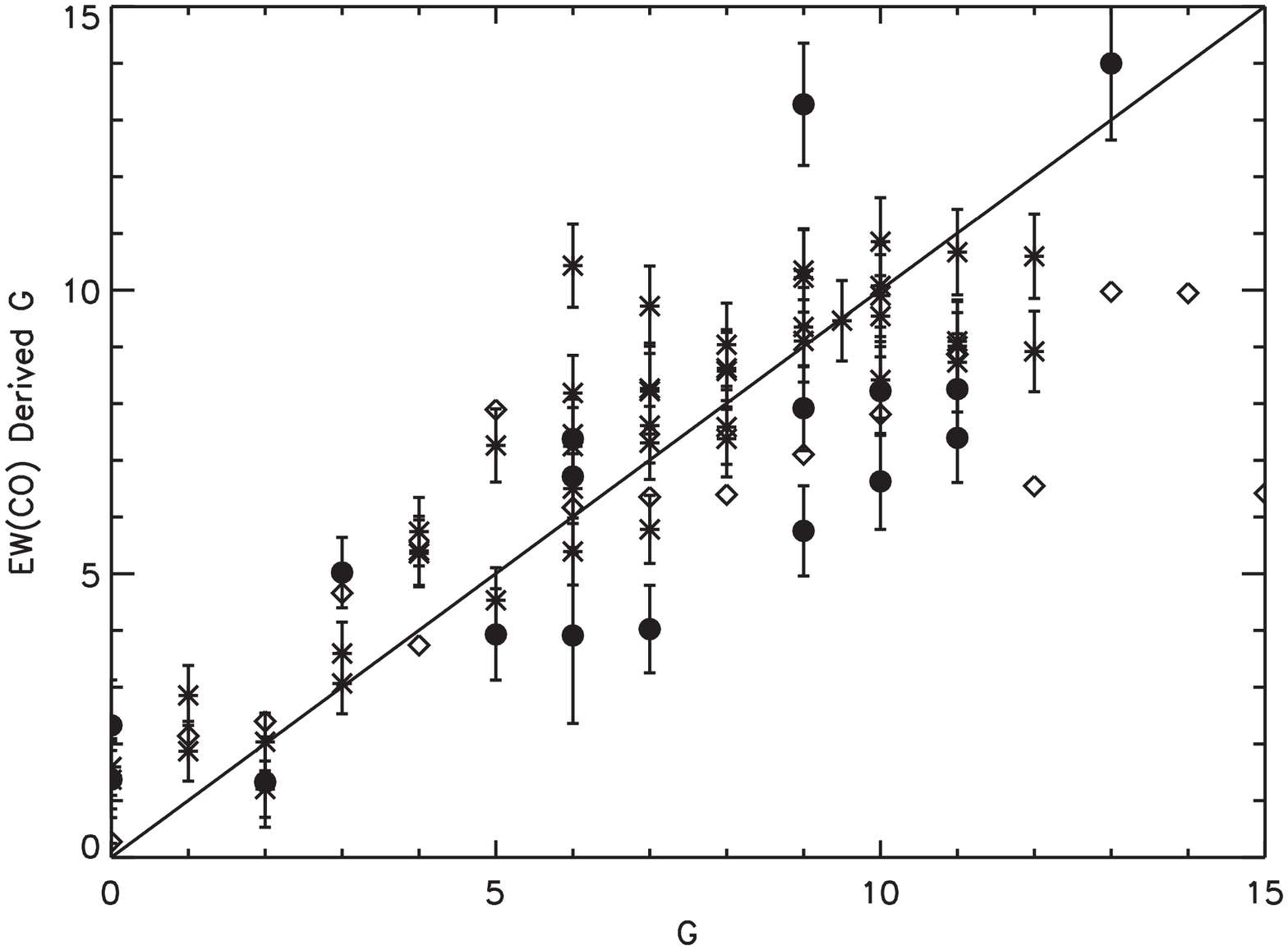,height=6cm}}
		\caption{Comparison between the spectral types of the standard stars, obtained by means of the CO equivalent width, with respect to their known spectral types (both spectral type definitions are described by the integer number G), for the standard stars of our sample (filled circles), the data from  Ram\'{\i}rez et al. (1997) (asterisks) and the calculations for Pickles' library stars (diamonds). The solid line marks the identity.}
		\label{comparisonG}
	\end{figure}

	We have tested this relationship with the sample of standard stars, and compared their spectral types with those obtained using the previous expression. The comparison is plotted in Fig. \ref{comparisonG}, and the standard deviation of the fit gives to a typical uncertainty in G of $\pm$1.2, well enough to identify the spectral type of a given star.

\begin{figure}[!h]
		\centering
		\resizebox*{8cm}{6cm}{\epsfig{file=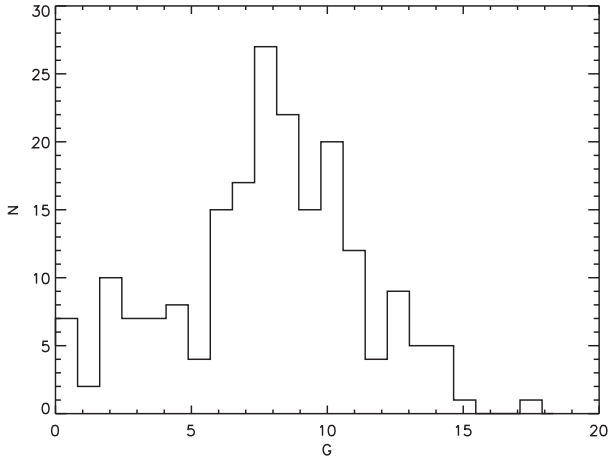,height=6cm}}
		\caption{Spectral type distribution for all observed stars.}
		\label{histoG}
	\end{figure}

	As can be seen in Fig. \ref{comparisonG}, there is good agreement between the empirical estimates based upon the EW(CO) and the tabulated spectral types for those stars, within the error of the estimations. However, for spectral types later than M5 (G=11) there is a larger deviation between the CO-calculated and the tabulated spectral types. This could be explained if we take into account that for the later spectral types there are several factors that affect the line width calculations, mainly the effect of the metallic line blending, which lowers the overall level of the continuum and the contributions of wide molecular bands, such as H$_2$O or CN, that become stronger with decreasing temperature (see for example Vanhollebeke et al. 2006 or Lan\c{c}on \& Wood 2000) and can contaminate the spectral features we are measuring. For values of EW(CO) higher than 23.5$\pm$0.25 nothing could be said for the spectral type of the star, apart from its being later than M5. However, by inspection of  Fig. \ref{COTeff} we observe that these values are the ones that are slightly below the Ramirez et al. relationship (in the lower $T_{eff}$ range), so this can explain the differences observed in Figure \ref{comparisonG}. Assuming those values can be ignored, we have applied eq. (\ref{eqCO}) to the whole sample of stars, obtaining the distribution shown in Figure \ref{histoG}.

\begin{figure}[!h]
		\centering
		\resizebox*{8cm}{6cm}{\epsfig{file=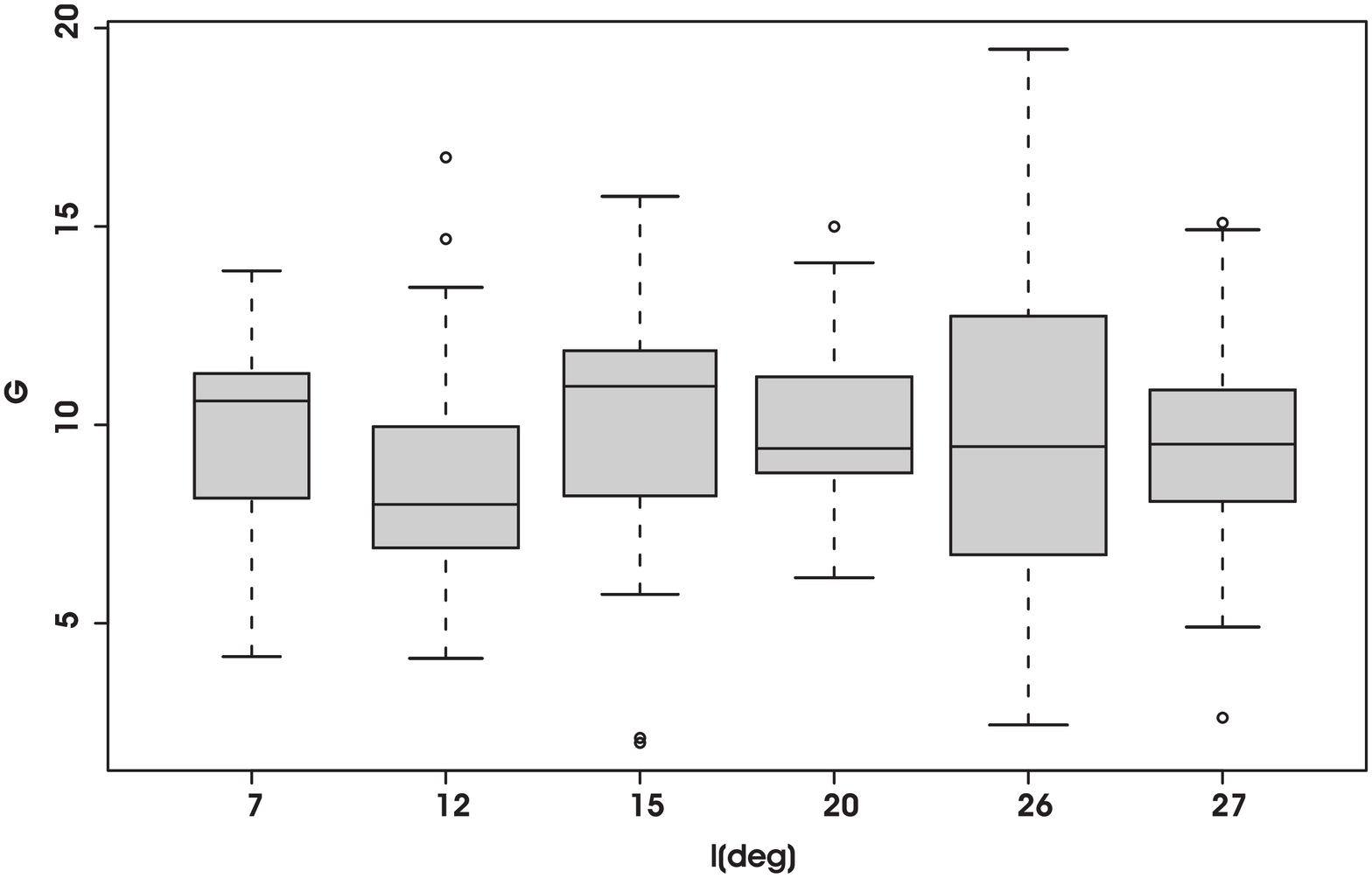,height=6cm}}
		\caption{Distribution (see Fig. \ref{boxlg}) of spectral types (derived from the CO equivalent width) at each considered longitude. Open circles denote the outliers.}
		\label{Gvslog}
	\end{figure}

	As shown in Fig. \ref{histoG}, the majority ($\sim$75\%) of the stars correspond to M spectral types, with $\sim$48\% of M0--4 stars,  $\sim$13\% of M4--6 types. We obtain only 20\% of stars in the sample with K spectral types consistent with their location in the CMDs (Fig. \ref{CMDs}). Little can be said about the distribution of spectral types with galactic longitude (Fig. \ref{Gvslog}), except that fields near the Galactic centre (l=7$^\circ$) seem to have older populations that the outer ones.

\subsection{Metallicity}
	Ram\'{\i}rez et al. (2000) and Frogel et al. (2001) obtained a metallicity scale for luminous red giants based on equivalent width measurements of the CO bandhead, the NaI doublet and the CaI triplet. Their calibration is based on giants in globular clusters for -1.8 $<$ $[Fe/H]$ $<$ -0.1. Therefore, we have used the following relation (Frogel et al. 2001):

	\begin{eqnarray}
		[Fe/H]=-1.811 + 0.389 \times EW (Na) - 0.047 \times EW (Na)^2  \nonumber \\
		- 0.030 \times EW (Ca) + 0.024 \times EW (Ca)^2 \\
		+ 0.043 \times EW (CO) - 0.001 \times EW (CO)^2 \nonumber
		\label{metalicidad1}
	\end{eqnarray}
	
	where EW (Na), EW(Ca) and EW (CO) are the equivalent widths of NaI, and CaI and $^{12}$CO (2,0) (defined as in Table \ref{tabla2} while Table B1 gives the derived values). As described in Ram\'{\i}rez et al. (2000) and Frogel et al. (2001), typical errors in $[Fe/H]$ are of the order of $\sim$0.1 dex, while from error propagation with the uncertainties for our equivalent width a value of about $\sim$0.2 dex is derived. There is another possible calibration that incorporates the dereddened $(J-K)$ colour and the absolute K magnitude of the star. However, as can be seen in Table 11 from Frogel et al. (2001), including the quadratic terms for intrinsic colour and magnitude yields  an increase in the accuracy of the calibration of 0.01 dex, as the quadratic solution for the equivalent widths gives a dispersion (when comparing the derived metallicities with independent measurements of each cluster [Fe/H]) of 0.11 dex, while the full solution with 11 terms results in a dispersion of 0.10 dex. Although we could implement the magnitude dependence in our calculations, the improvement of the result would be a meager gain, still well below the estimated error in metallicity. Even more, to obtain absolute magnitudes we need to estimate the interstellar extinction. While it can be derived from the CMDs, there is an intrinsic error associated with this procedure large enough to dampen the benefits of a more accurate calibration.

	We computed the mean value of $[Fe/H]$ and the equivalent widths of the main spectral features for each field by averaging the results of the individual stars. Values obtained are listed in Table \ref{tabfehl}, while the metallicity distribution of the sources with galactic longitude is shown in Fig. \ref{boxfeh}.
	
	\begin{table}[!h]
		\caption[]{Number of measures and mean and dispersion of the metallicity at each Galactic longitude.}
		\label{tabfehl}
		\begin{center}
		\begin{tabular}{cccc}
		\hline
		Galactic Longitude& N & $<[Fe/H]>$ & $\sigma_{[Fe/H]}$\\
		\hline
		\hline
		7 &  13 & -0.32 & 0.34 \\
		12 & 21 & -0.38 & 0.31 \\
		15 & 23 & -0.37 & 0.36 \\
		20 & 24 & -0.32 & 0.18 \\
		26 & 24 & -0.53 & 0.32 \\
		27 & 21 & -0.31 & 0.28 \\
		\hline
		\end{tabular}
		\end{center}
	\end{table}

	\begin{figure}[!h]
		\centering
		\resizebox*{8cm}{6cm}{\epsfig{file=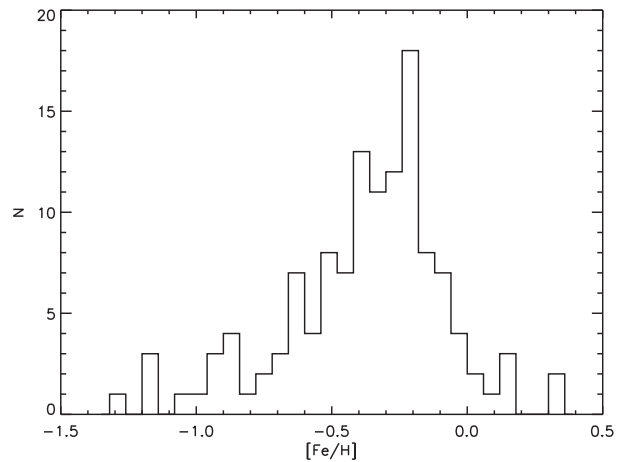,height=6cm}}
		\caption{Metallicity distribution of all sources.}
		\label{metalicidad}
	\end{figure}

\section{Discussion. Metallicity distribution in the inner Milky Way}

\begin{figure}[!h]
		\centering
		\resizebox*{8cm}{6cm}{\epsfig{file=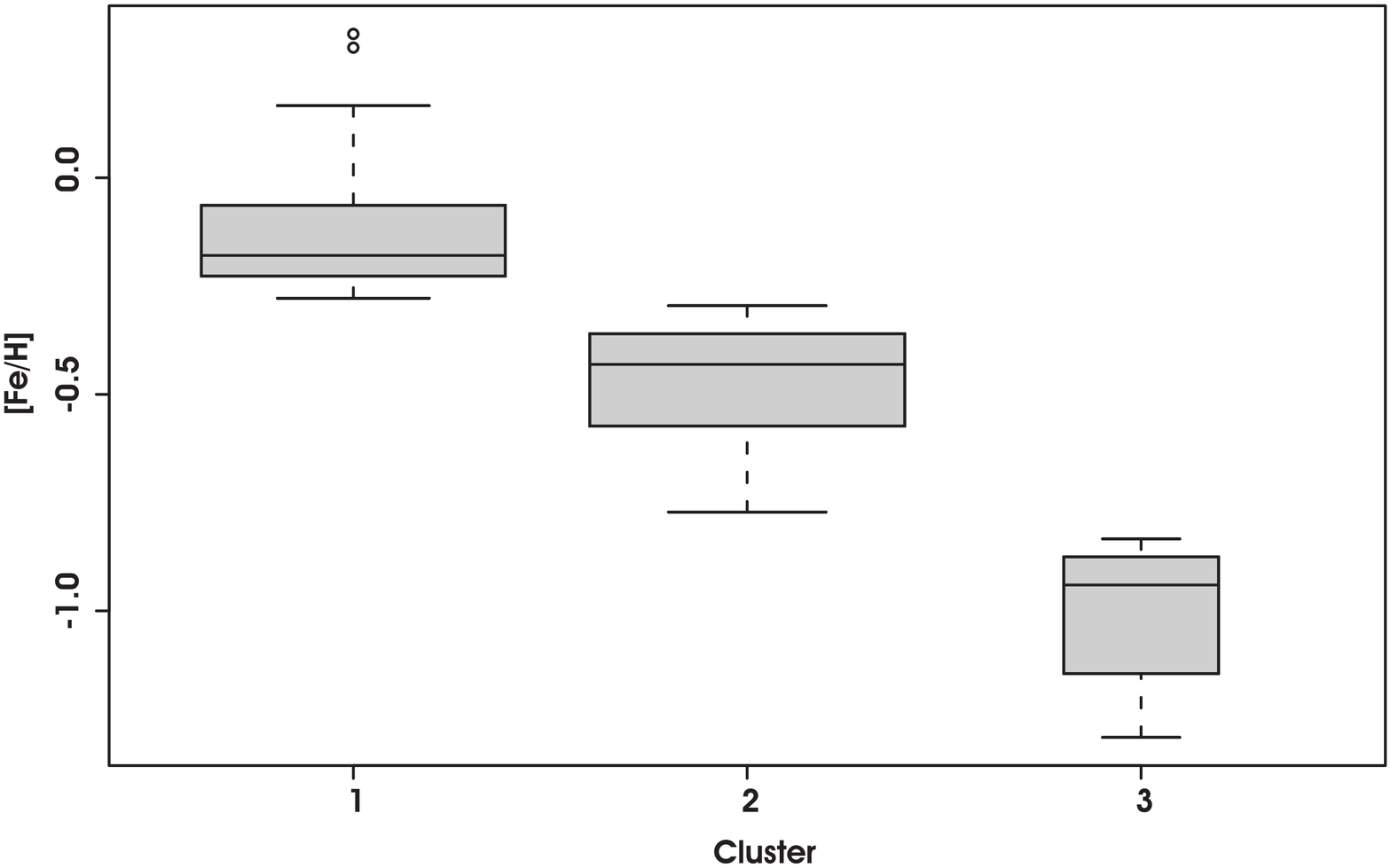,height=6cm}}
		\caption{Metallicity clusters on the final sample.}
		\label{metclus}
	\end{figure}

	We have derived metallicities from a sample of 129 stars at different galactic longitudes. If we repeat the cluster analysis over the distribution in Fig. \ref{metalicidad}, three clear metallicity clusters emerge, as can be seen in Figure \ref{metclus}.

	The first cluster encompasses stars with [Fe/H]$>$-0.28 dex and with a median value of -0.19 dex, which agrees with metallicity estimates for the Galactic bulge (such as those of Molla et al. 2000, or Schultheis et al. 2003). The members of the second cluster verify -0.29$<$[Fe/H]$<$0.77 dex, and present a median value of -0.43 dex, similar to the value that Rocha-Pinto et al. (2006) find for the inner disc ($\sim$-0.38 dex at R=5.5kpc). The third cluster has [Fe/H]$<$-0.83 dex, with a median of -0.94 dex and only 10\% of the stars. A possible relationship of this cluster with the thick disc component cannot be discarded, as its metallicity distribution and the relative frequency of these stars agrees with typical estimates for the local normalization of the thick disc component, about 5--10\% of the thin disc population in the Galactic Plane (e.g.\ Cabrera-Lavers et al. 2005 and references therein).

\begin{figure}[!h]
		\centering
		\resizebox*{8cm}{6cm}{\epsfig{file=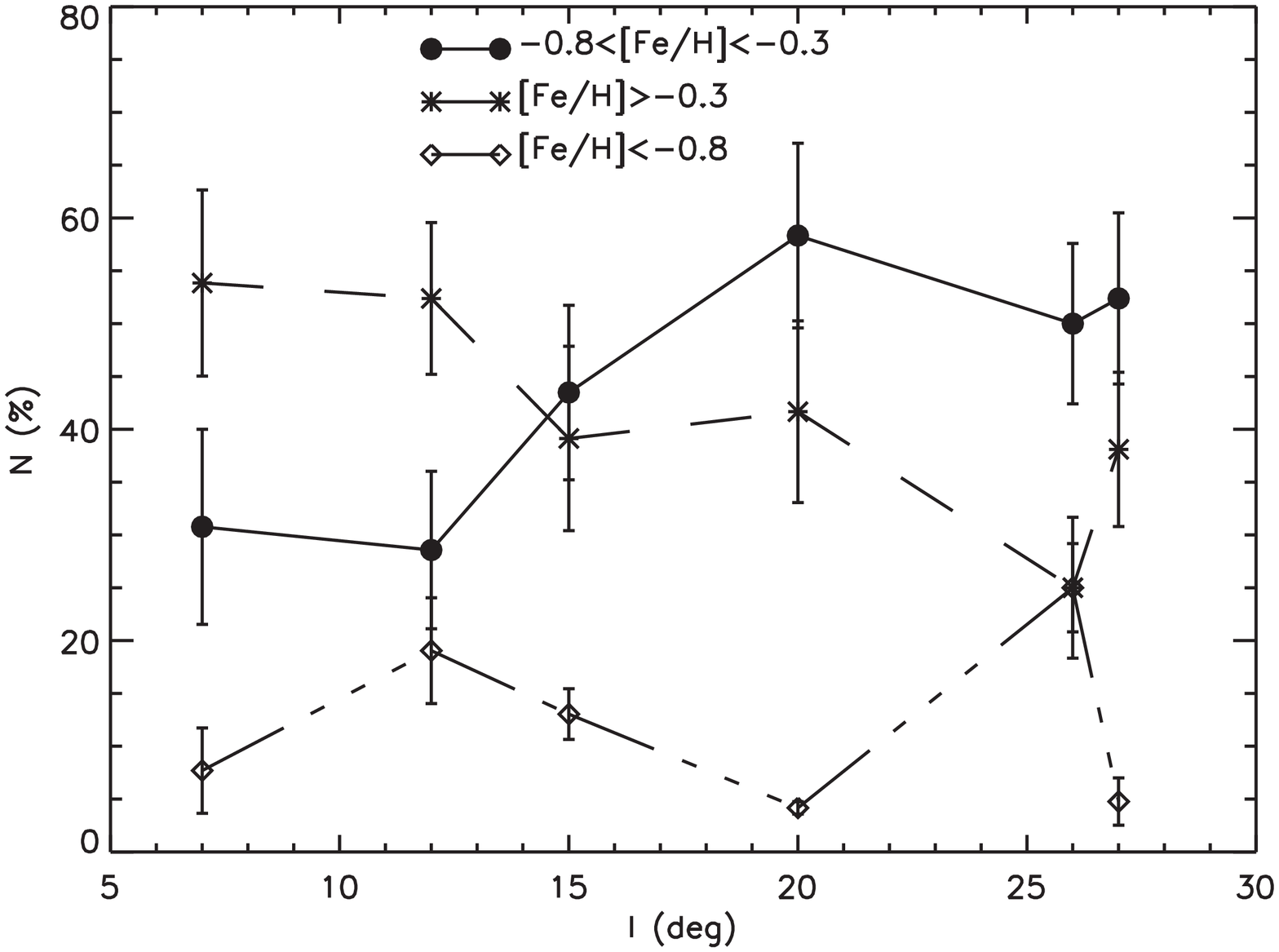,height=6cm}}
		\caption{Fraction of stars pertaining to each cluster along Galactic longitude. Note how there is a significant number of bulge-like stars even at l$>$20$^{\circ}$.}
		\label{cluslon}
	\end{figure}

\begin{figure}
		\centering
		\resizebox*{8cm}{6cm}{\epsfig{file=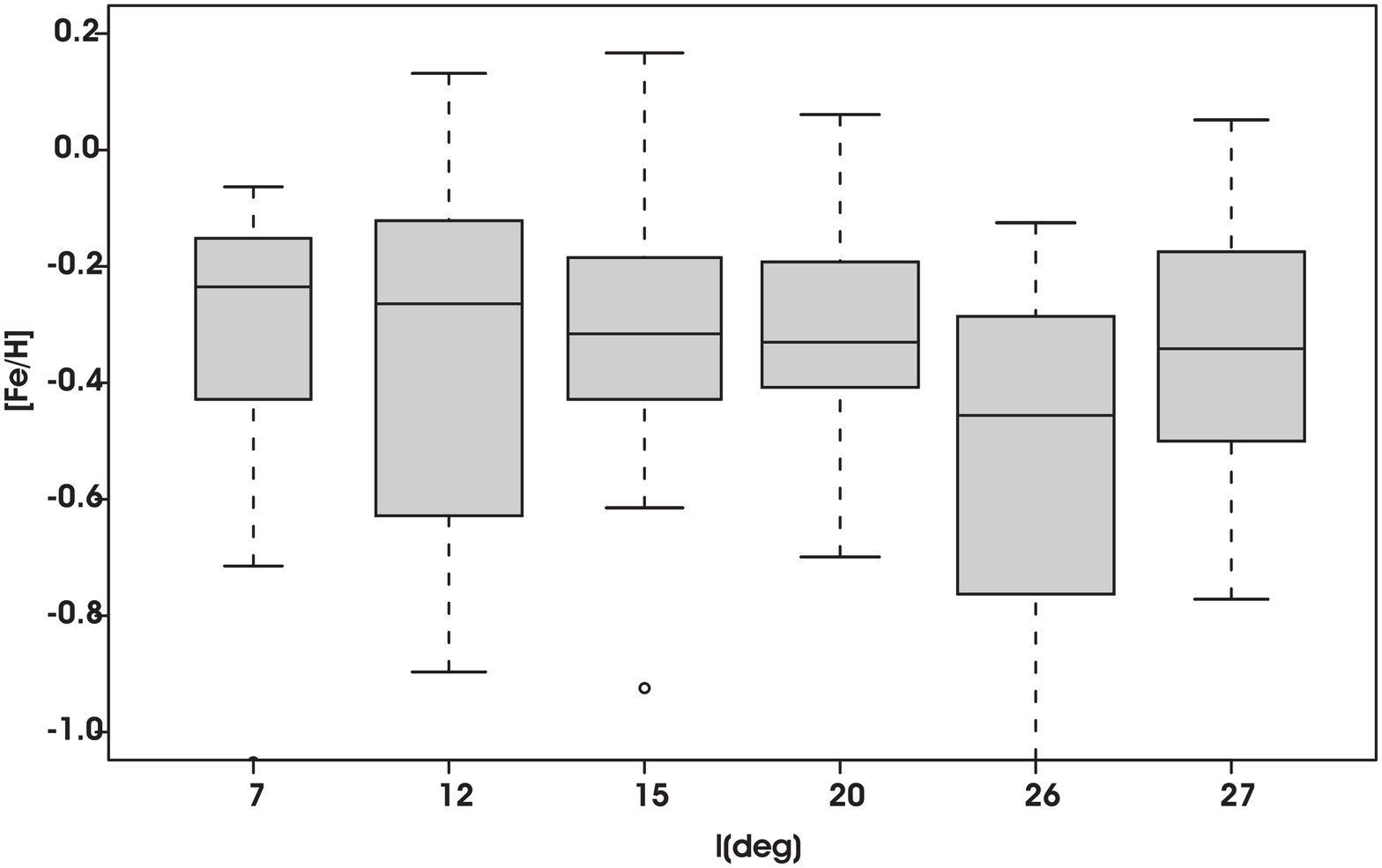,height=6cm}}
		\caption{Boxplot (see Figs. \ref{boxlg} and \ref{Gvslog}) of the final metallicity distribution over the different Galactic longitudes.}
		\label{boxfeh}
	\end{figure}

	We can study the distribution of these groups at different Galactic longitudes. As can be seen in Fig. \ref{cluslon}, bulge-like stars (those with higher metallicities) dominate the inner fields, while disc-like stars (those with [Fe/H] around -0.4 dex) become predominant at higher longitudes. The fraction of stars with the lower metallicities (those of cluster 3) seems to remain more or less constant.

	In the inner parts of the Galaxy there is, as expected, a dominance of bulge-like stars, and as we move away from the centre the disc-like stars become predominant. It should be noted that there is a substantial and continuous presence of these more metallic stars even at l=27$^{\circ}$. It could be argued that we do not penetrate deep enough into the Milky Way, and that at least a significant fraction of these objects are in fact disc stars. But since we are working mainly with a set of objects selected over a CMD (the spurious field stars comprise $\sim$30\% of the final filtered sample, and removing them leads to similar results), and if we take into account that acording to studies such as those of N\"{o}rdstrom et al. (2004) and Rocha-Pinto et al. (2006), the disc reaches metallicities of $\sim$-0.3 dex around a galactocentric radius of 6--6.5 kpc, the magnitude expected for a giant star at 2 kpc from Earth should be bright enough to be left out of our CMD selection.

	The same reasoning can be applied to the less metallic disc-like stars. Even at lower longitudes, there is a presence of these bodies that is again hard to explain by recurring to disc contamination. Therefore, in the inner 5 kpc of our galaxy there should be a mechanism able to convey bulge-like stars outwards and disc-like stars inwards, over the range 7$^{\circ}<$l$<$27$^{\circ}$.

\begin{figure}
		\centering
		\resizebox*{8cm}{6cm}{\epsfig{file=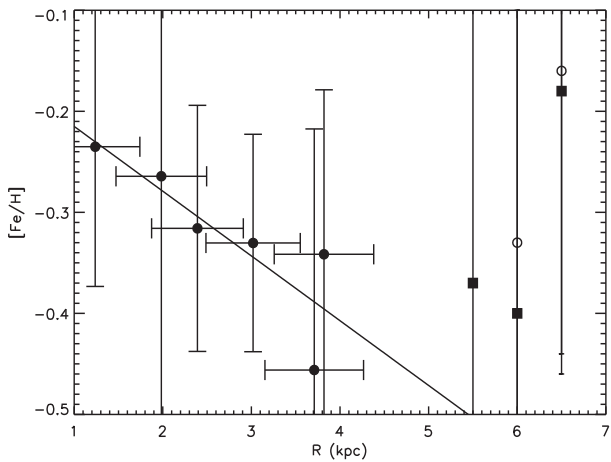,height=6cm}}
		\caption{Variation of the mean metallicity with galactrocentic distance, assuming an angle for the bar of $45^{\circ}$ with the line Sun-Galactic Centre. For our data, we plot with filled circles the median at each longitude. The vertical error bars denote the interquarticular range, and the horizontal ones the width of a 1 kpc bar. The filled squares are data from the Geneva-Copenhagen survey (N\"{o}rdstrom et al. 2004) and the open circles data from Rocha-Pinto et al. (2006). The solid line represents the best linear fit to our data, with a -0.06 dex$\cdot$kpc$^{-1}$ slope.}
		\label{bar45}
	\end{figure}

	If we assume this mechanism to be a long bar forming  an angle of $\sim$45$^{\circ}$ with the line of sight, as proposed by Hammersley et al. (2000), we can assume that at each galactic longitude, the majority of our stars belong to the overdensity that this bar generates. This being so, we can translate our angular coordinates into galactocentric distances, as shown in Figure \ref{bar45}.

\section{Conclusion}
	We have obtained 126 low resolution (R=500) NIR spectra, 80 of selected sources and 46 of serendipitous ones, in six inner in-plane Galactic fields at $l$=7$^\circ$ (13 stars), 12$^\circ$ (21 stars), 15$^\circ$ (23 stars), 20$^\circ$ (24 stars), 26$^\circ$ (24 stars) and 27$^\circ$ (21 stars). From the equivalent widths of the main features of the K band spectra (NaI, CaI and CO bandhead) we have derived the metallicities of the sources by means of the scale obtained by Ram\'{\i}rez et al. (2000) and Frogel et al. (2001) for luminous red giants.

	Our results show how the possible populations present, segregated according to their metallicities, vary along Galactic longitude in a continuous way; this could be indicative of the existence of a component apart from the disc, that could be a Galactic bar, such as that detected in previous works as an overdensity of stars located in the same Galactic coordinates (Hammersley et al. 1994, 2000; Picaud et al. 2003). However, the small-number statistics makes it difficult to be conclusive in this statement, and more work is needed to complete a significant sample of stars in those strongly reddened regions of the innermost Milky Way.

	To overcome this problem, our group is planning to continue the observing campaigns with existing facilities to augment the database. This initial catalogue will form the bulk of the preparatory observing programme of a much more ambitious project aiming at obtaining NIR spectra of a significant sample of the stellar population in the inner Galaxy, the GALEP project, with the forthcoming EMIR instrument (Garz\'{o}n et al, 2006) at the GTC.

	Acknowledgments: The authors would like to thank the anonymous referee for a fruitful and detailed discussion of the topics addressed in this study, which has helped to improve the overall quality of the paper.

\newpage
\begin{onecolumn}
\begin{appendix}

\section{Sample Spectra}
	\begin{figure*}[!h]
	\resizebox*{7cm}{4cm}{\includegraphics{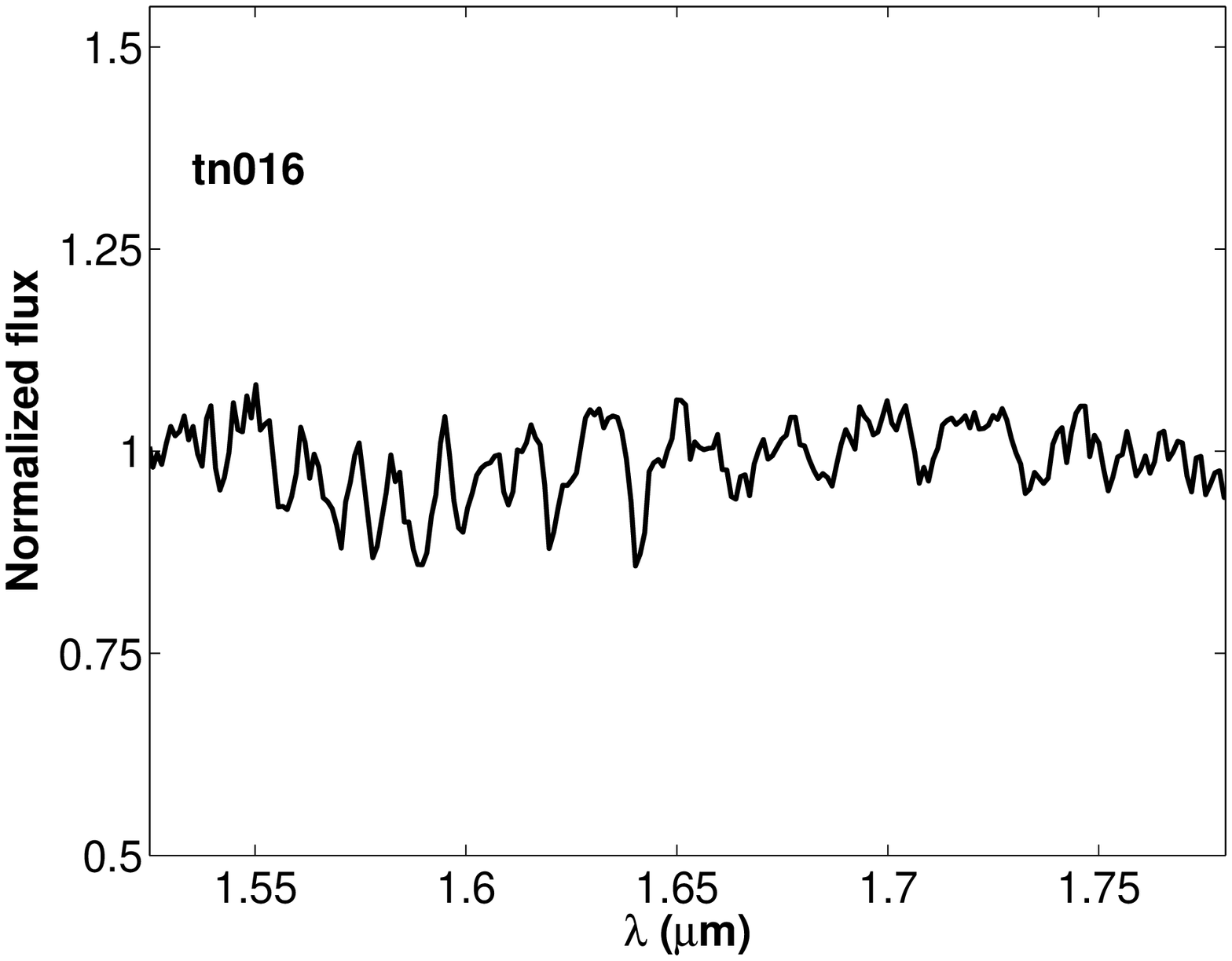}}\resizebox*{7cm}{4cm}{\includegraphics{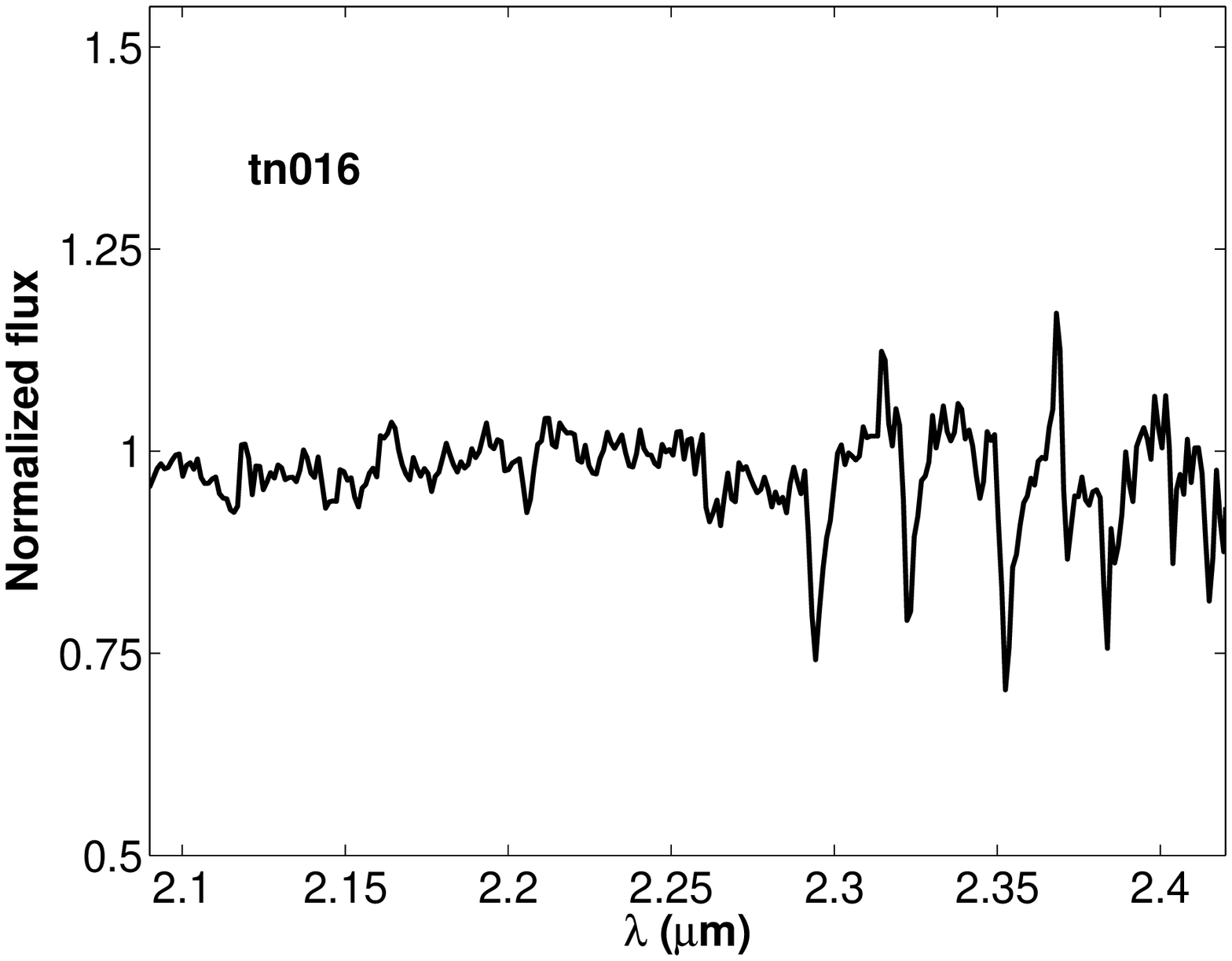}}
	\resizebox*{7cm}{4cm}{\includegraphics{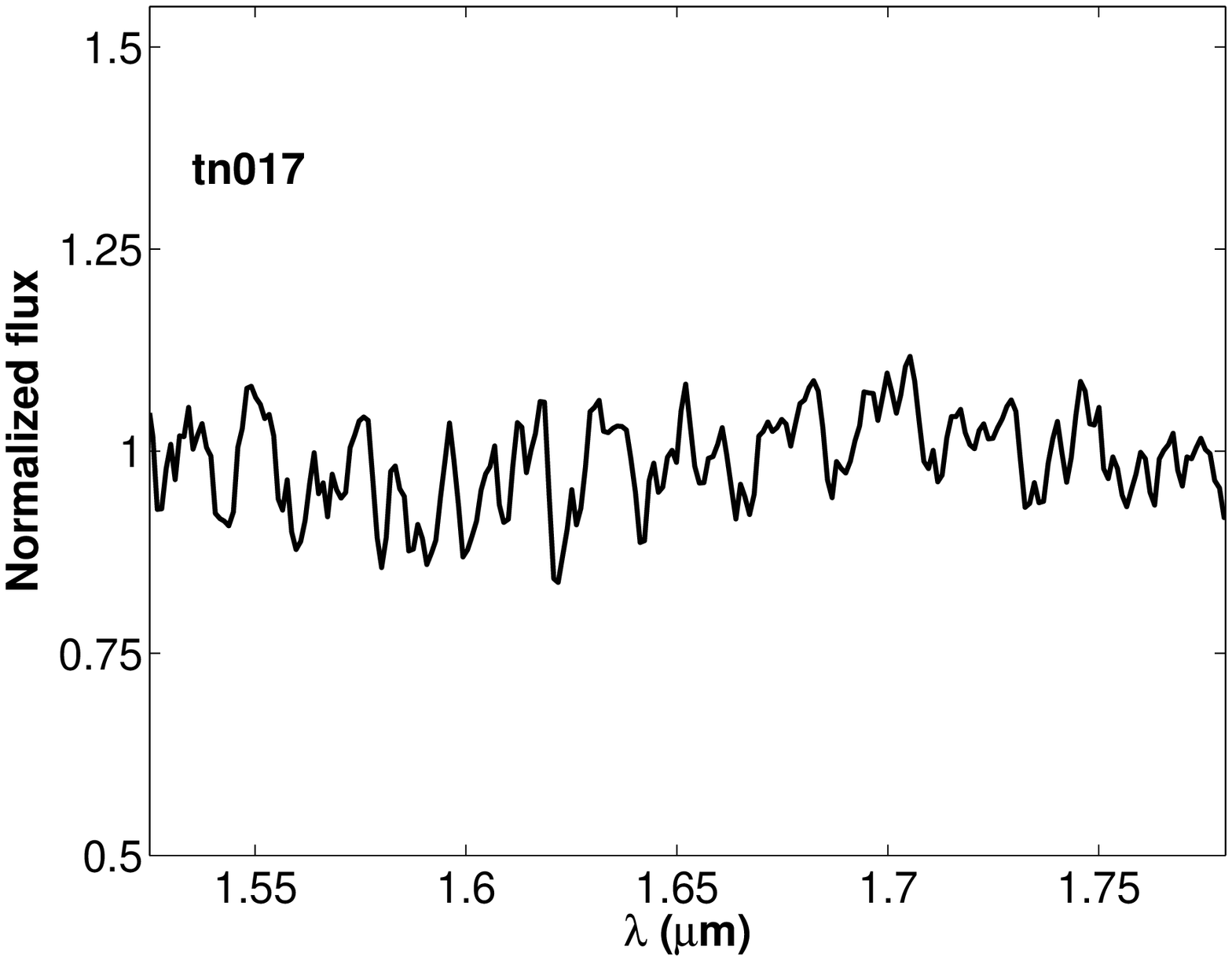}}\resizebox*{7cm}{4cm}{\includegraphics{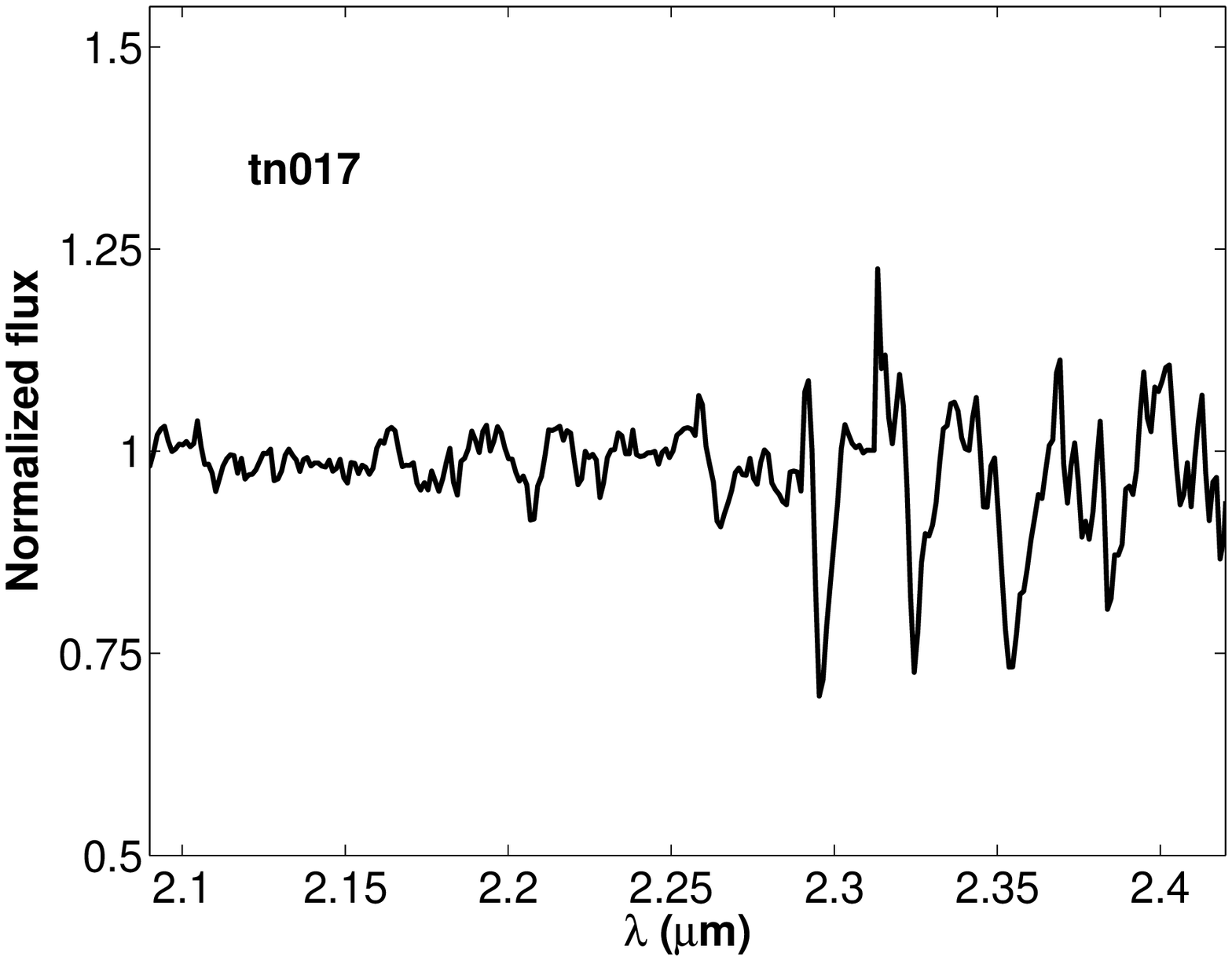}}
	\resizebox*{7cm}{4cm}{\includegraphics{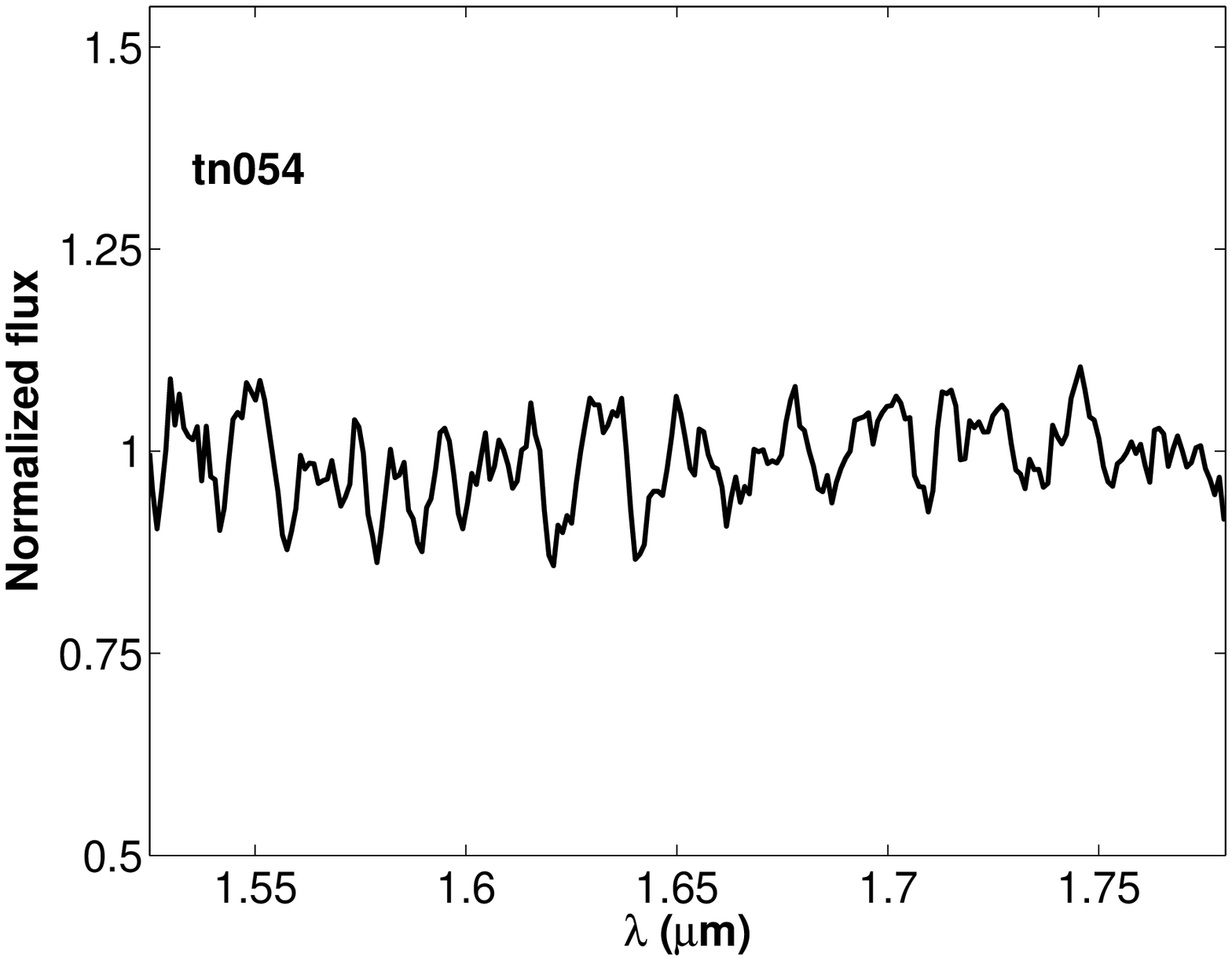}}\resizebox*{7cm}{4cm}{\includegraphics{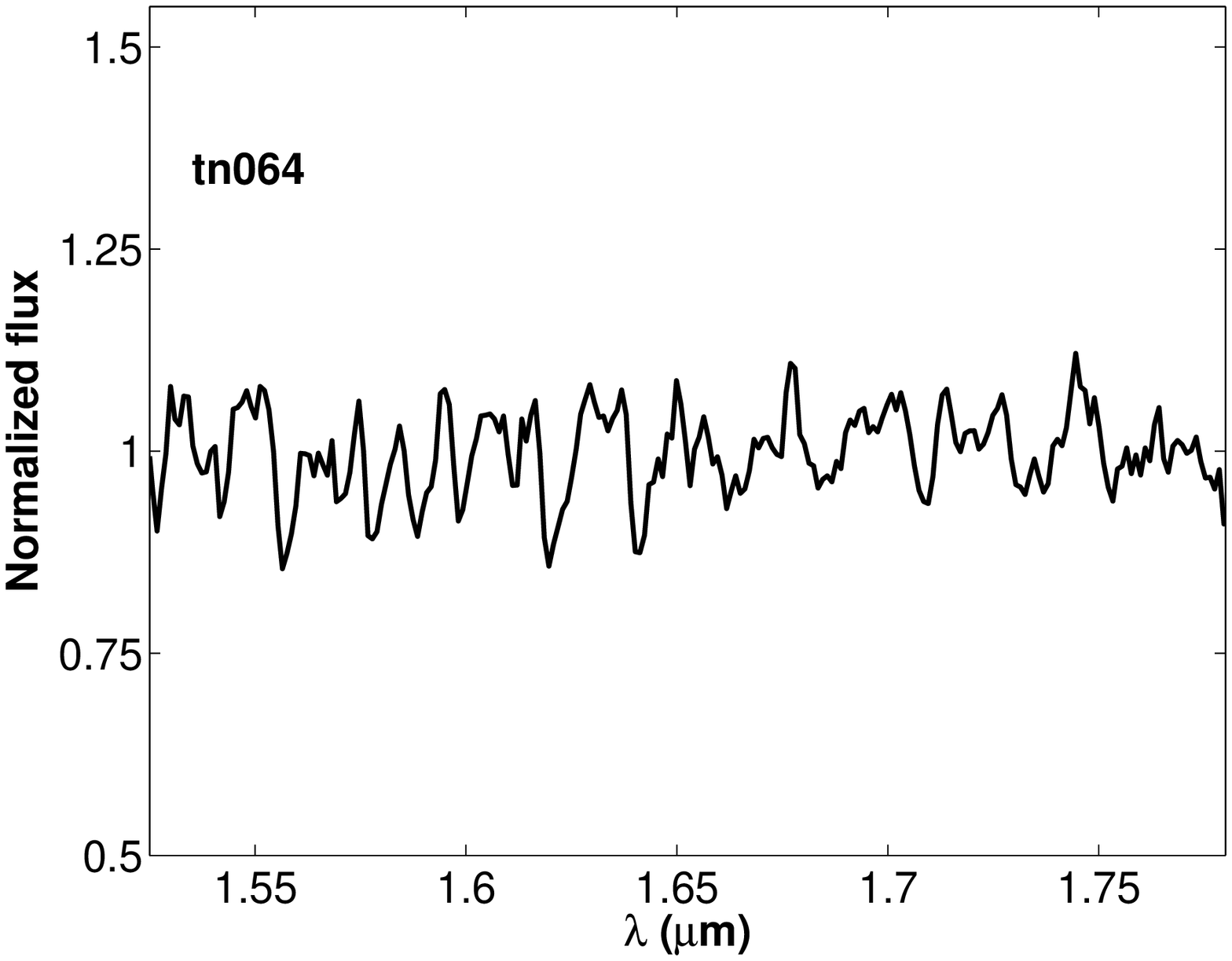}}
	\resizebox*{7cm}{4cm}{\includegraphics{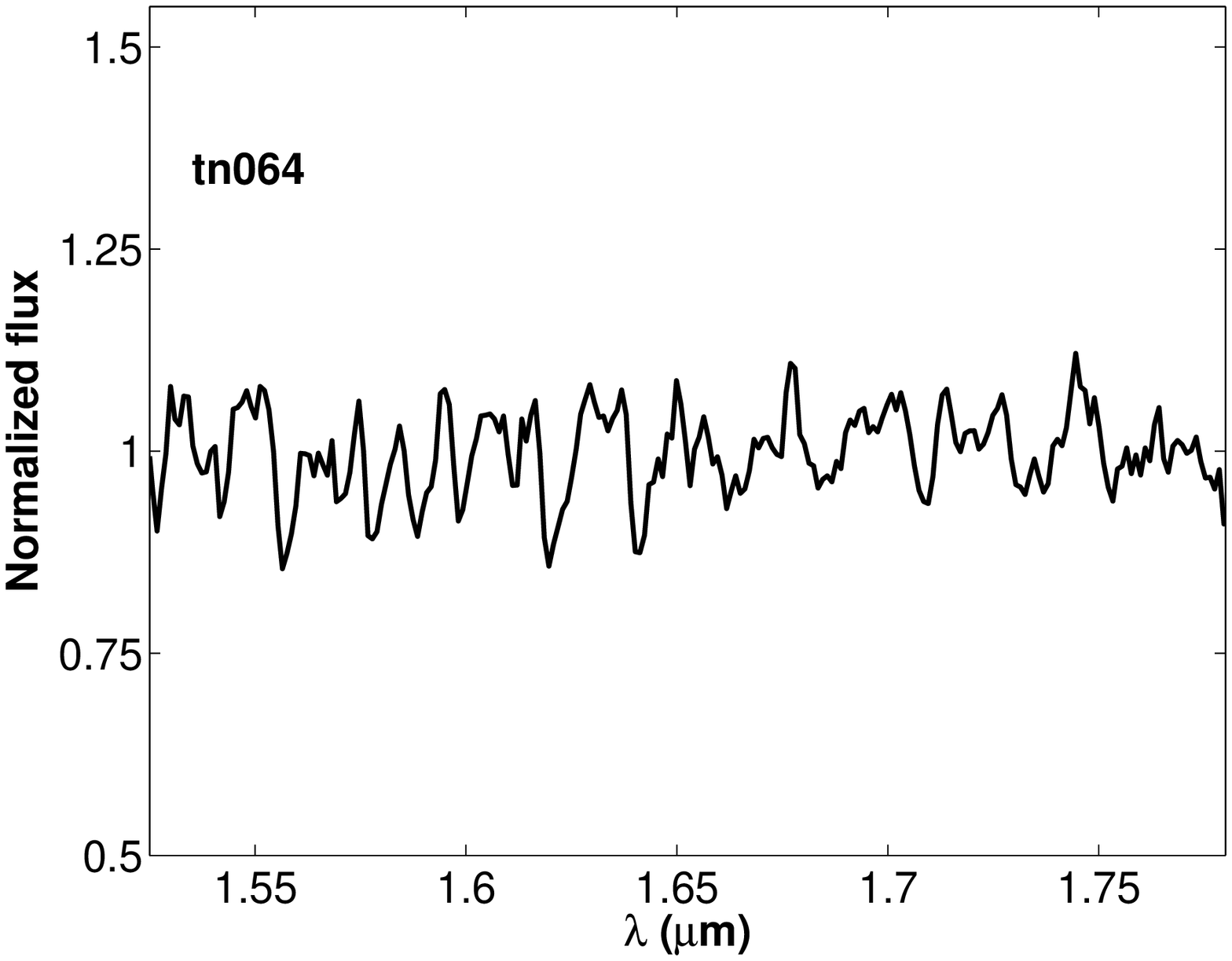}}\resizebox*{7cm}{4cm}{\includegraphics{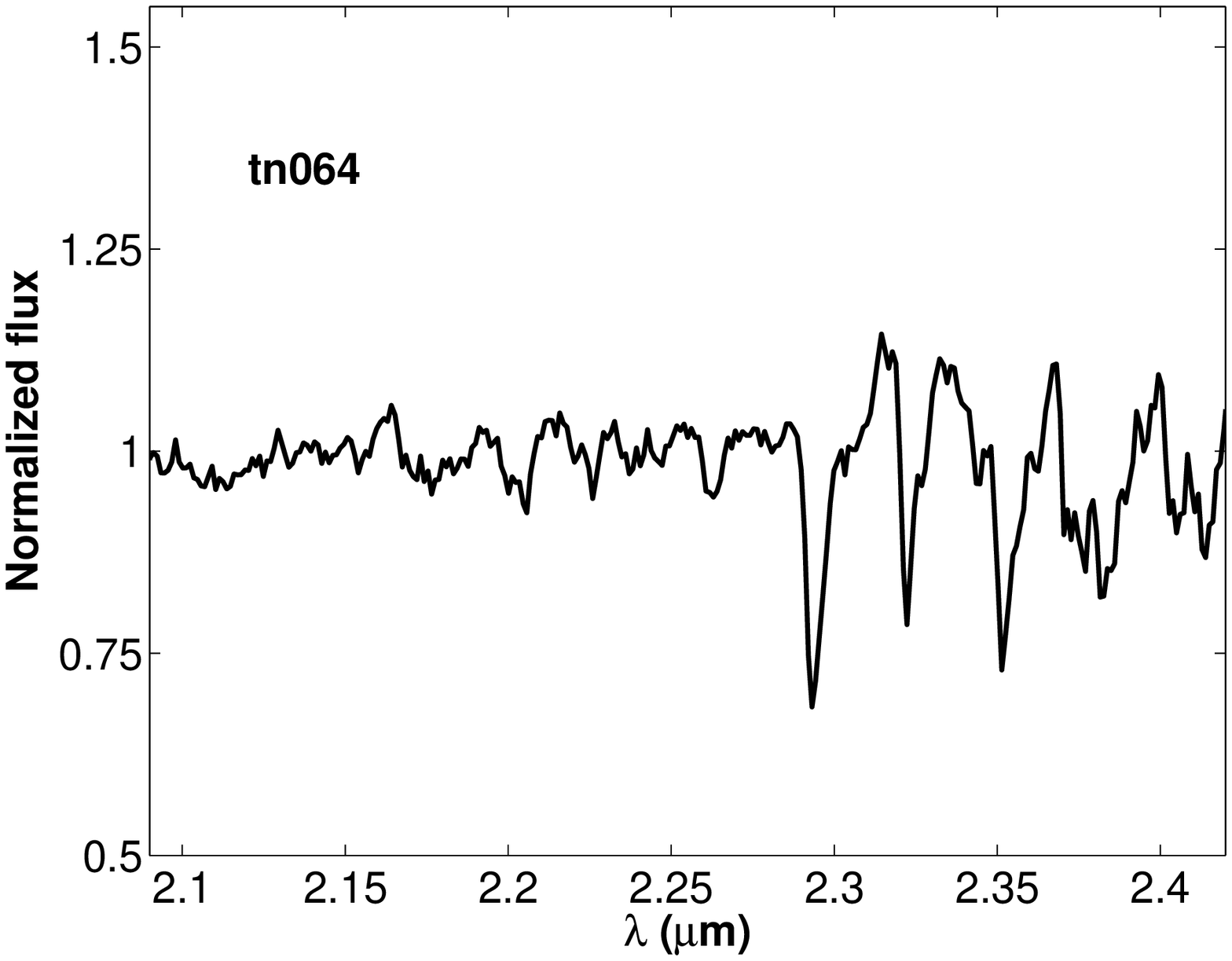}}
	\resizebox*{7cm}{4cm}{\includegraphics{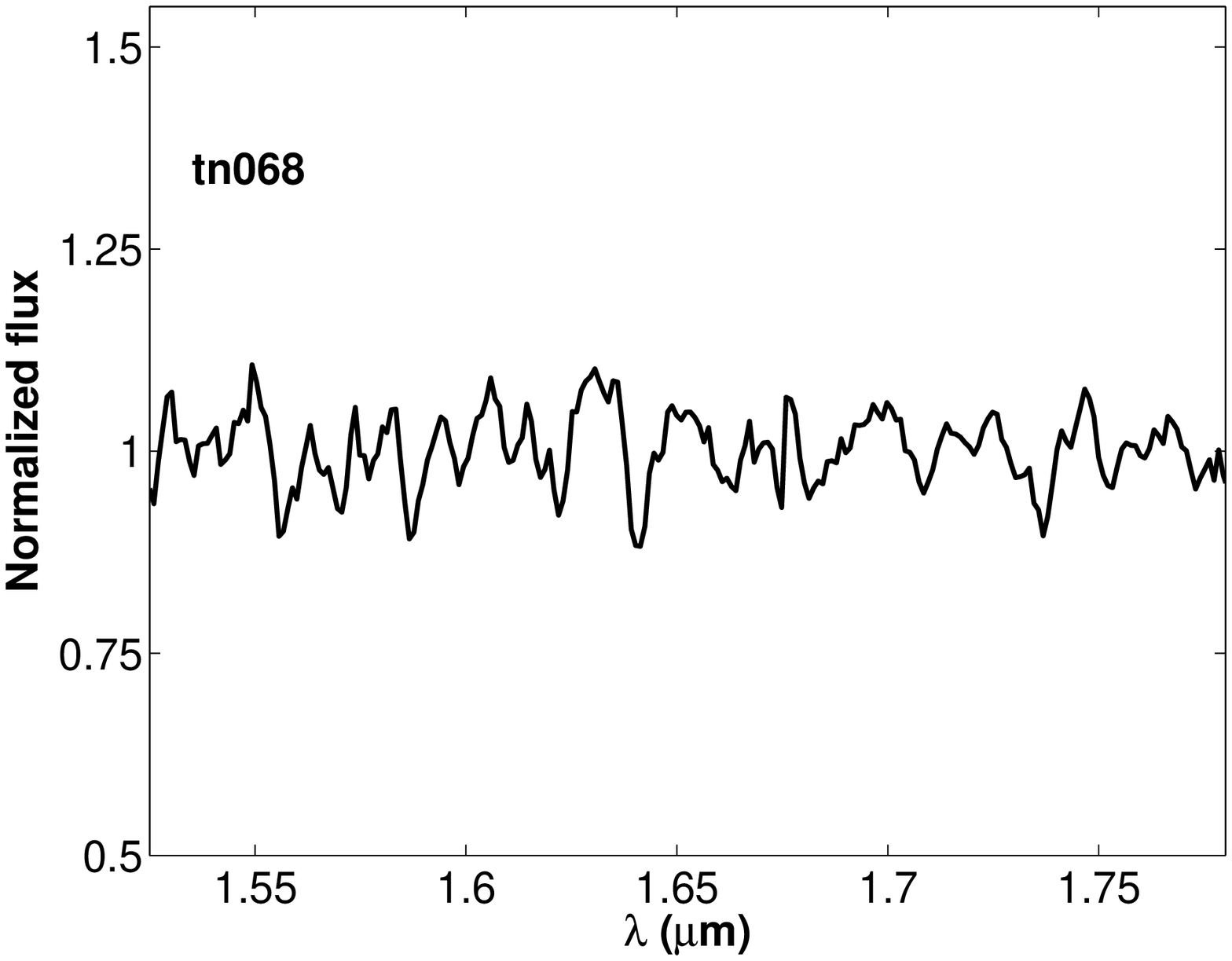}}\resizebox*{7cm}{4cm}{\includegraphics{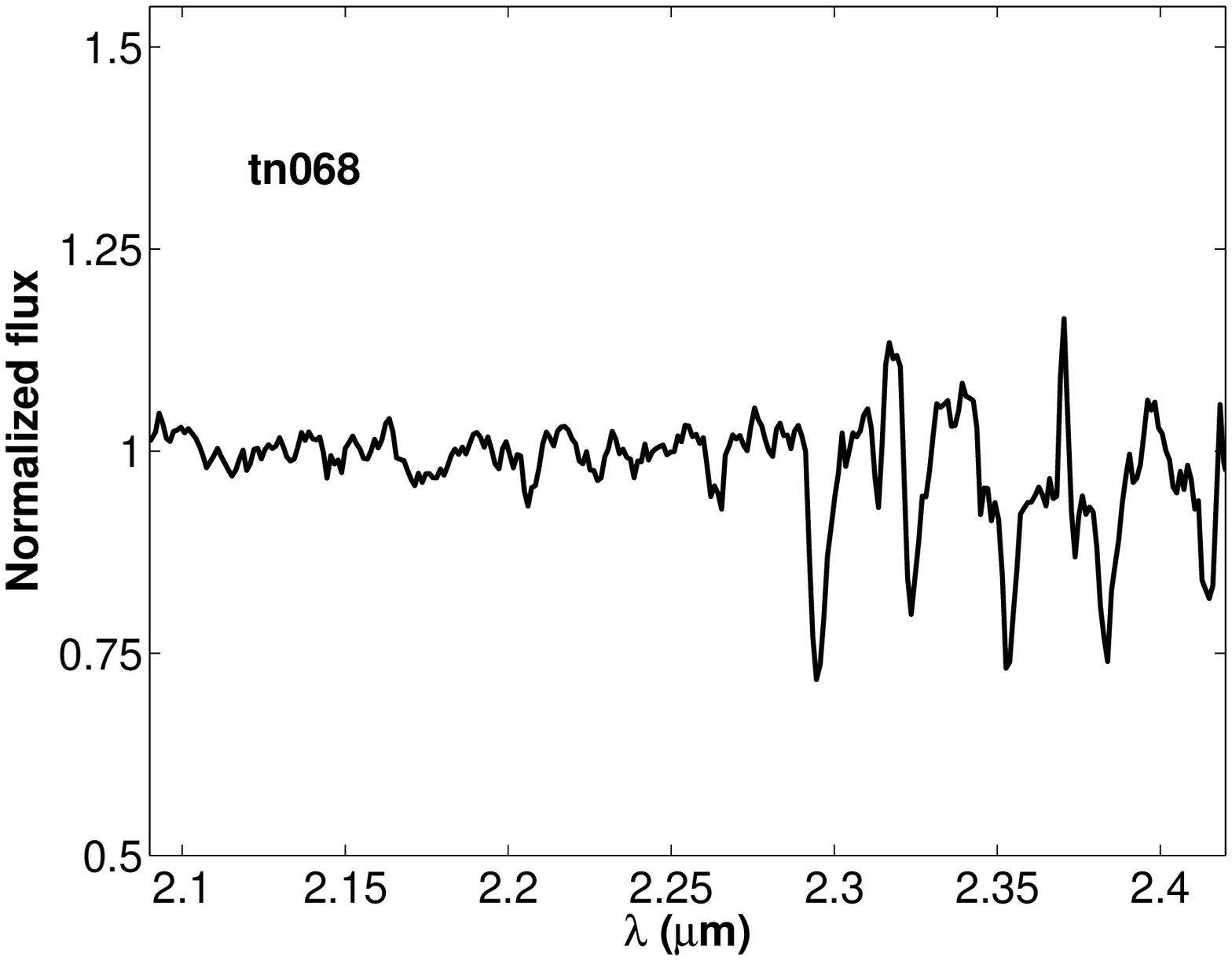}}
	\caption{Some examples of normalized spectra obtained for sources in the field l=7$^\circ$, b=0$^\circ$.}
	\label{spec}
	\end{figure*}
	\newpage
	
	\begin{figure*}[!h]
	\resizebox*{7cm}{4cm}{\includegraphics{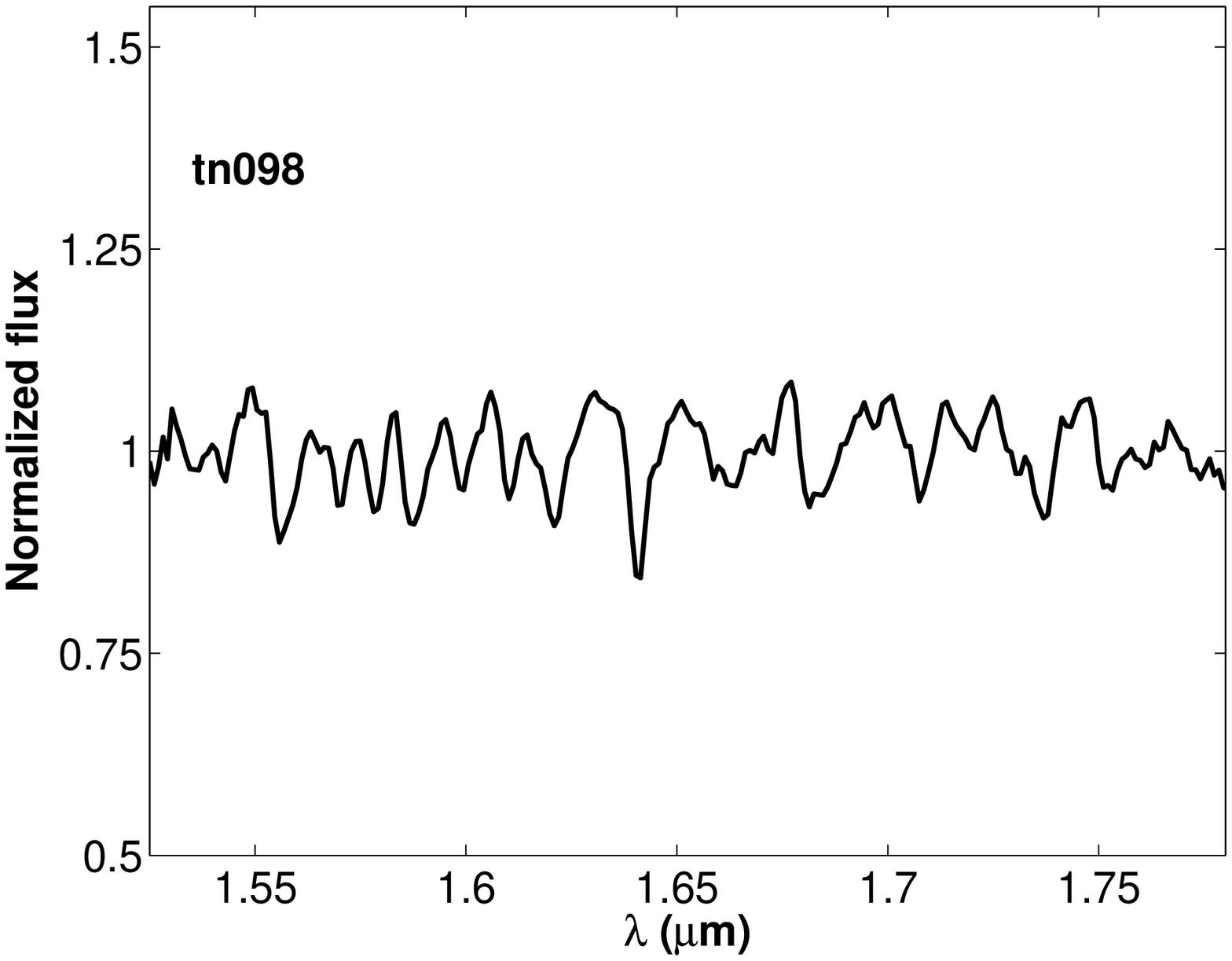}}\resizebox*{7cm}{4cm}{\includegraphics{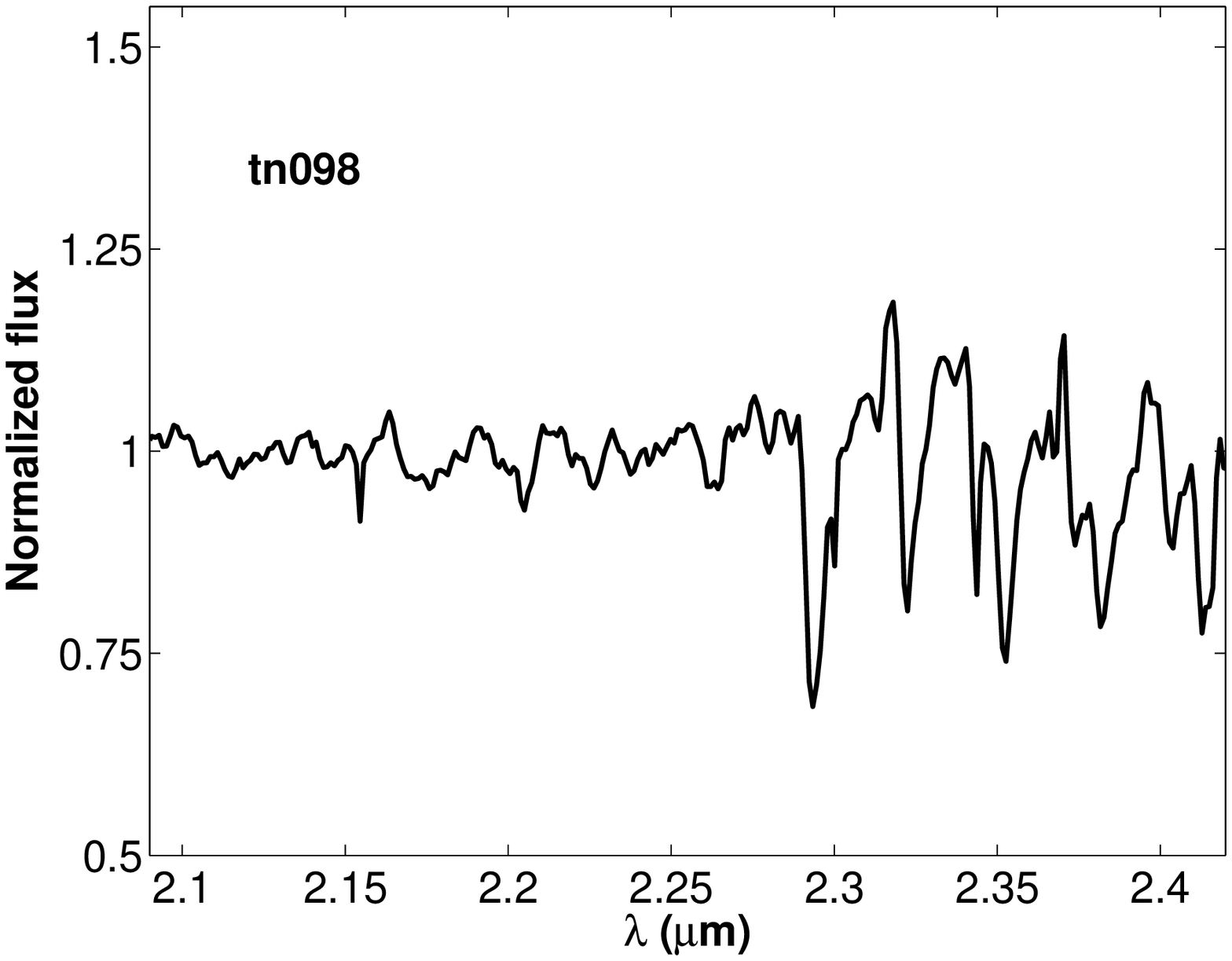}}
	\resizebox*{7cm}{4cm}{\includegraphics{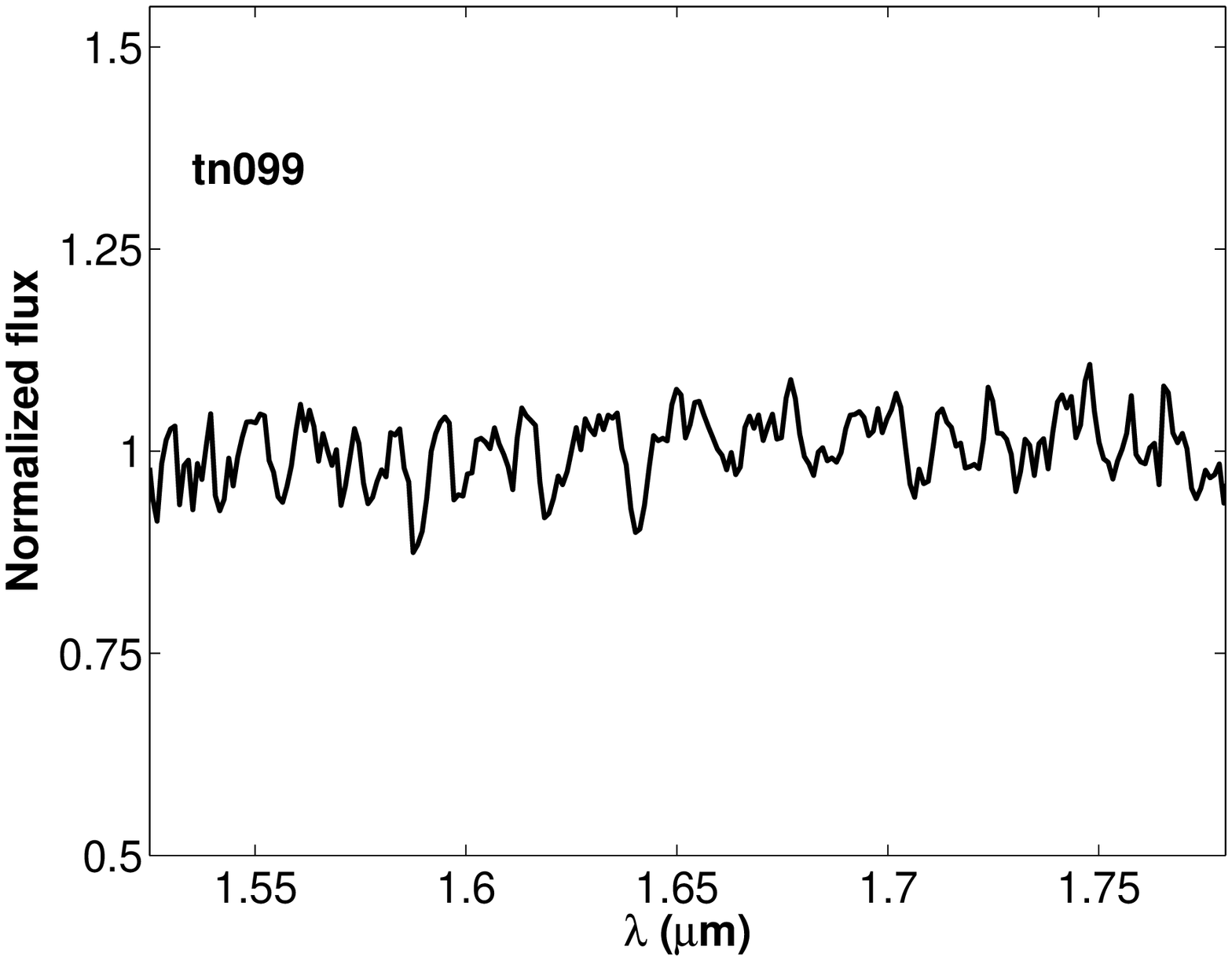}}\resizebox*{7cm}{4cm}{\includegraphics{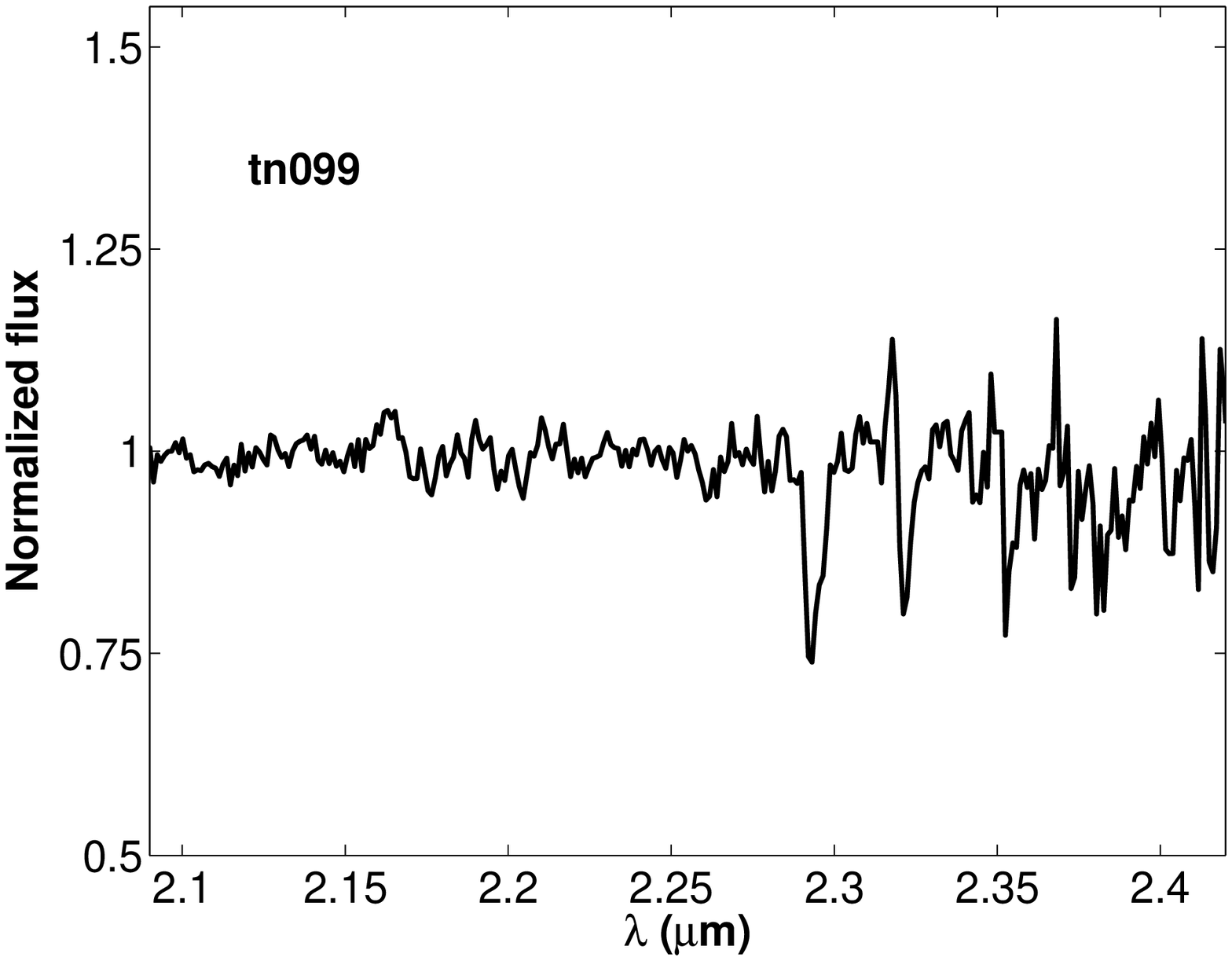}}
	\resizebox*{7cm}{4cm}{\includegraphics{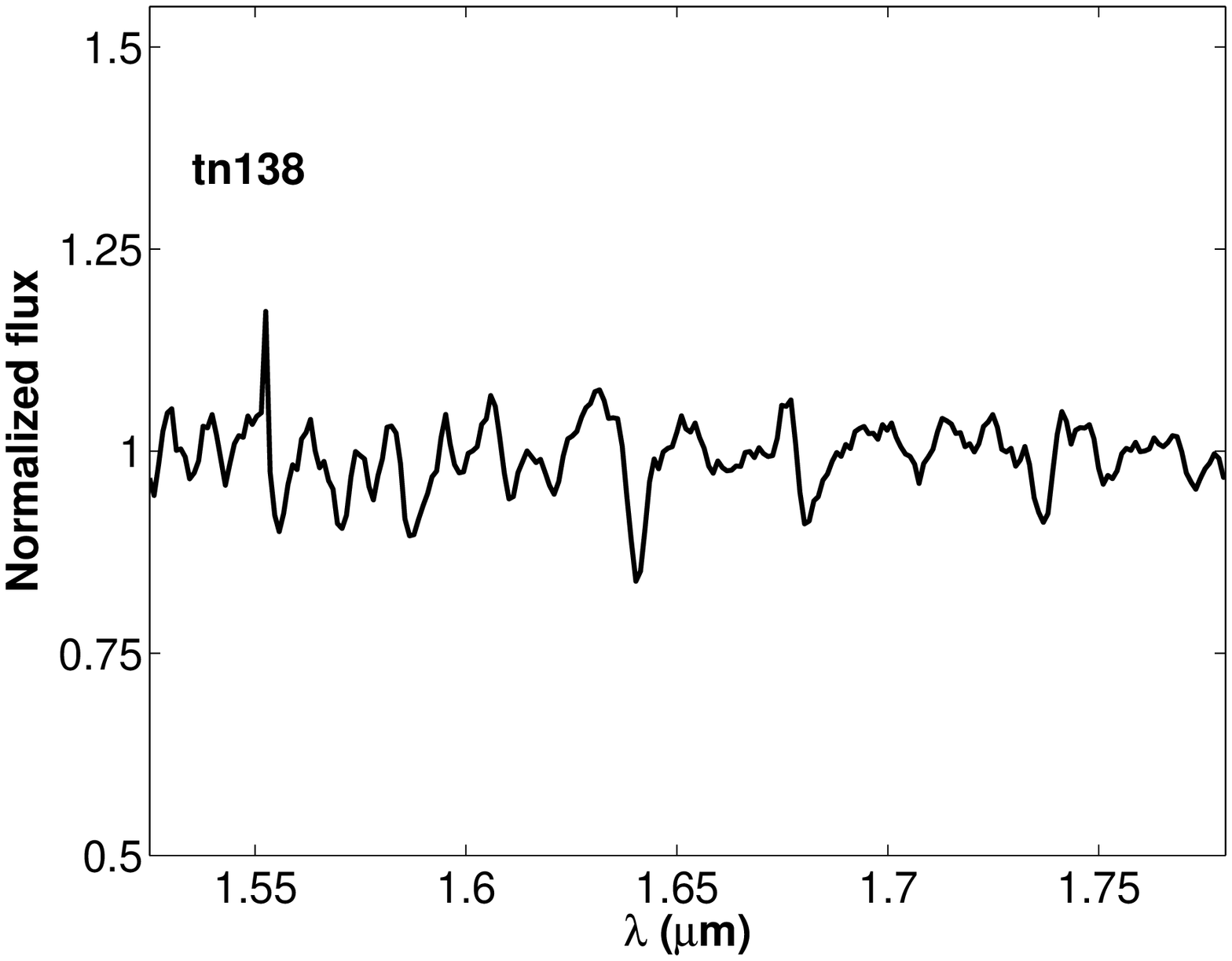}}\resizebox*{7cm}{4cm}{\includegraphics{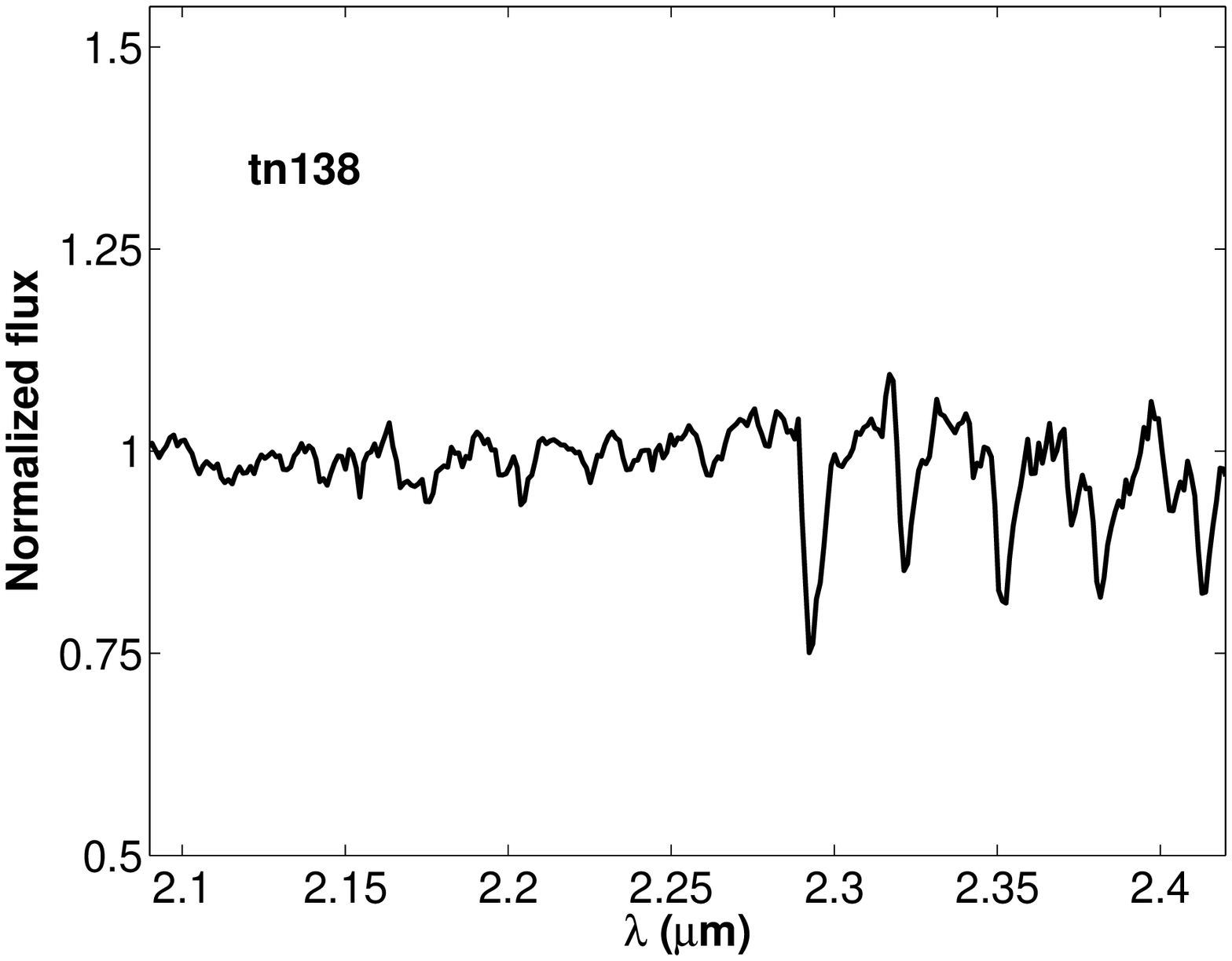}}
	\resizebox*{7cm}{4cm}{\includegraphics{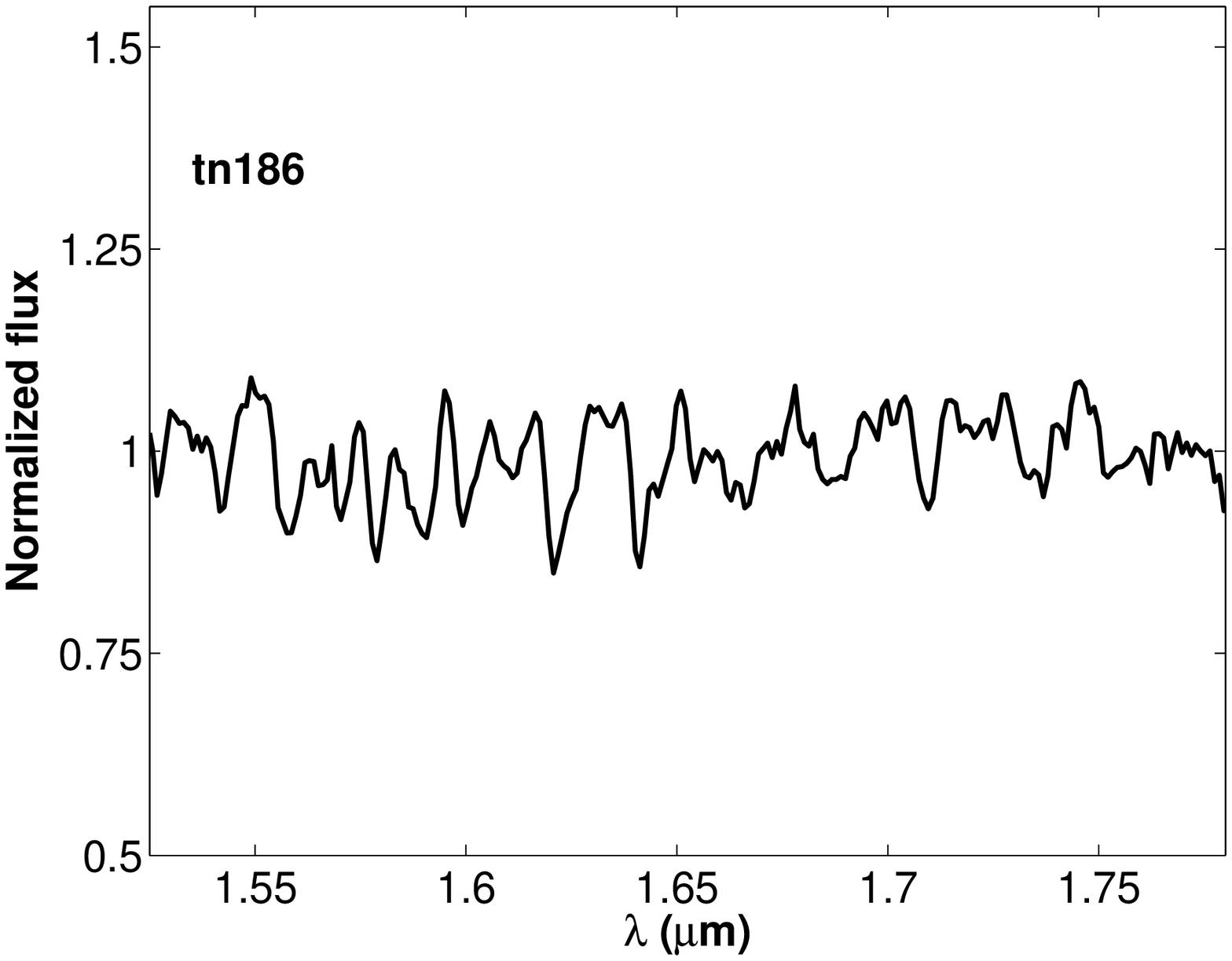}}\resizebox*{7cm}{4cm}{\includegraphics{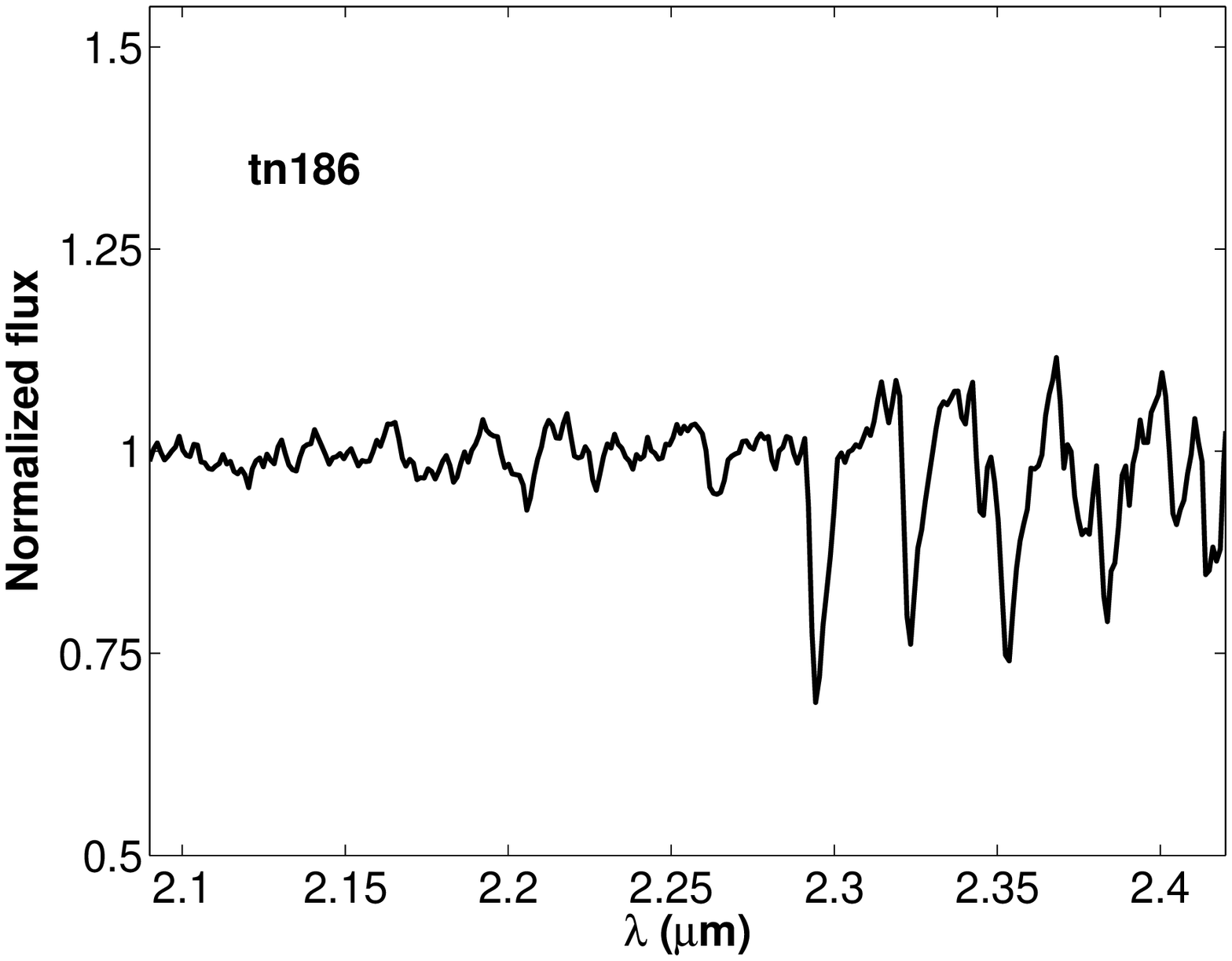}}
	\resizebox*{7cm}{4cm}{\includegraphics{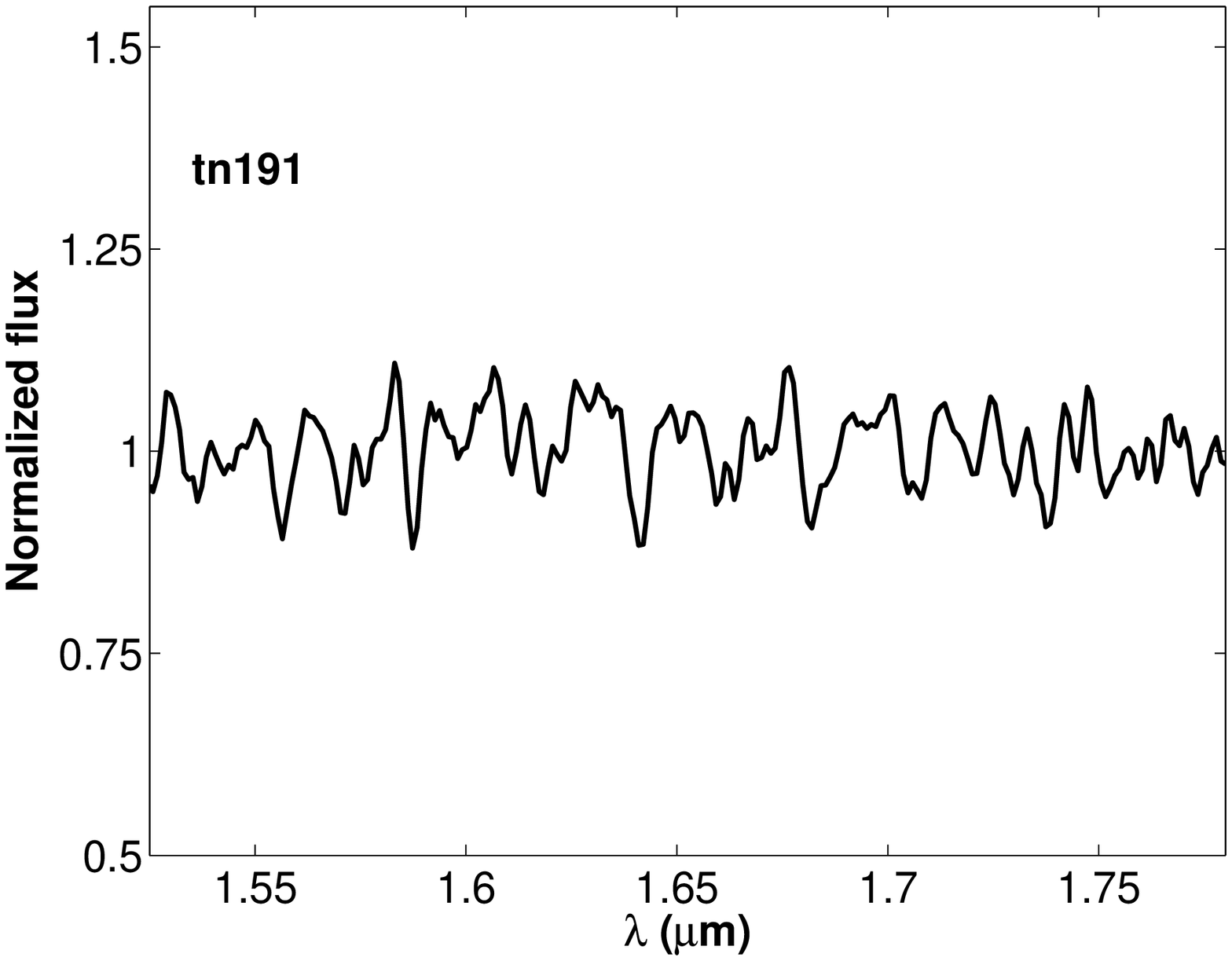}}\resizebox*{7cm}{4cm}{\includegraphics{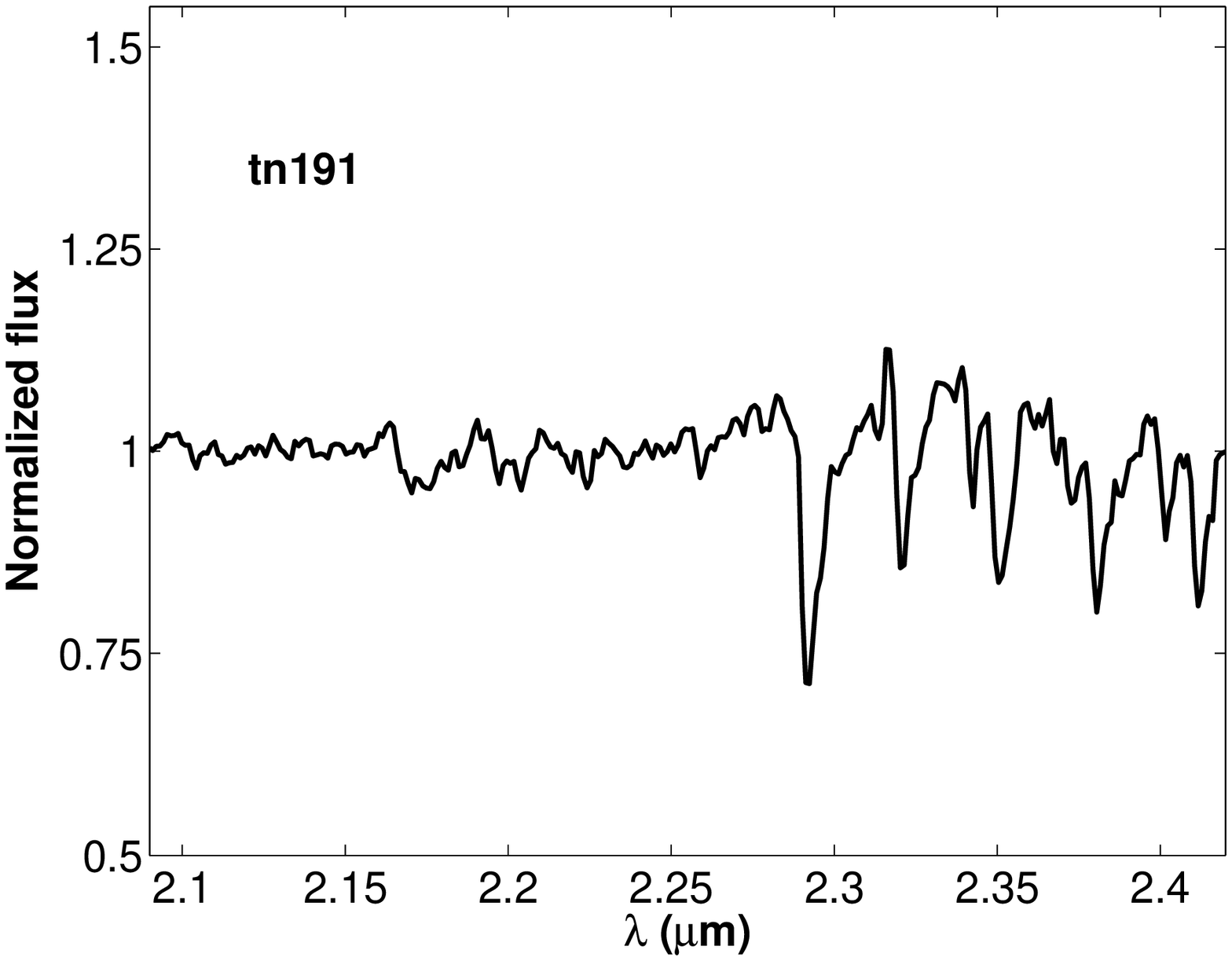}}
	\caption{Same as Fig. \ref{spec} but for l=12$^\circ$, b=0$^\circ$.}
	\end{figure*}
	\newpage
	
	\begin{figure*}[!h]
	\resizebox*{7cm}{4cm}{\includegraphics{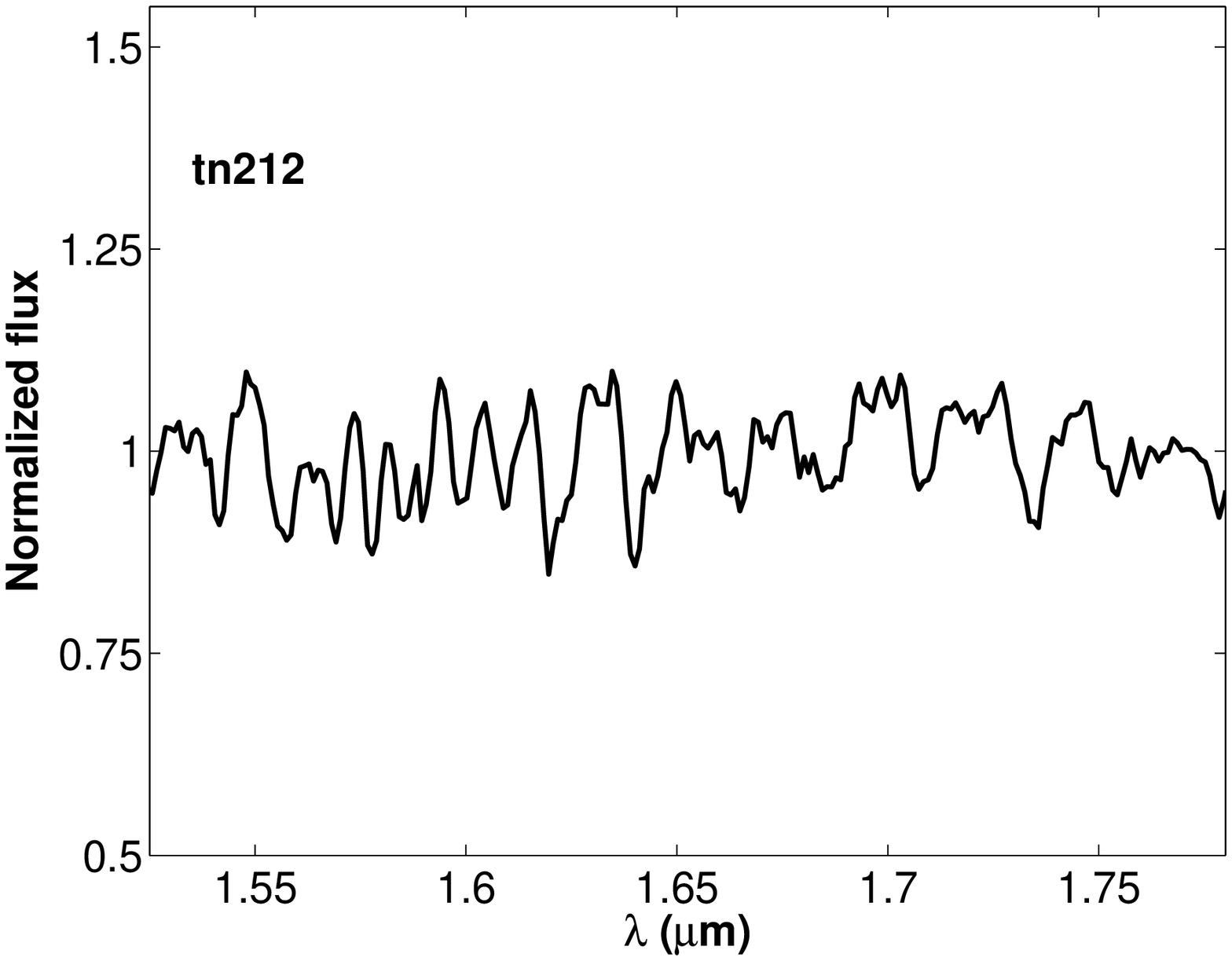}}\resizebox*{7cm}{4cm}{\includegraphics{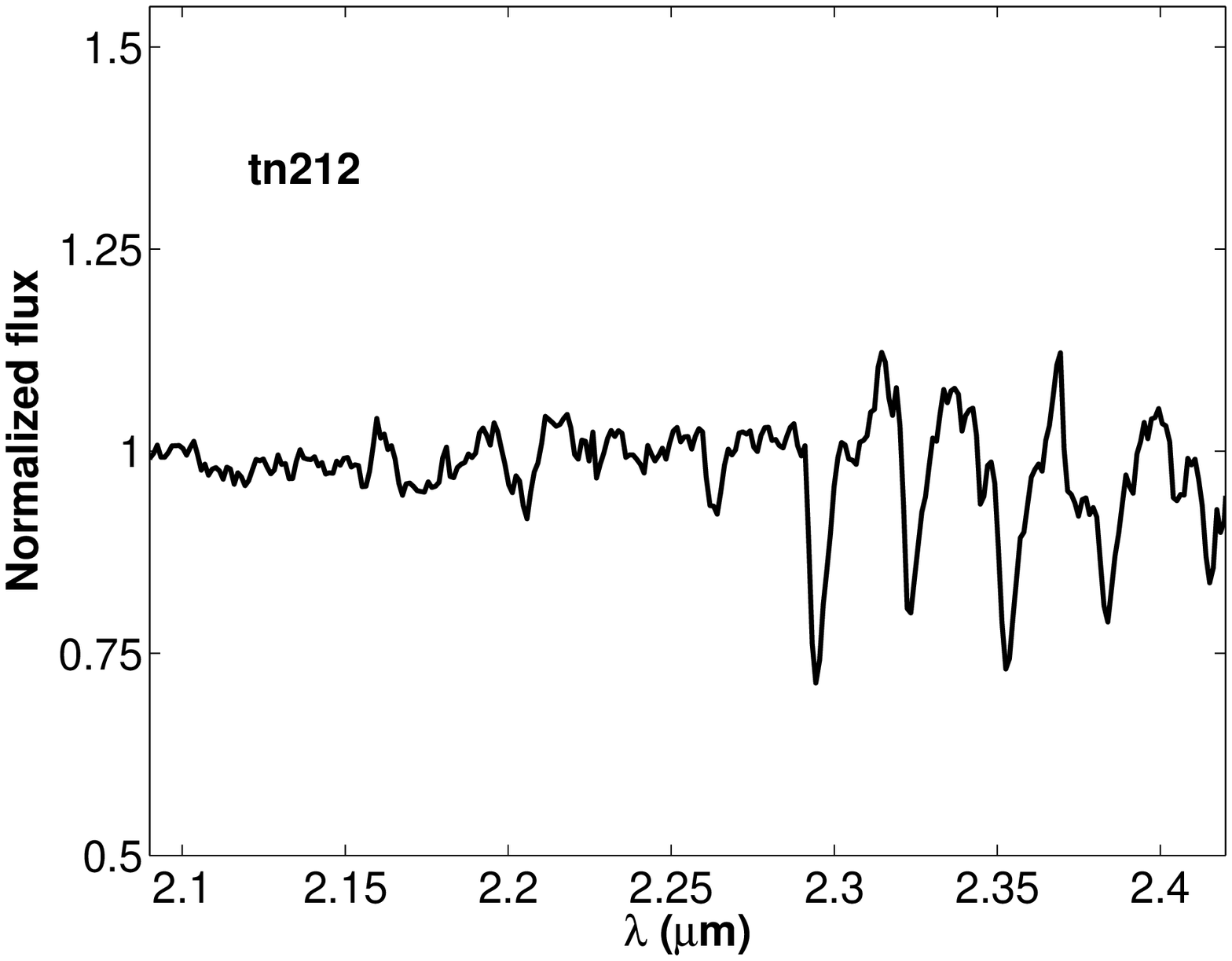}}
	\resizebox*{7cm}{4cm}{\includegraphics{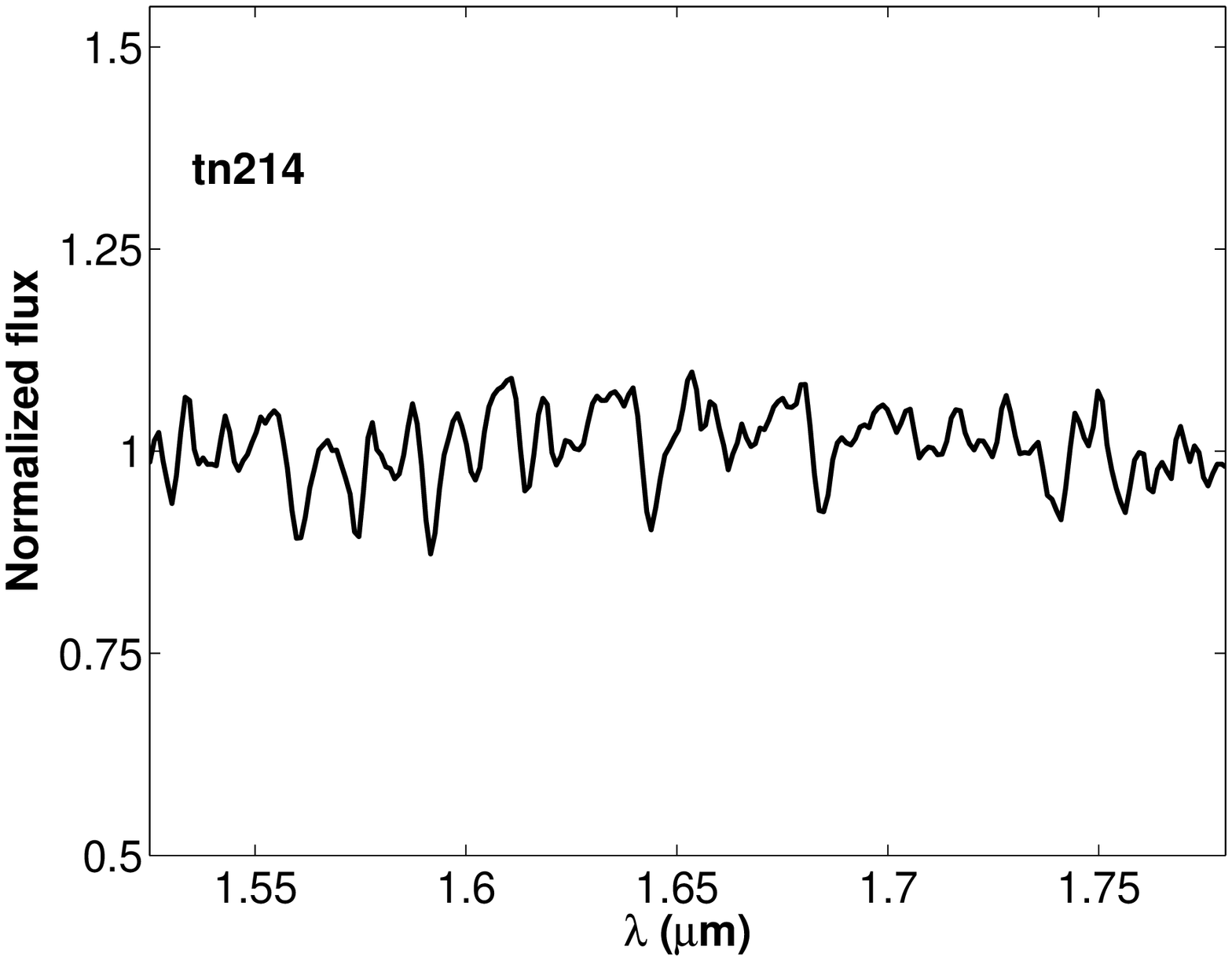}}\resizebox*{7cm}{4cm}{\includegraphics{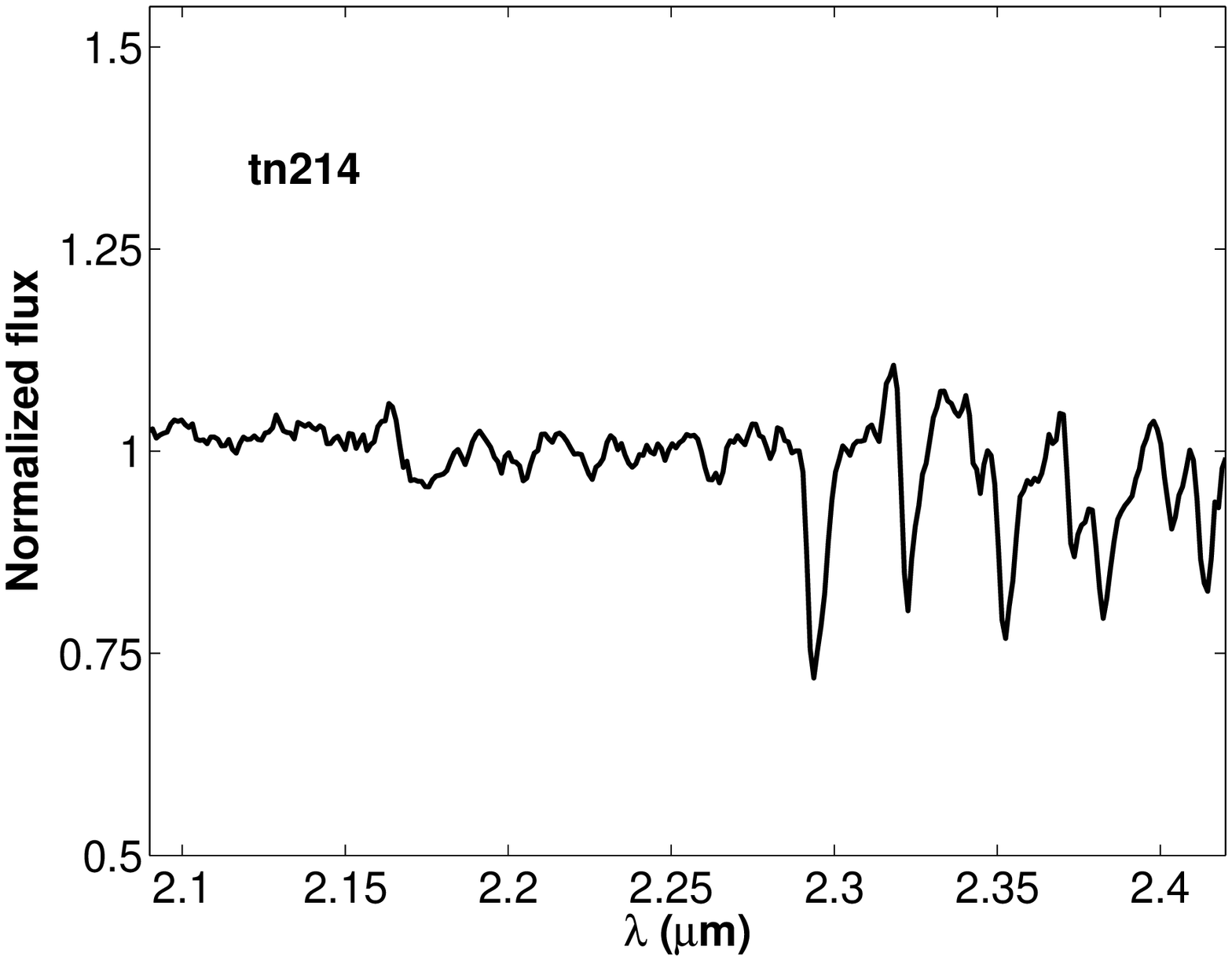}}
	\resizebox*{7cm}{4cm}{\includegraphics{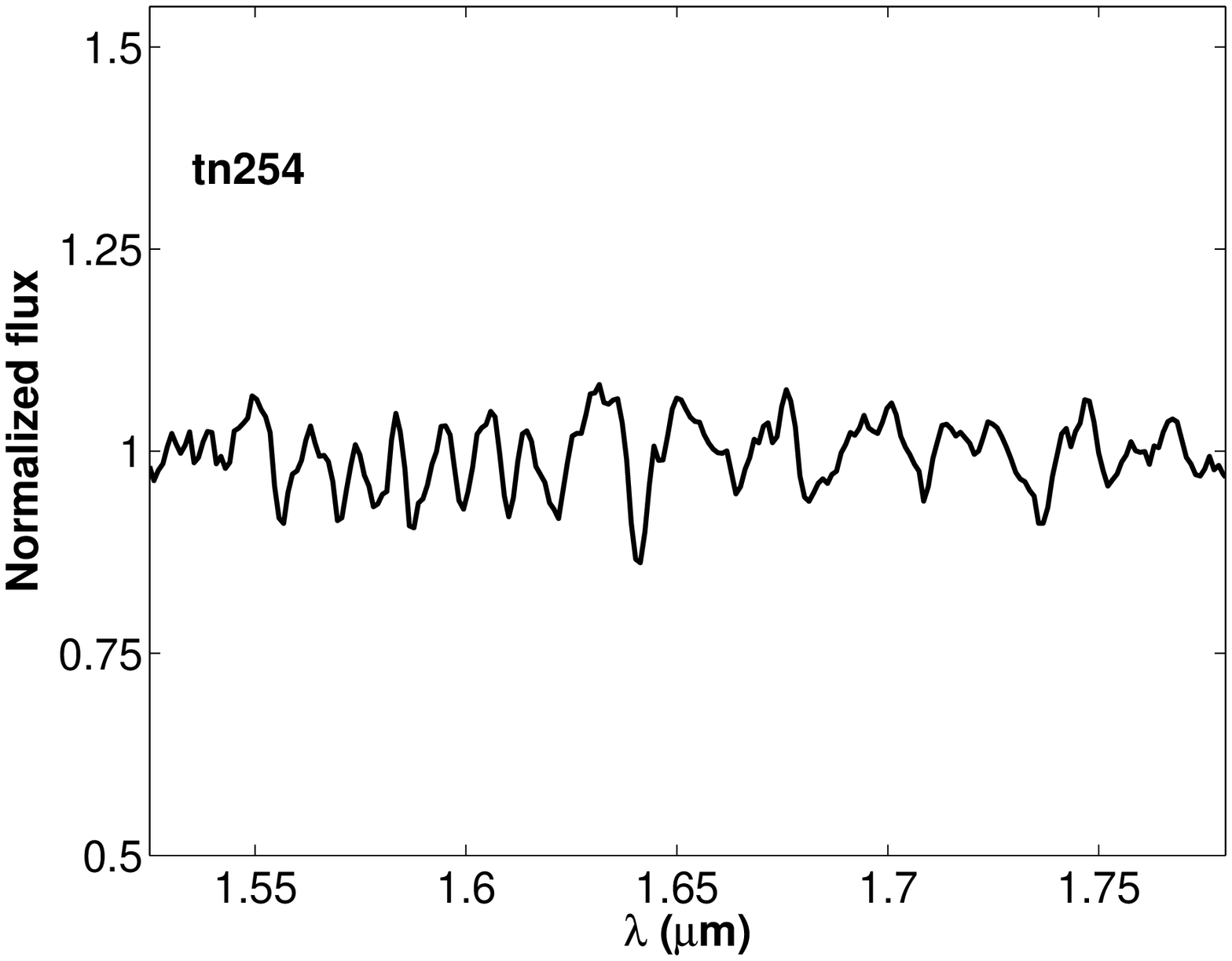}}\resizebox*{7cm}{4cm}{\includegraphics{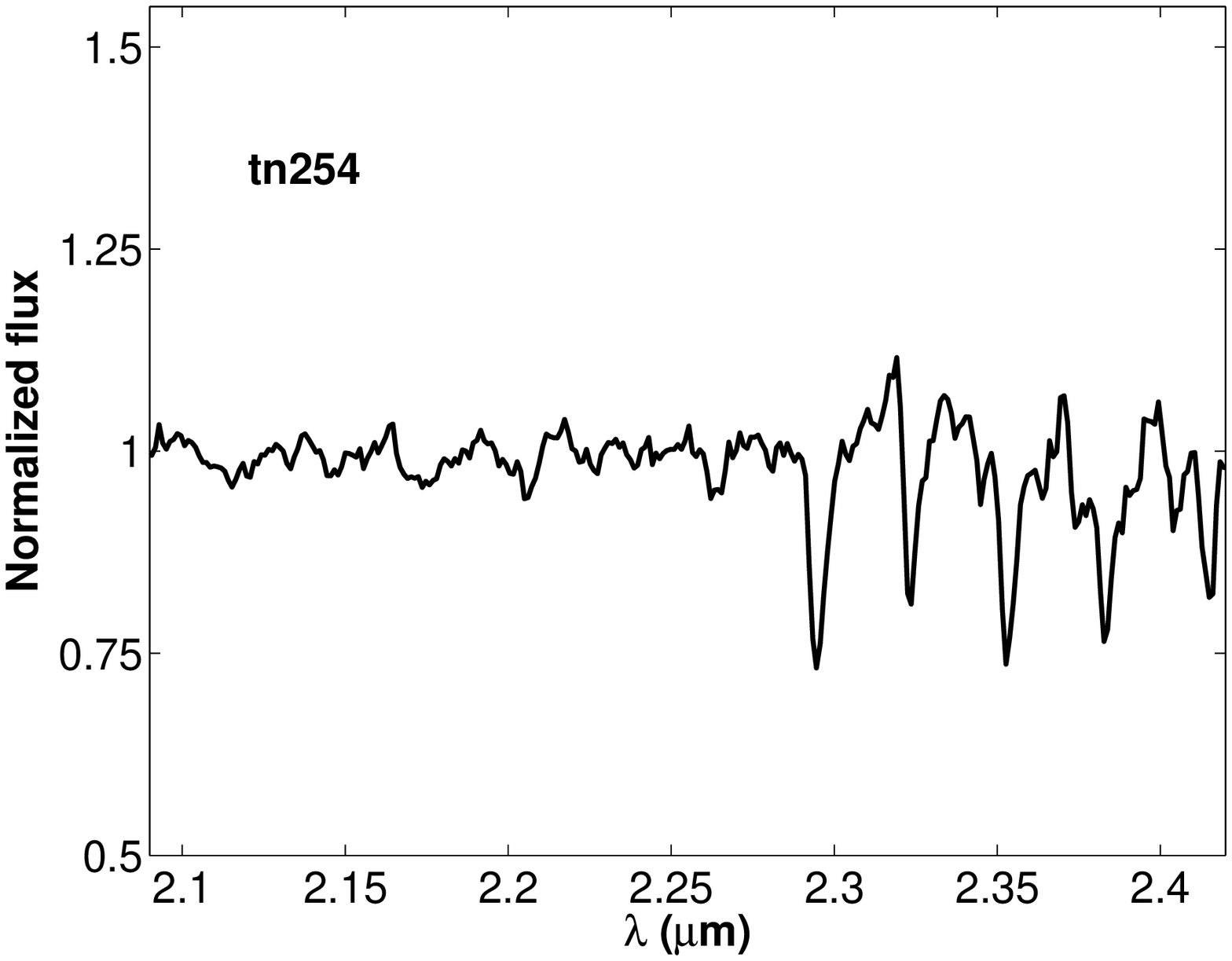}}
	\resizebox*{7cm}{4cm}{\includegraphics{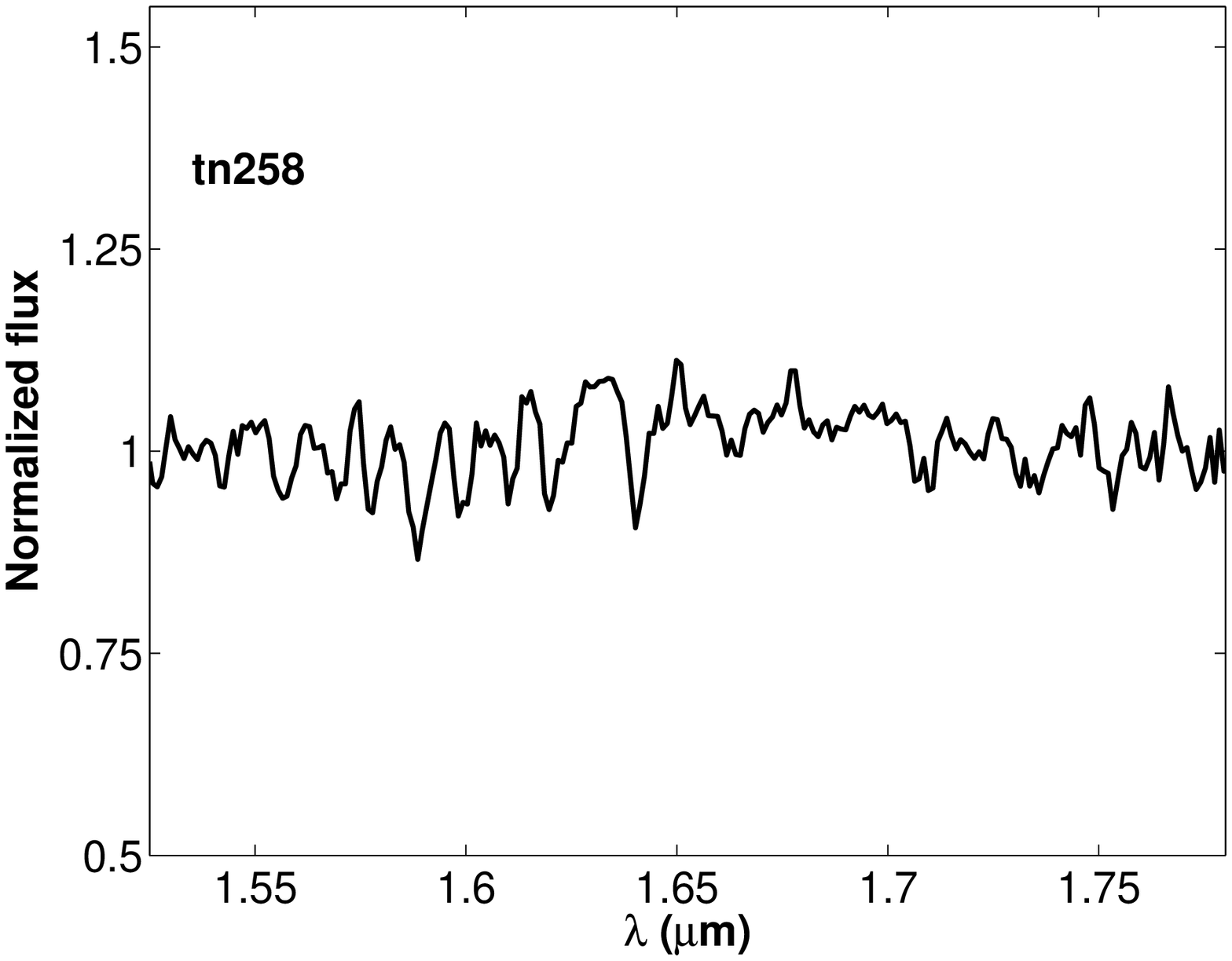}}\resizebox*{7cm}{4cm}{\includegraphics{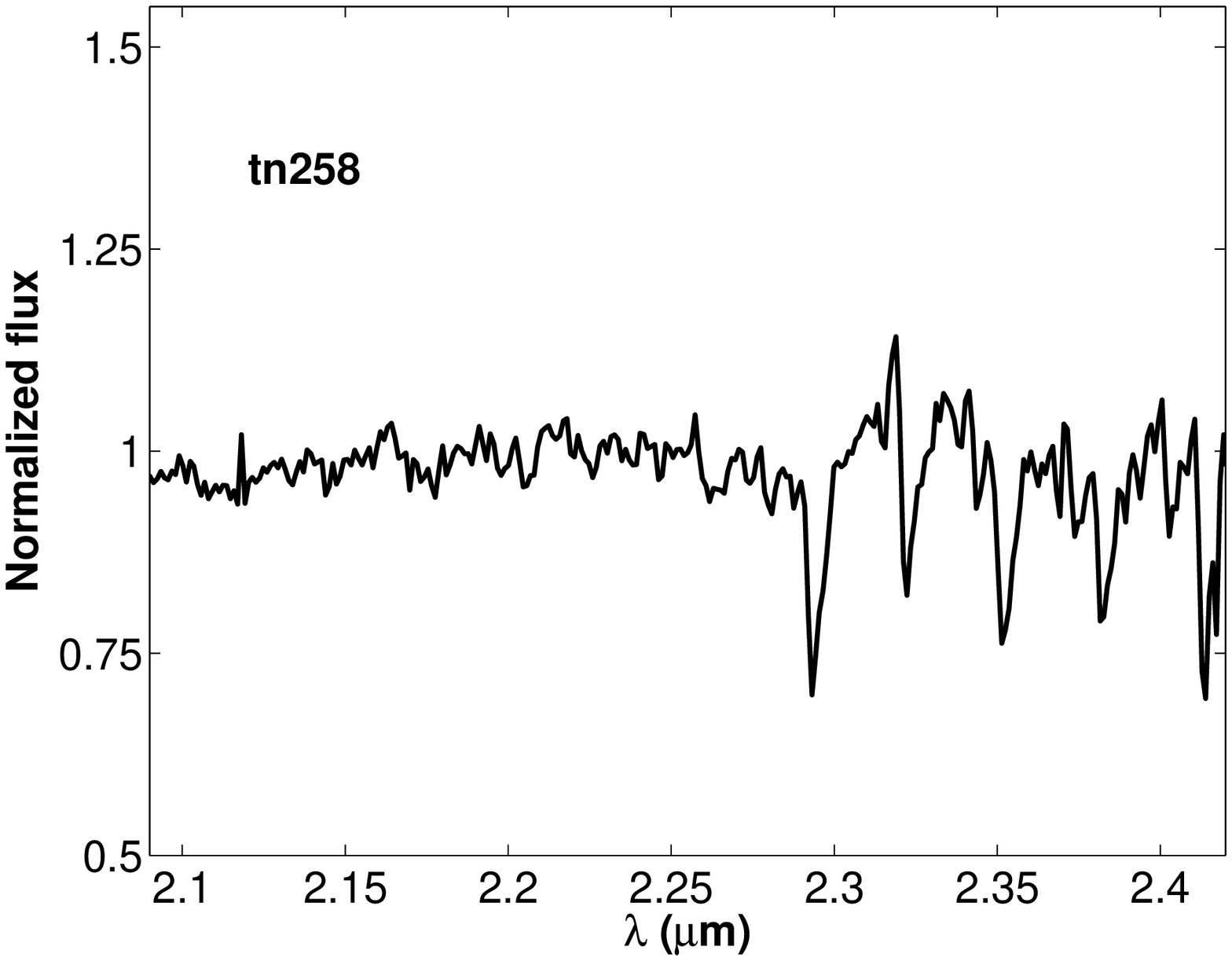}}
	\resizebox*{7cm}{4cm}{\includegraphics{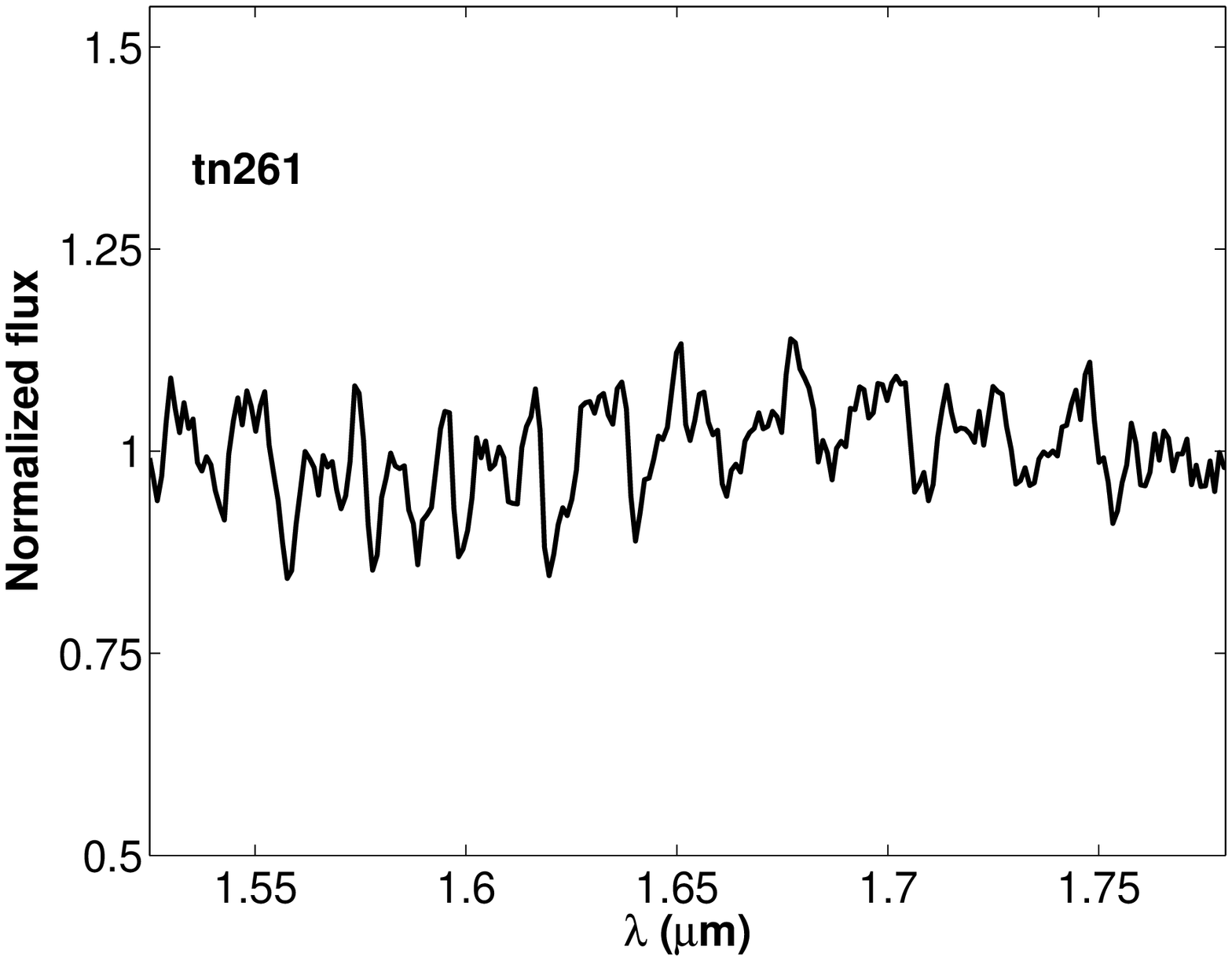}}\resizebox*{7cm}{4cm}{\includegraphics{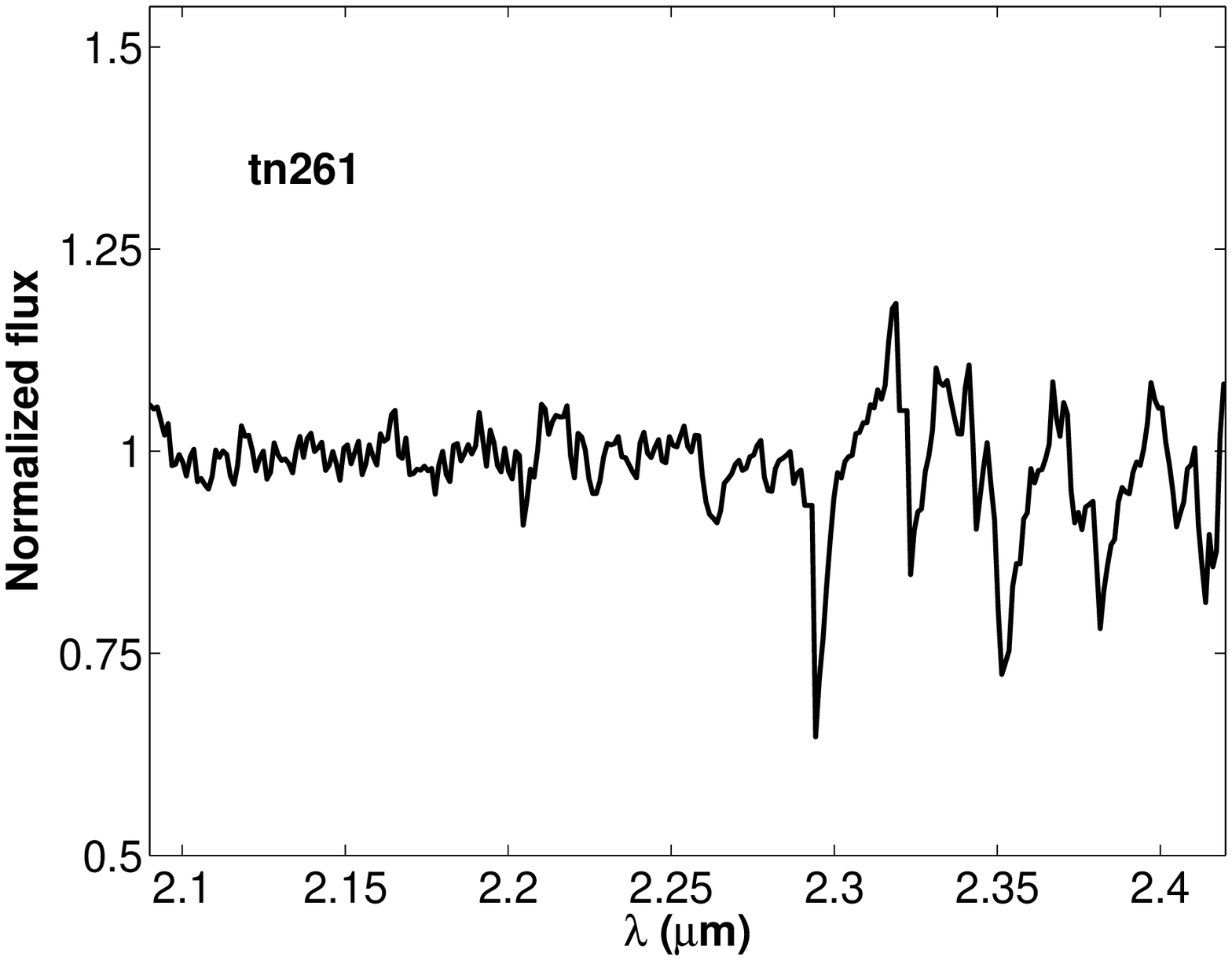}}
	\caption{Same as Fig. \ref{spec} but for l=15$^\circ$, b=0$^\circ$.}
	\end{figure*}
	\newpage
	
	\begin{figure*}[!h]
	\resizebox*{7cm}{4cm}{\includegraphics{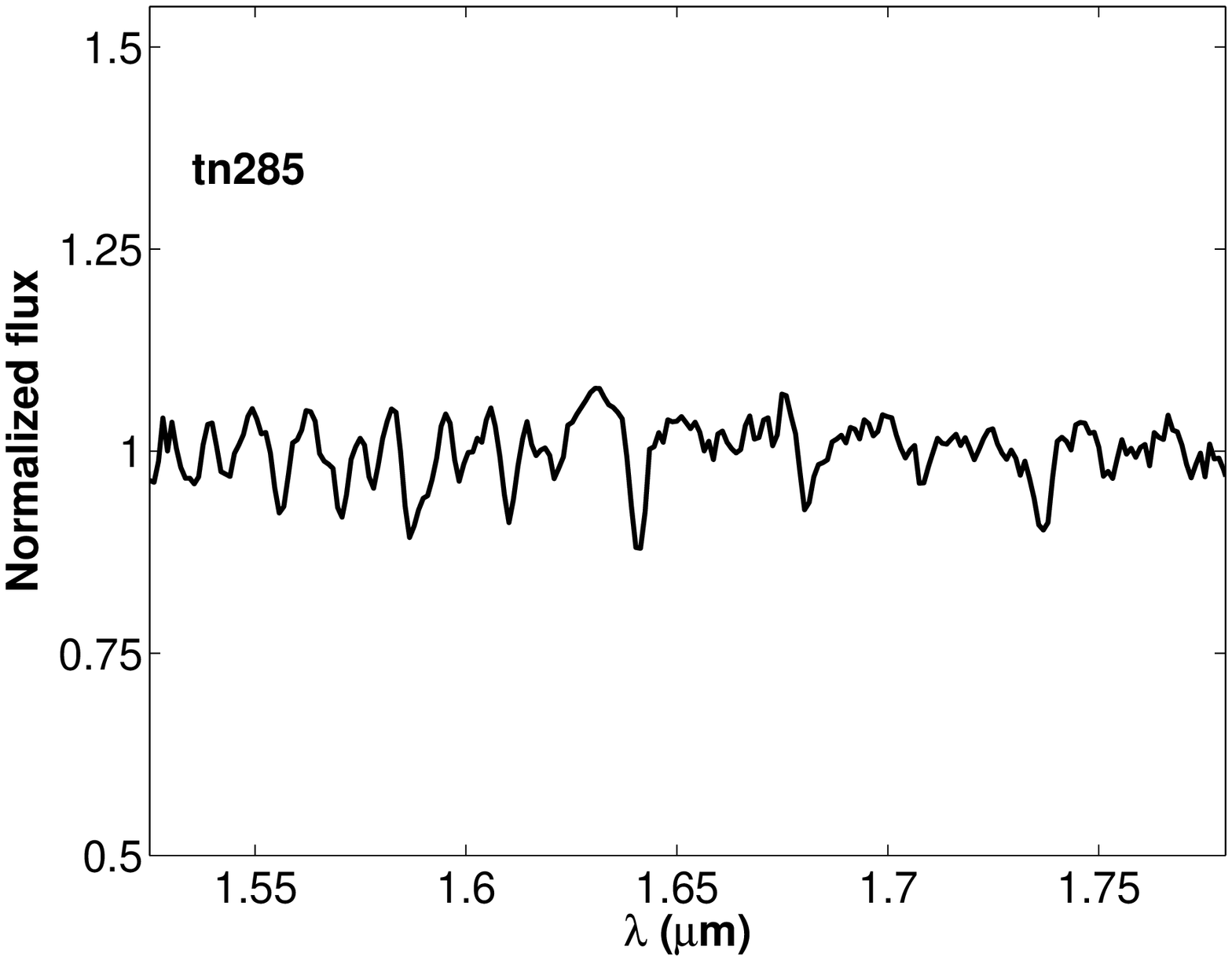}}\resizebox*{7cm}{4cm}{\includegraphics{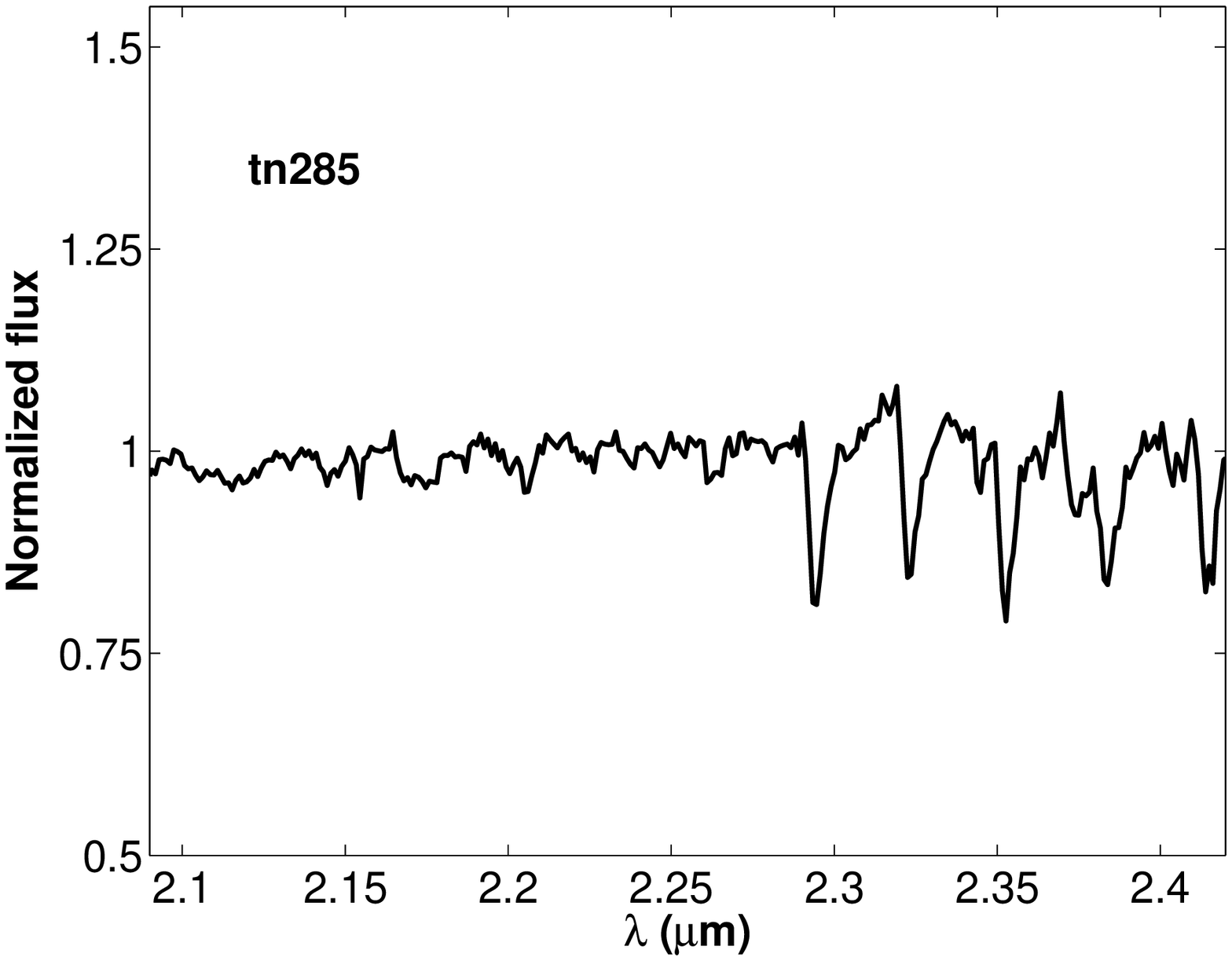}}
	\resizebox*{7cm}{4cm}{\includegraphics{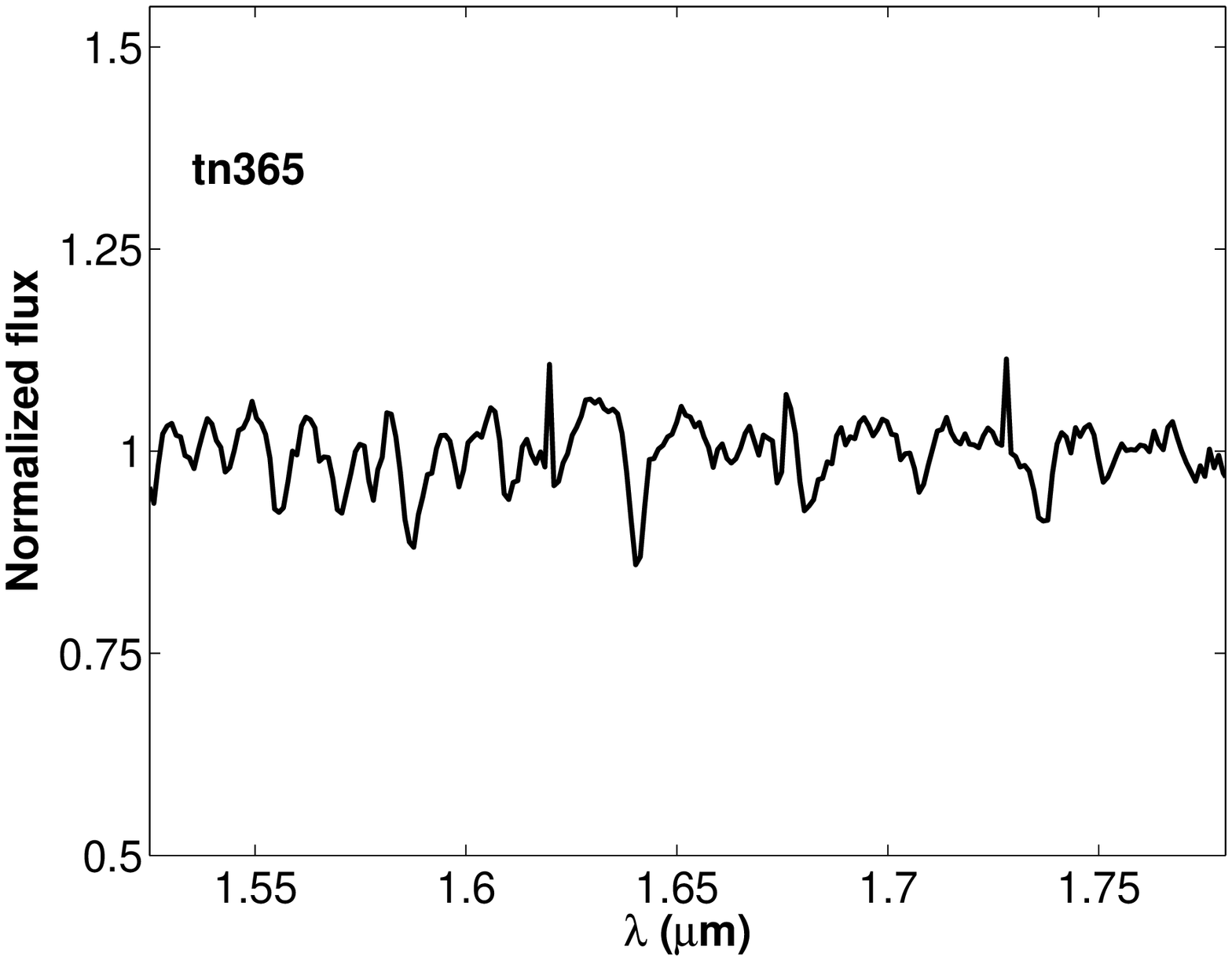}}\resizebox*{7cm}{4cm}{\includegraphics{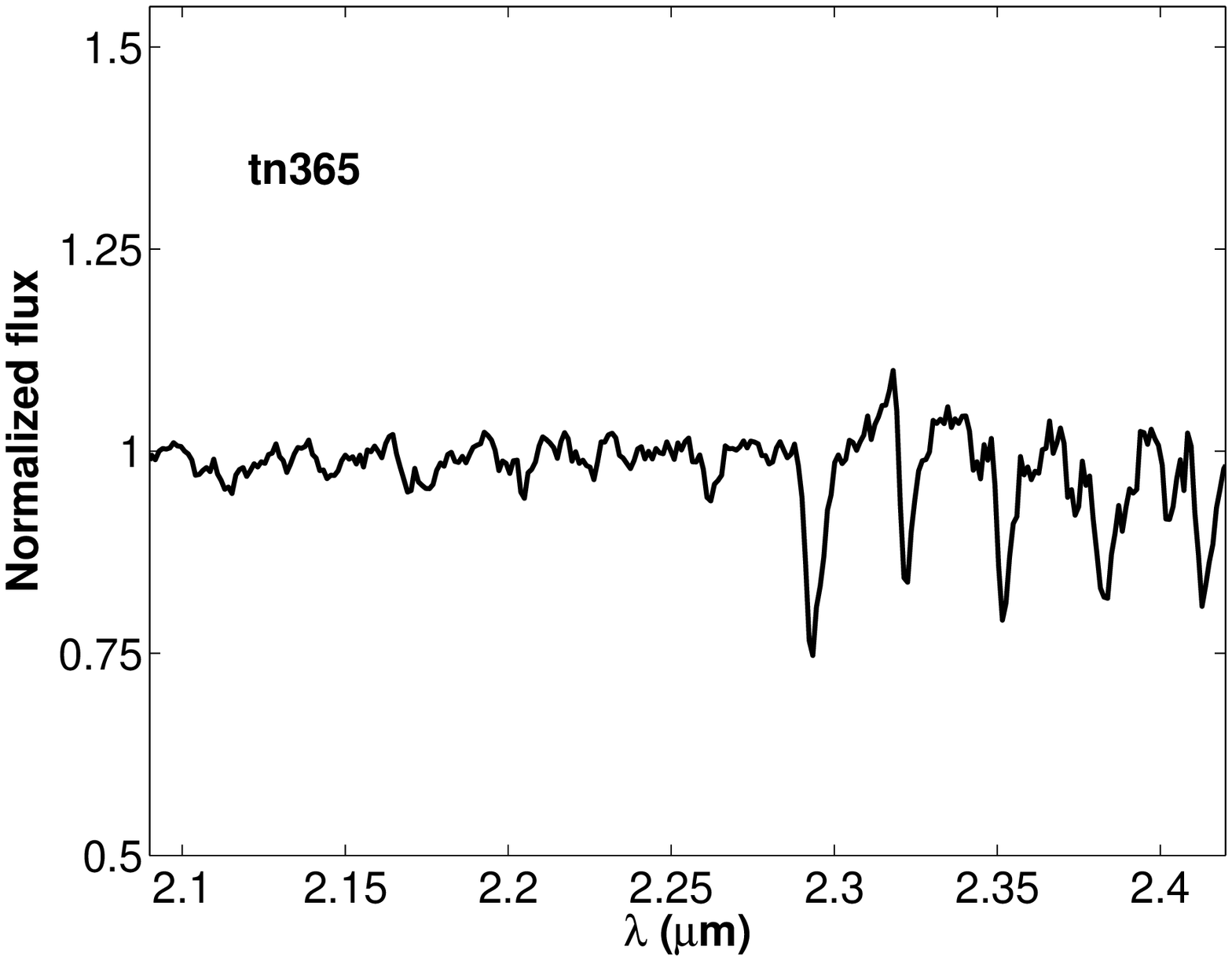}}
	\resizebox*{7cm}{4cm}{\includegraphics{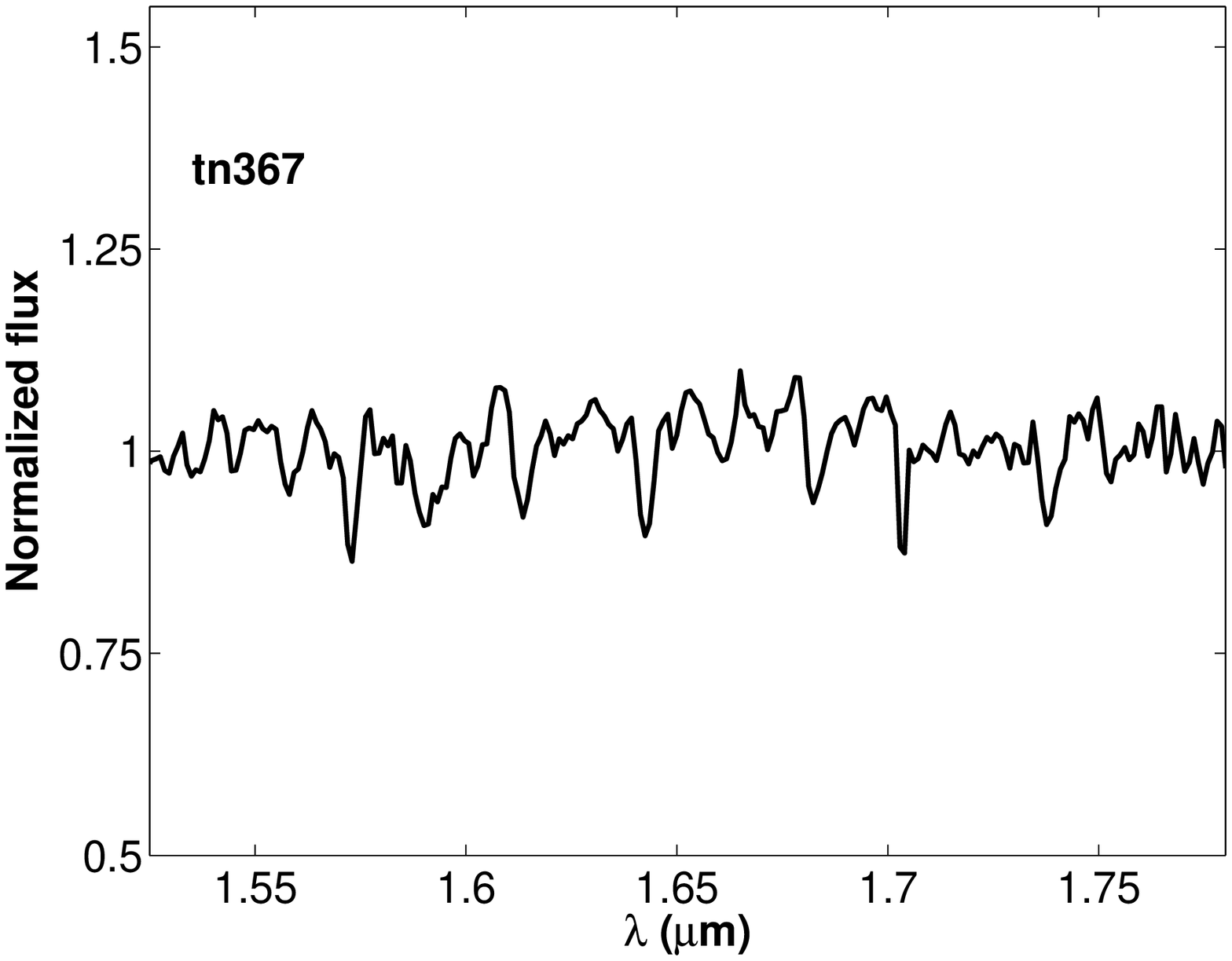}}\resizebox*{7cm}{4cm}{\includegraphics{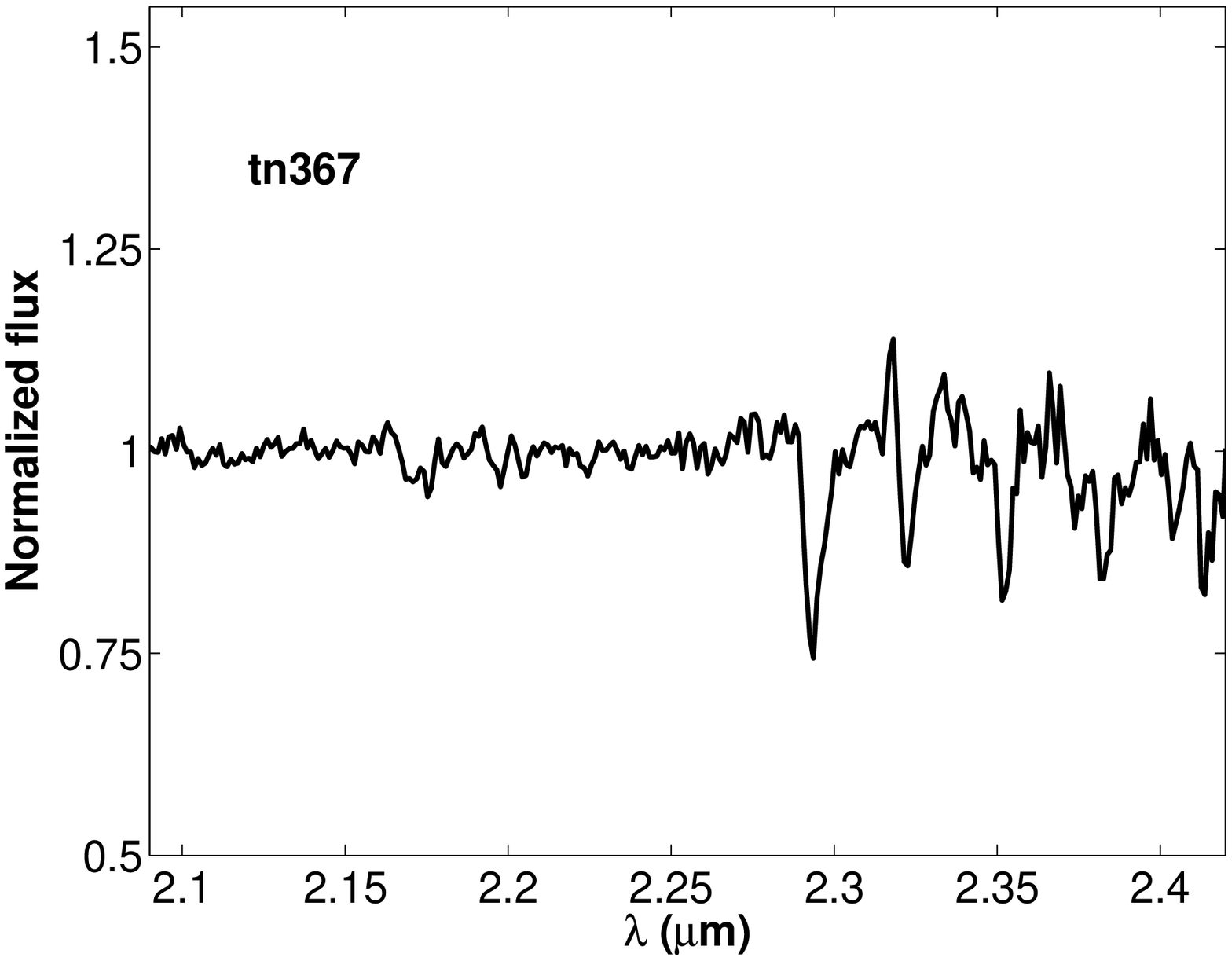}}
	\resizebox*{7cm}{4cm}{\includegraphics{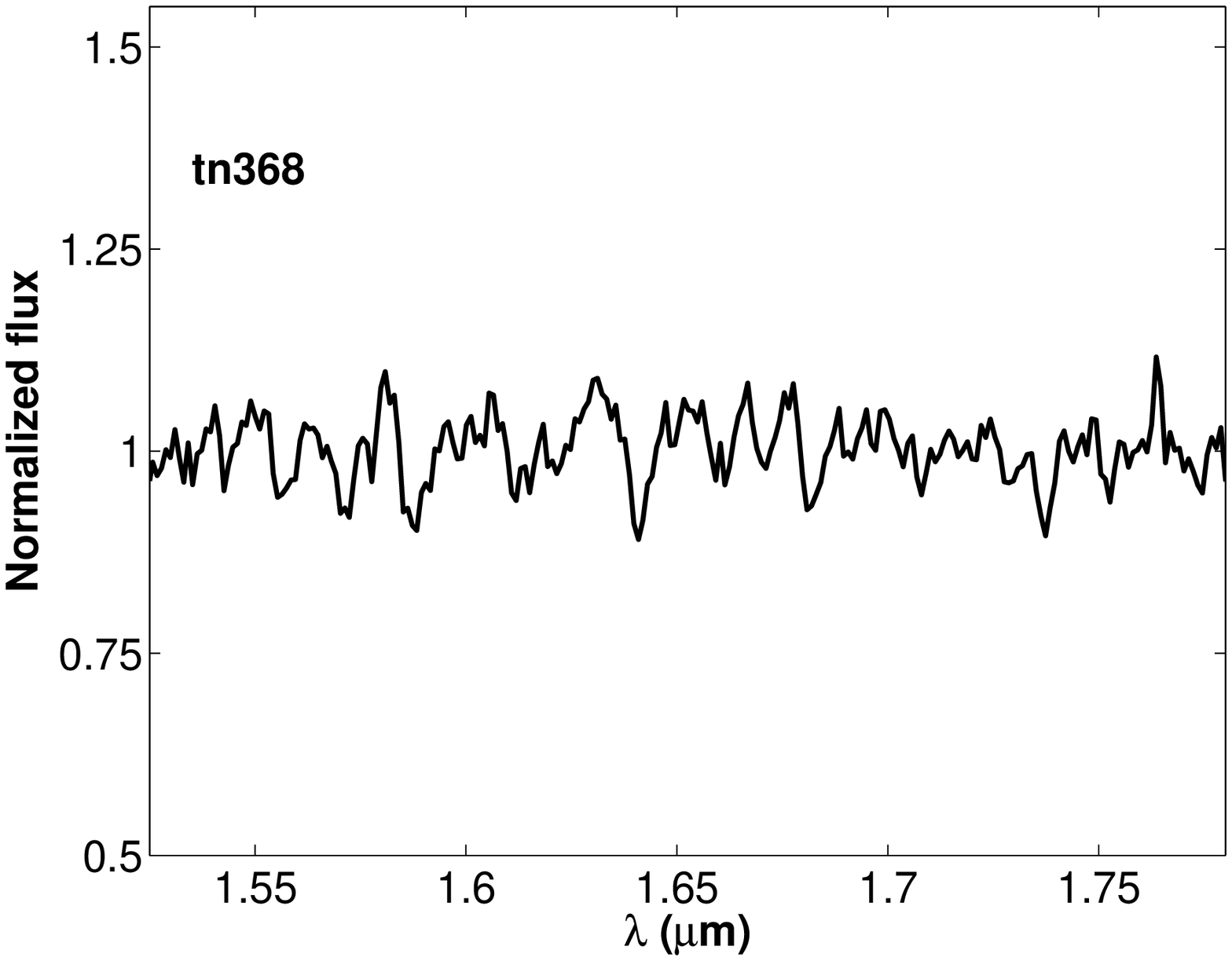}}\resizebox*{7cm}{4cm}{\includegraphics{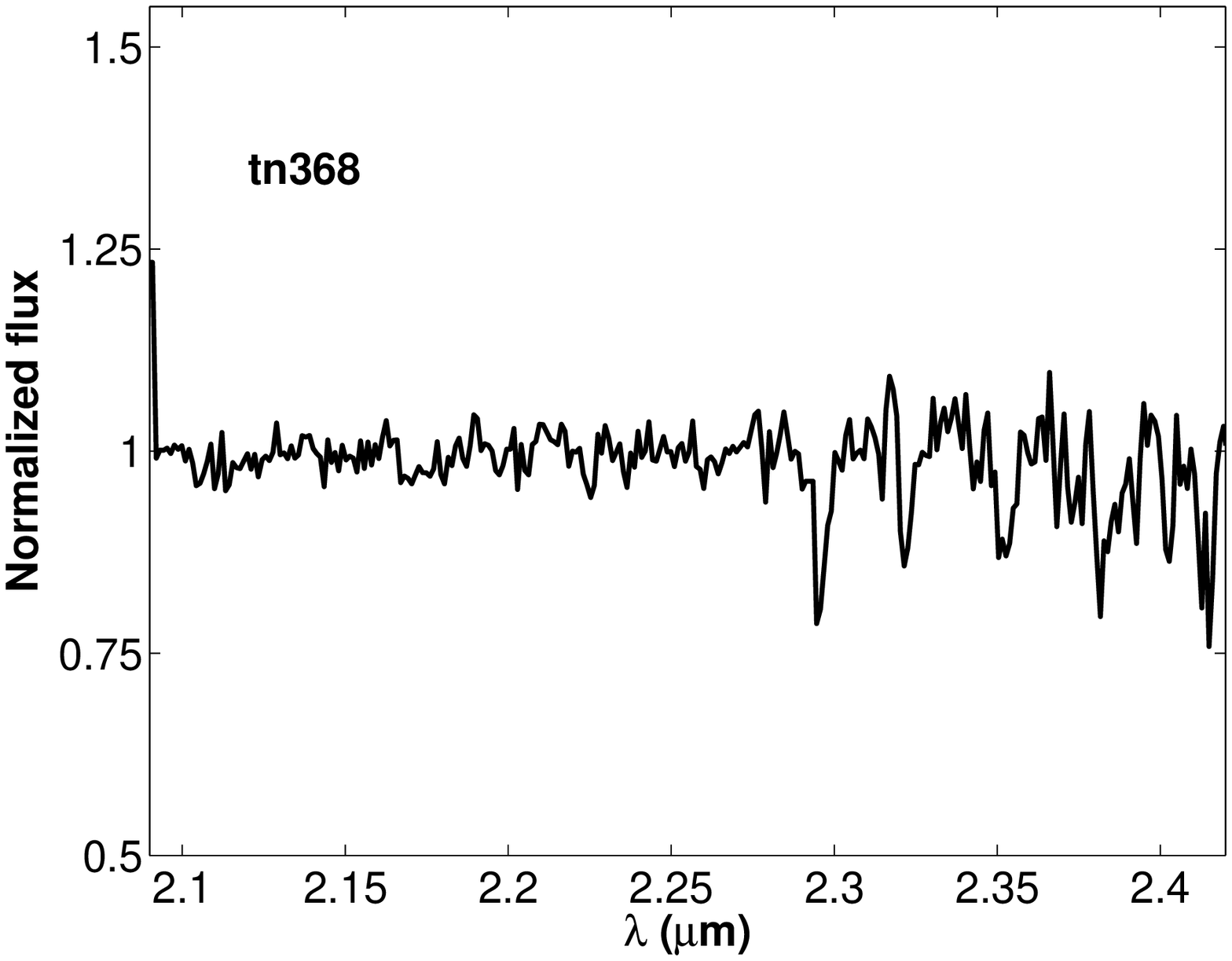}}
	\resizebox*{7cm}{4cm}{\includegraphics{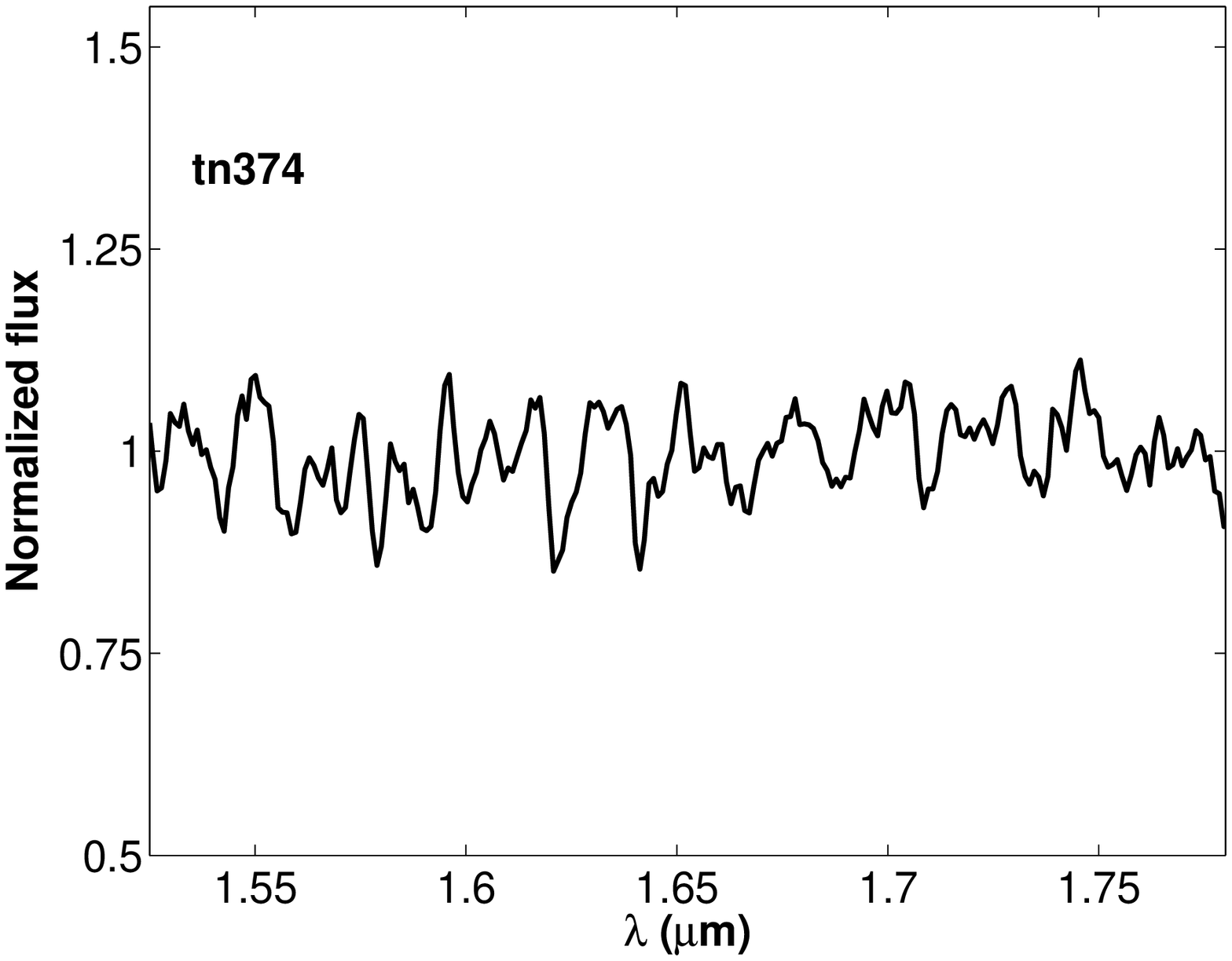}}\resizebox*{7cm}{4cm}{\includegraphics{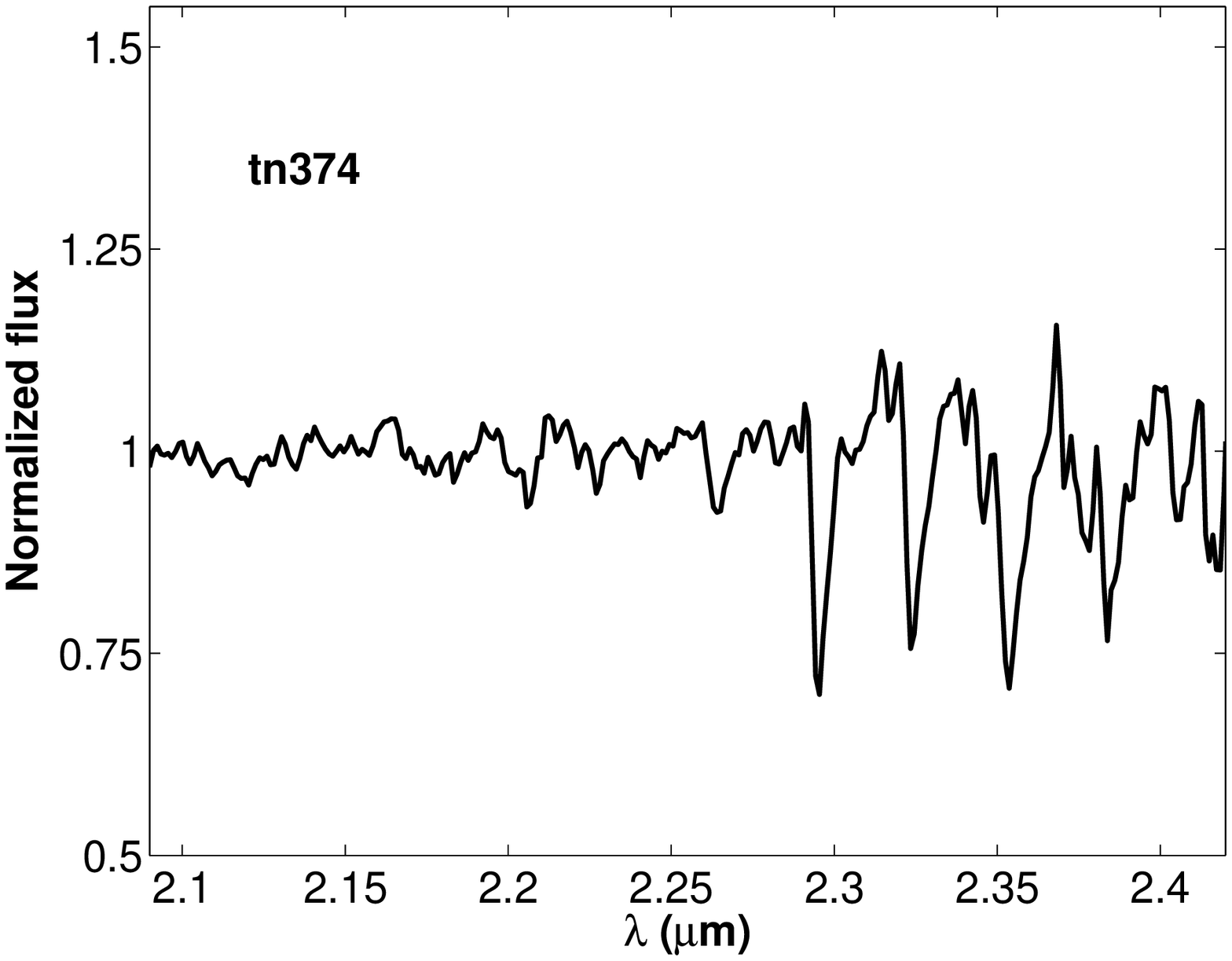}}
	\caption{Same as Fig. \ref{spec} but for l=20$^\circ$, b=0$^\circ$.}
	\end{figure*}
	\newpage
	
	\begin{figure*}[!h]
	\resizebox*{7cm}{4cm}{\includegraphics{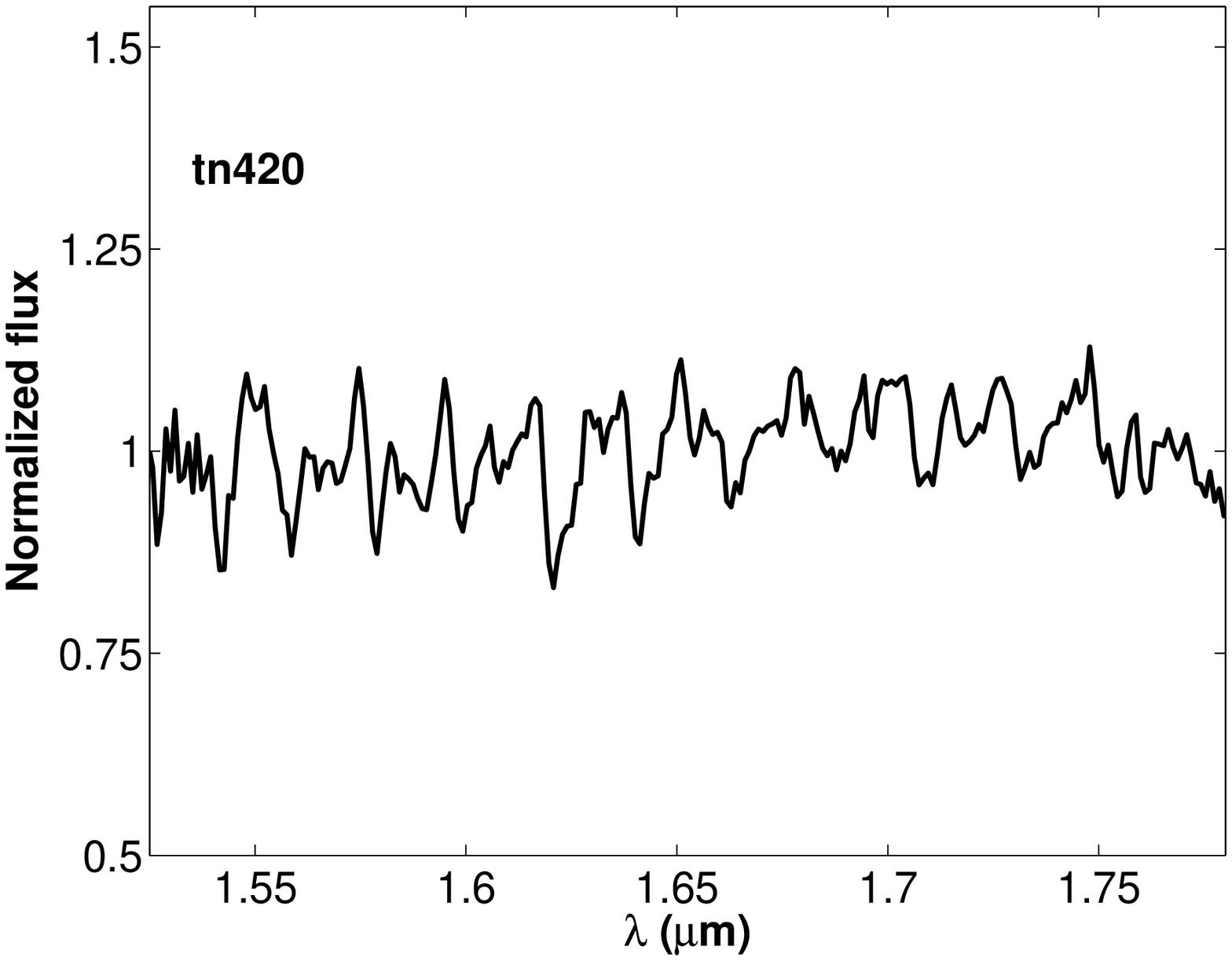}}\resizebox*{7cm}{4cm}{\includegraphics{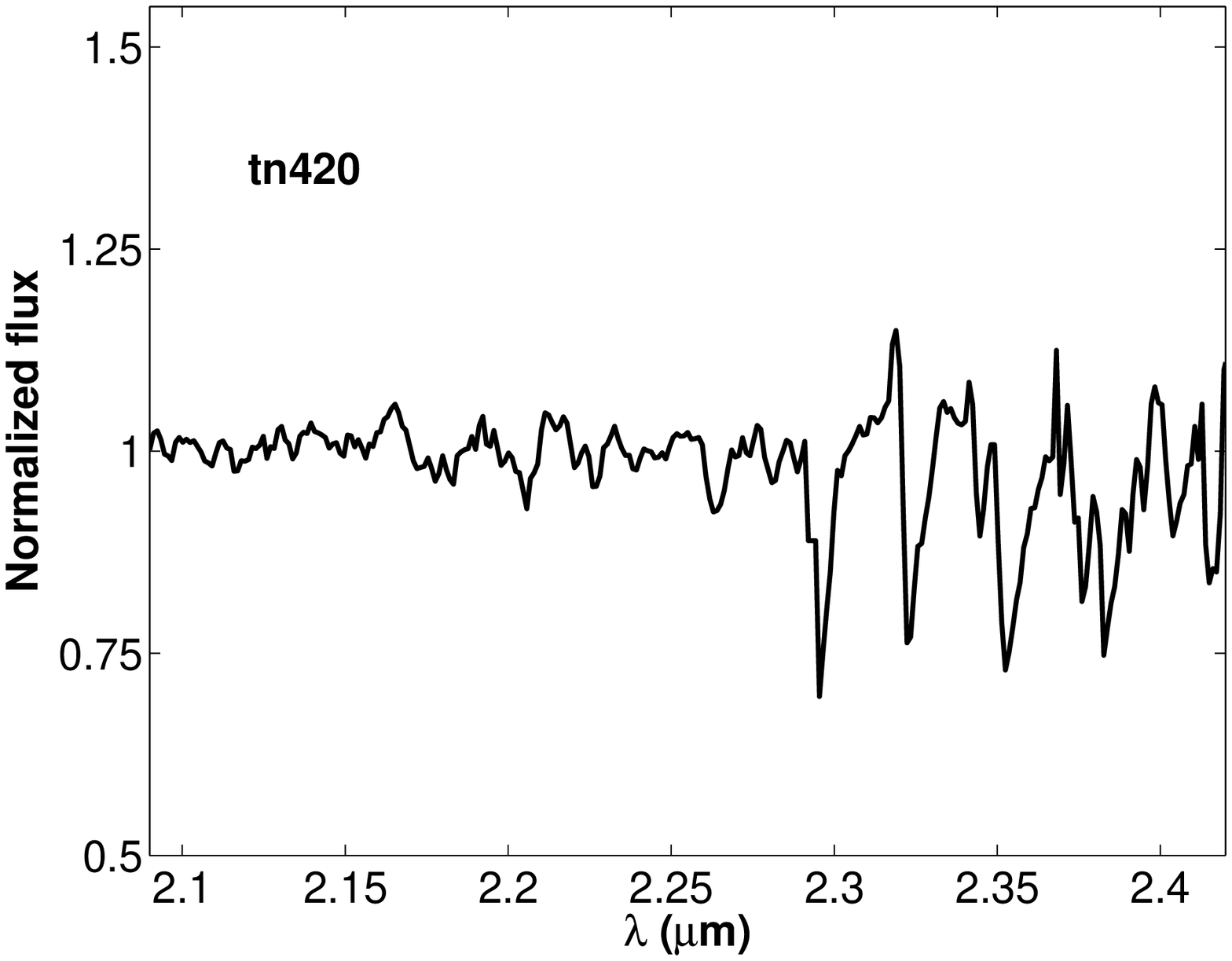}}
	\resizebox*{7cm}{4cm}{\includegraphics{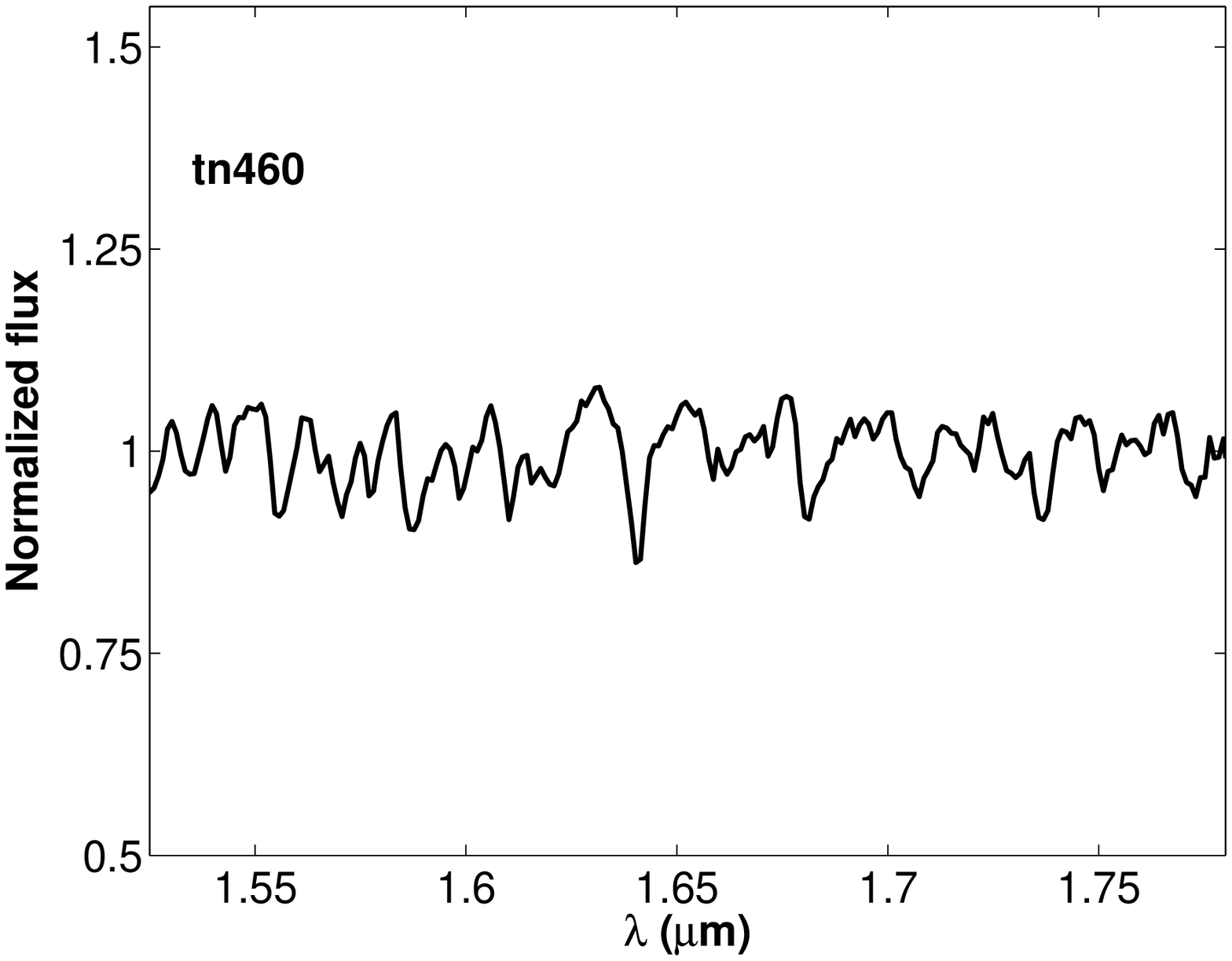}}\resizebox*{7cm}{4cm}{\includegraphics{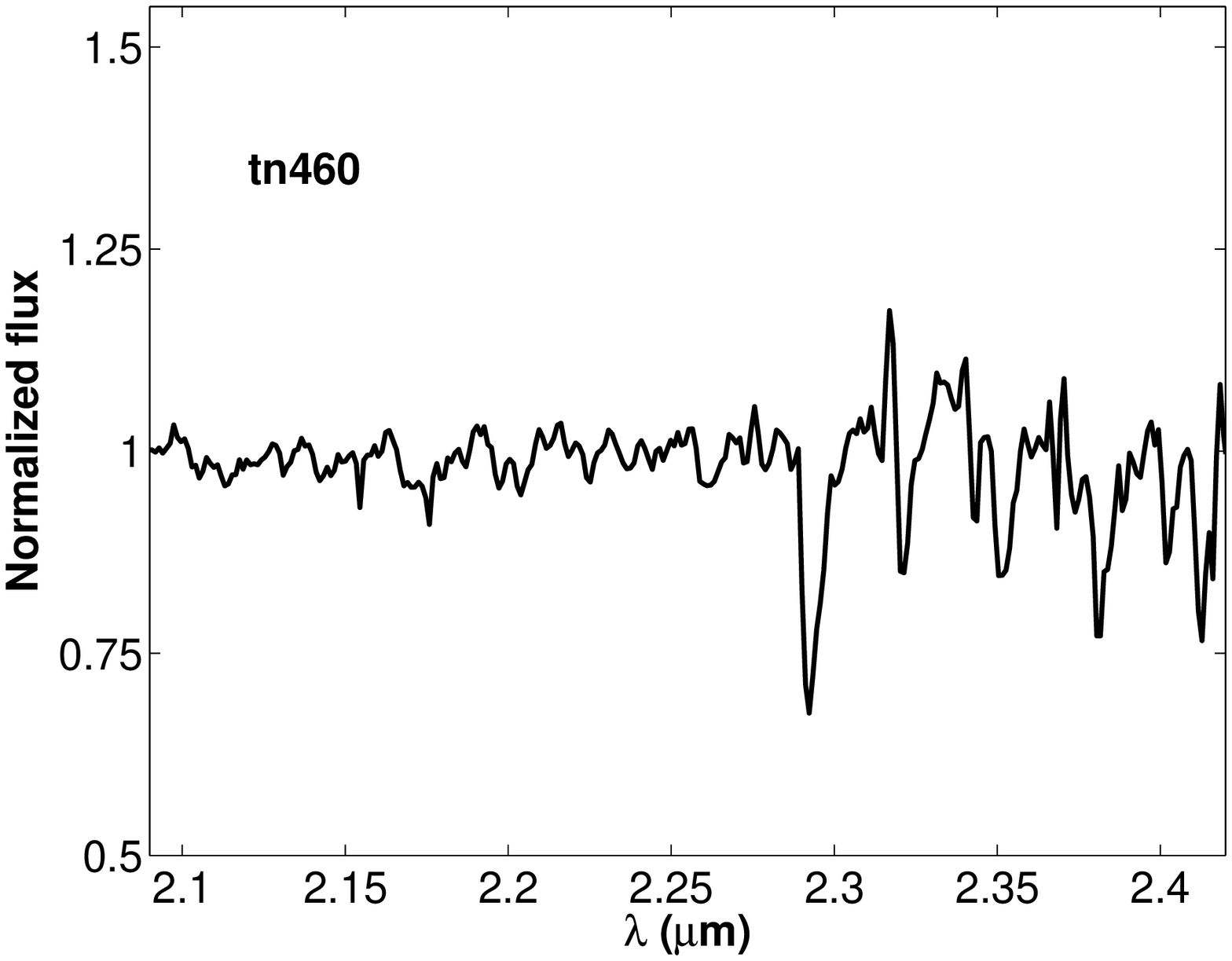}}
	\resizebox*{7cm}{4cm}{\includegraphics{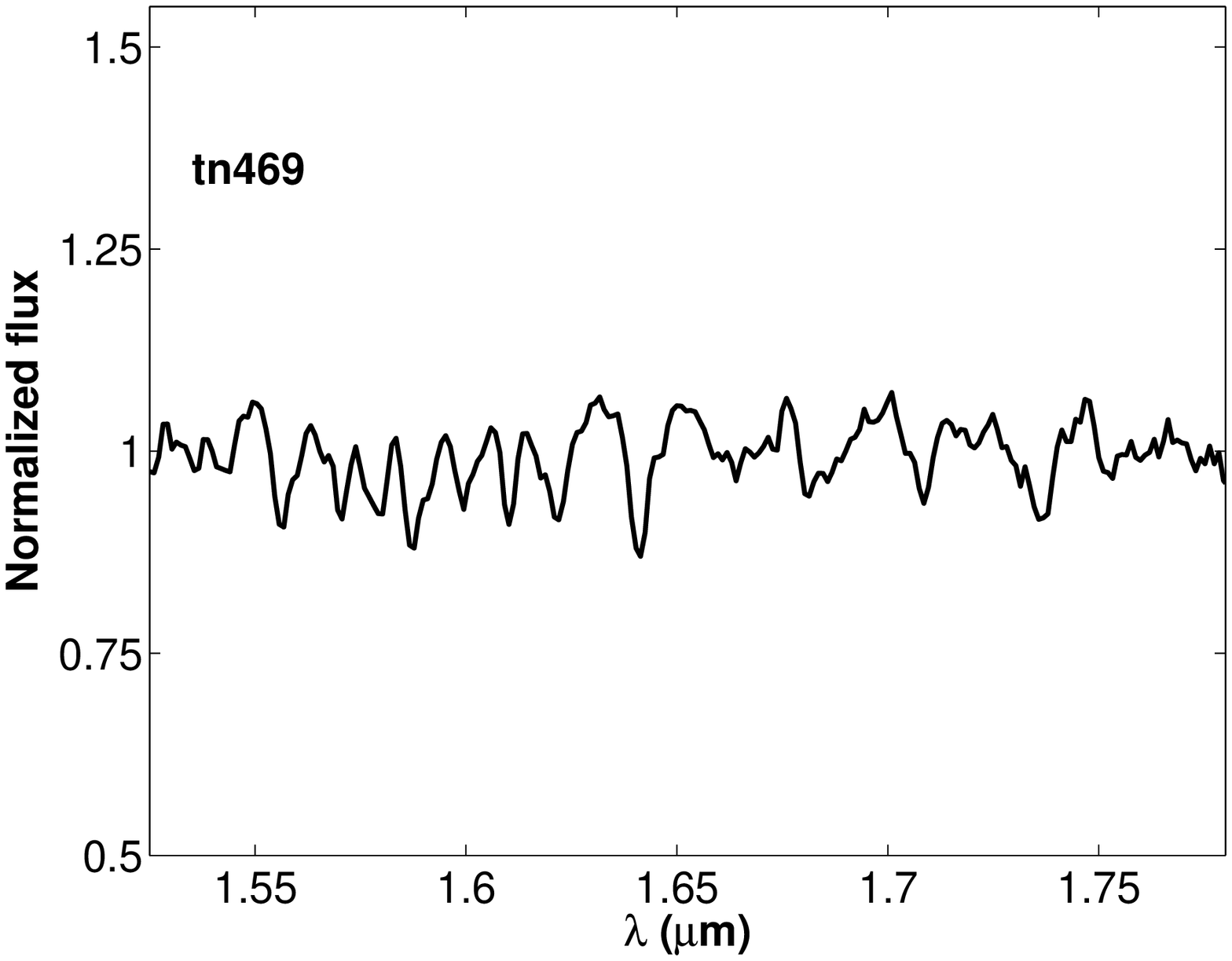}}\resizebox*{7cm}{4cm}{\includegraphics{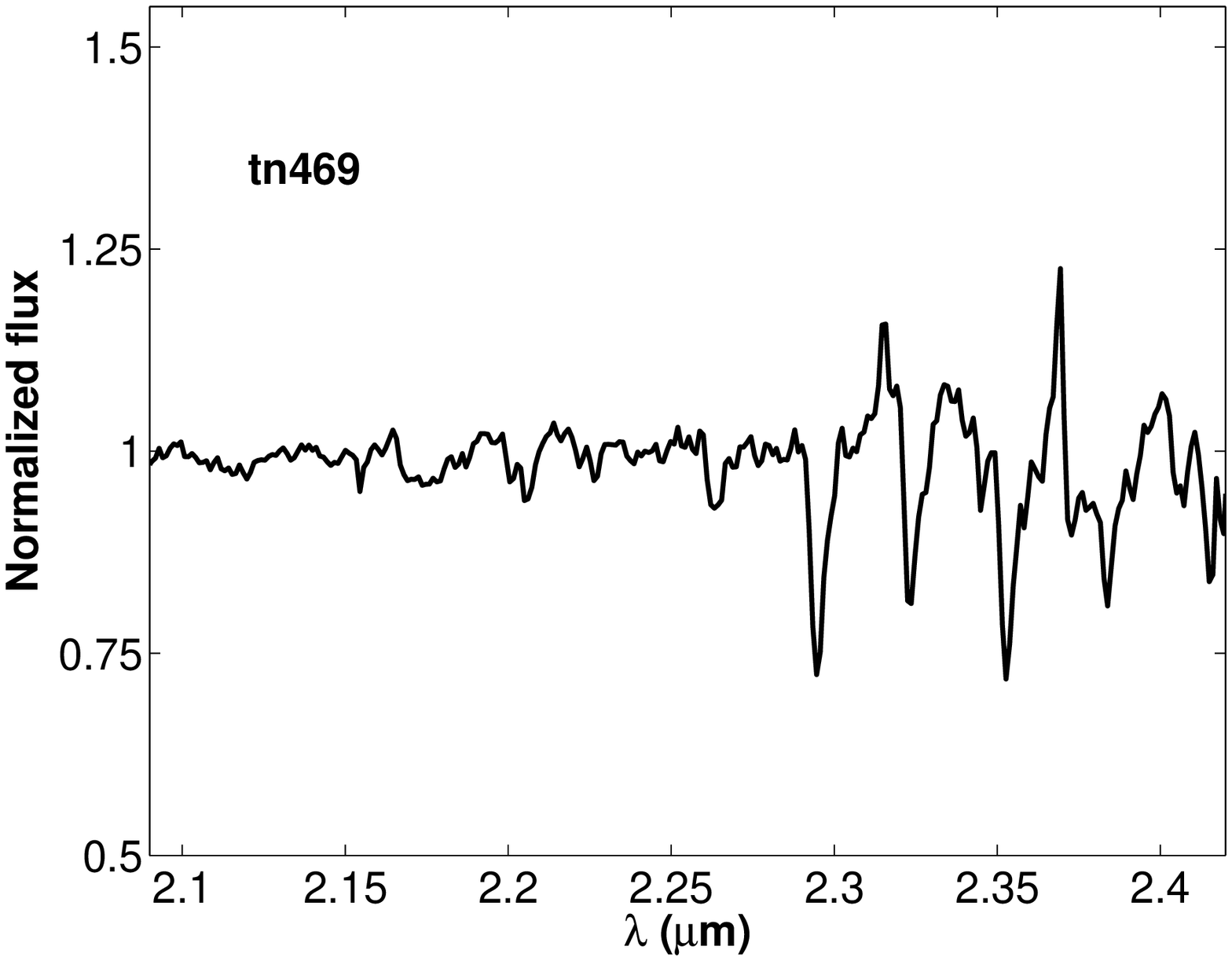}}
	\resizebox*{7cm}{4cm}{\includegraphics{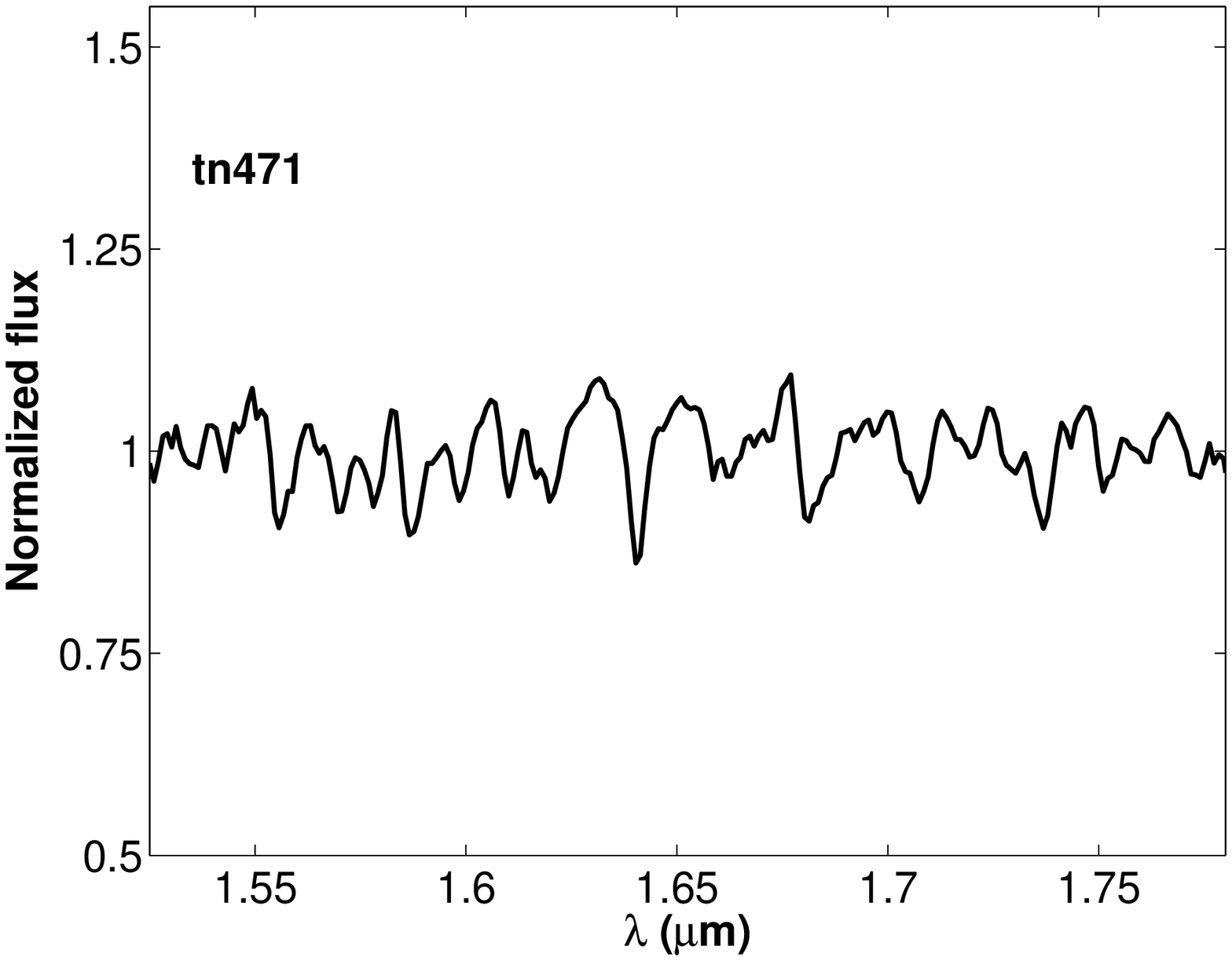}}\resizebox*{7cm}{4cm}{\includegraphics{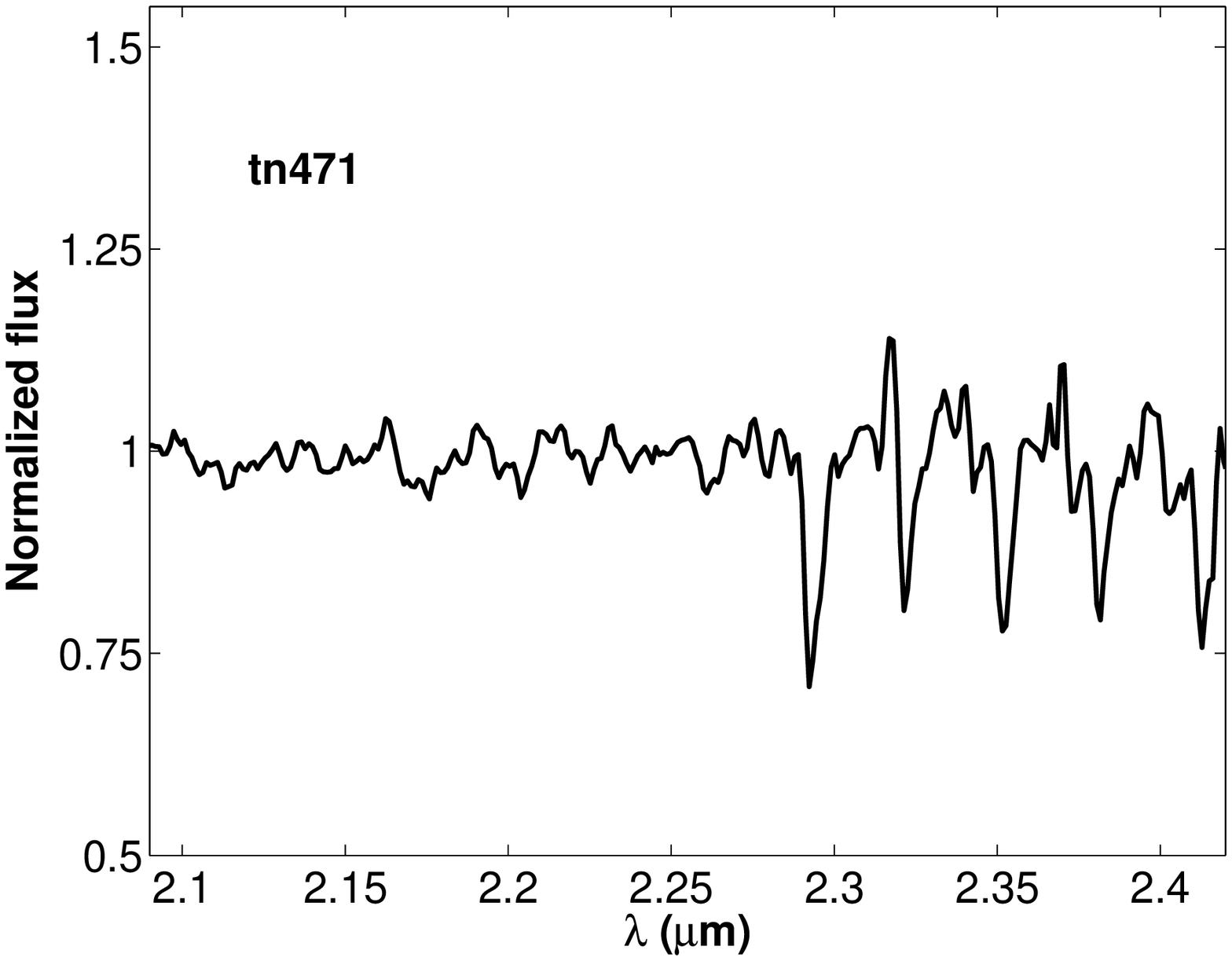}}
	\resizebox*{7cm}{4cm}{\includegraphics{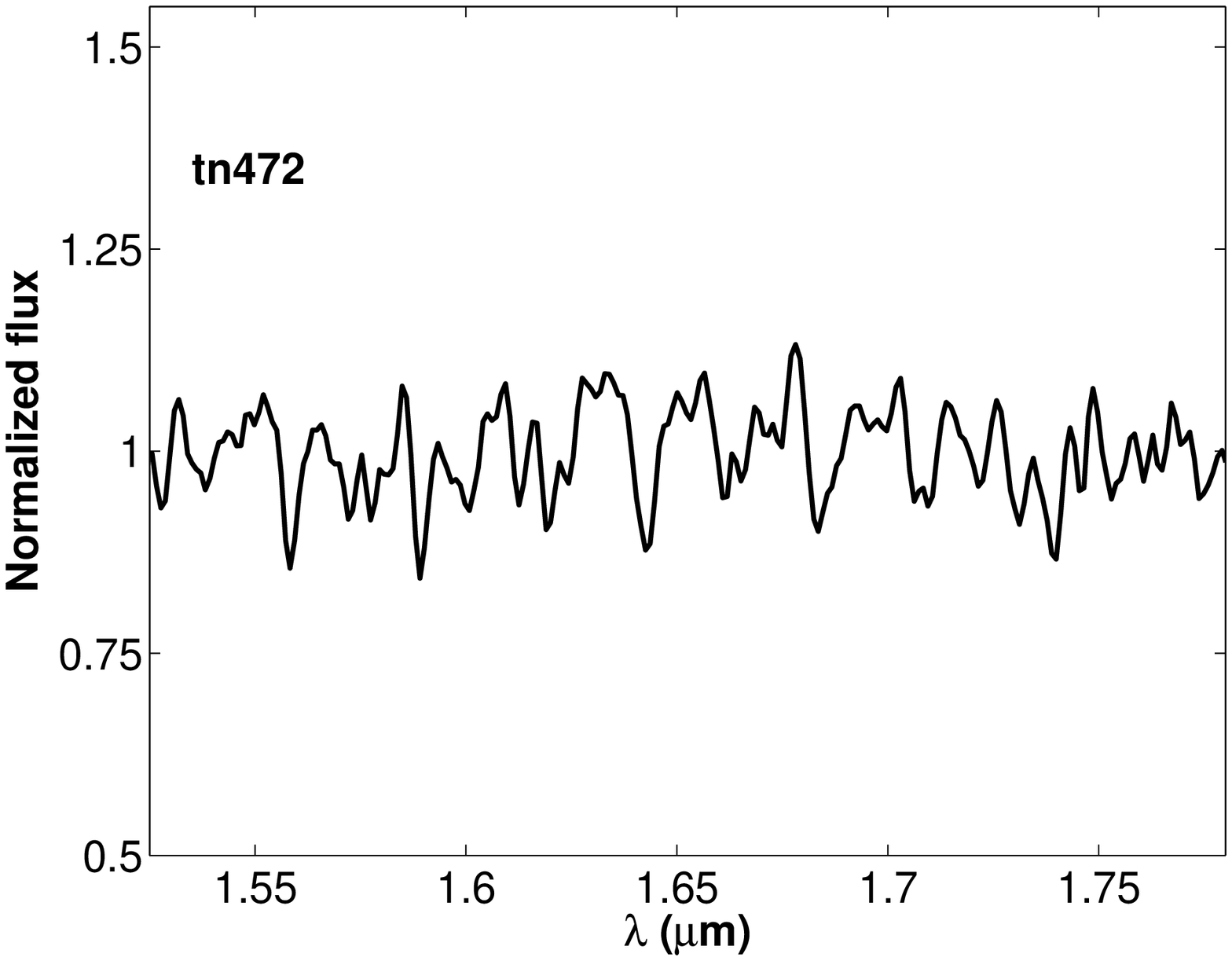}}\resizebox*{7cm}{4cm}{\includegraphics{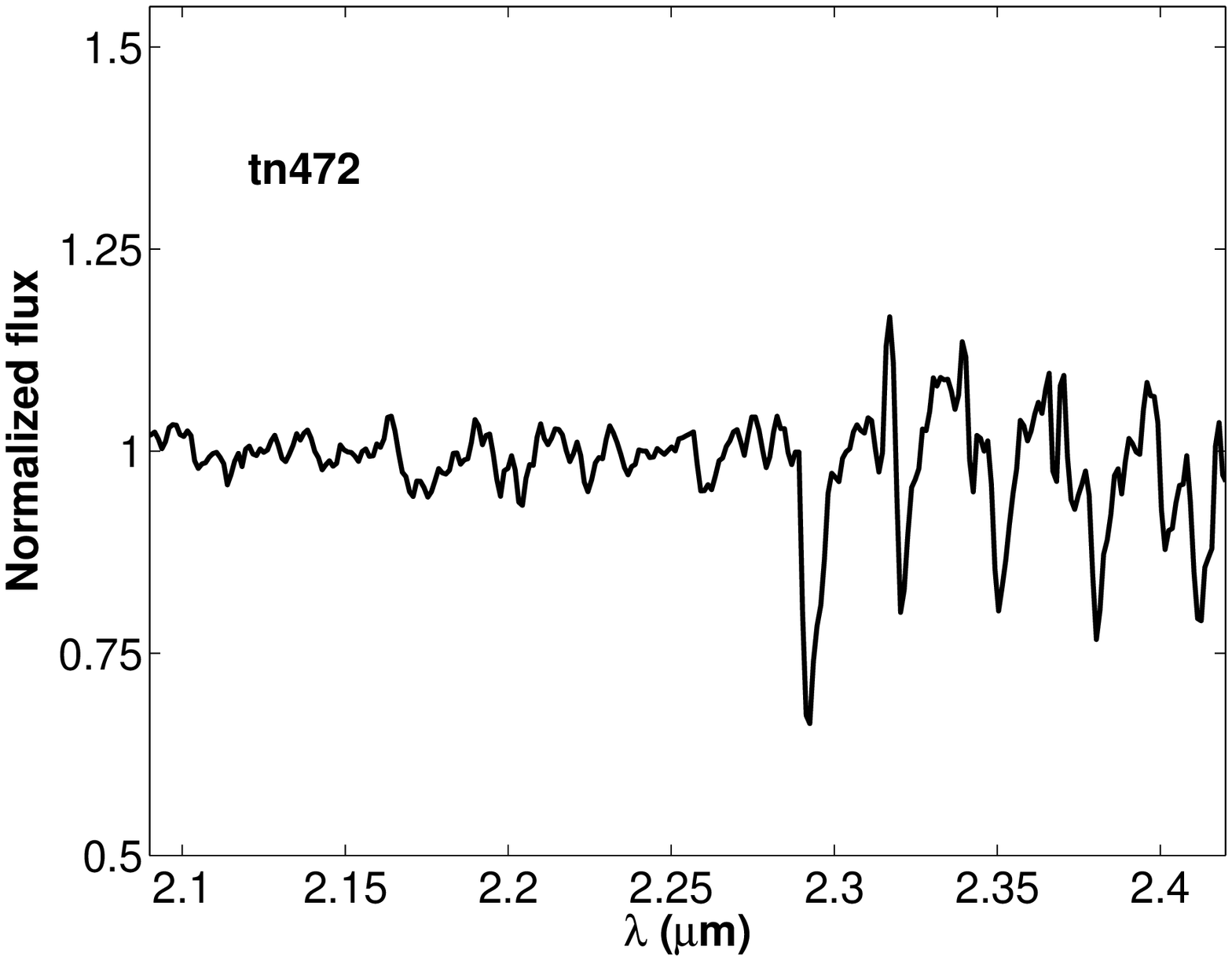}}
	\caption{Same as Fig. \ref{spec} but for l=26$^\circ$, b=0$^\circ$.}
	\end{figure*}
	\newpage
	
	\begin{figure*}[!h]
	\resizebox*{7cm}{4cm}{\includegraphics{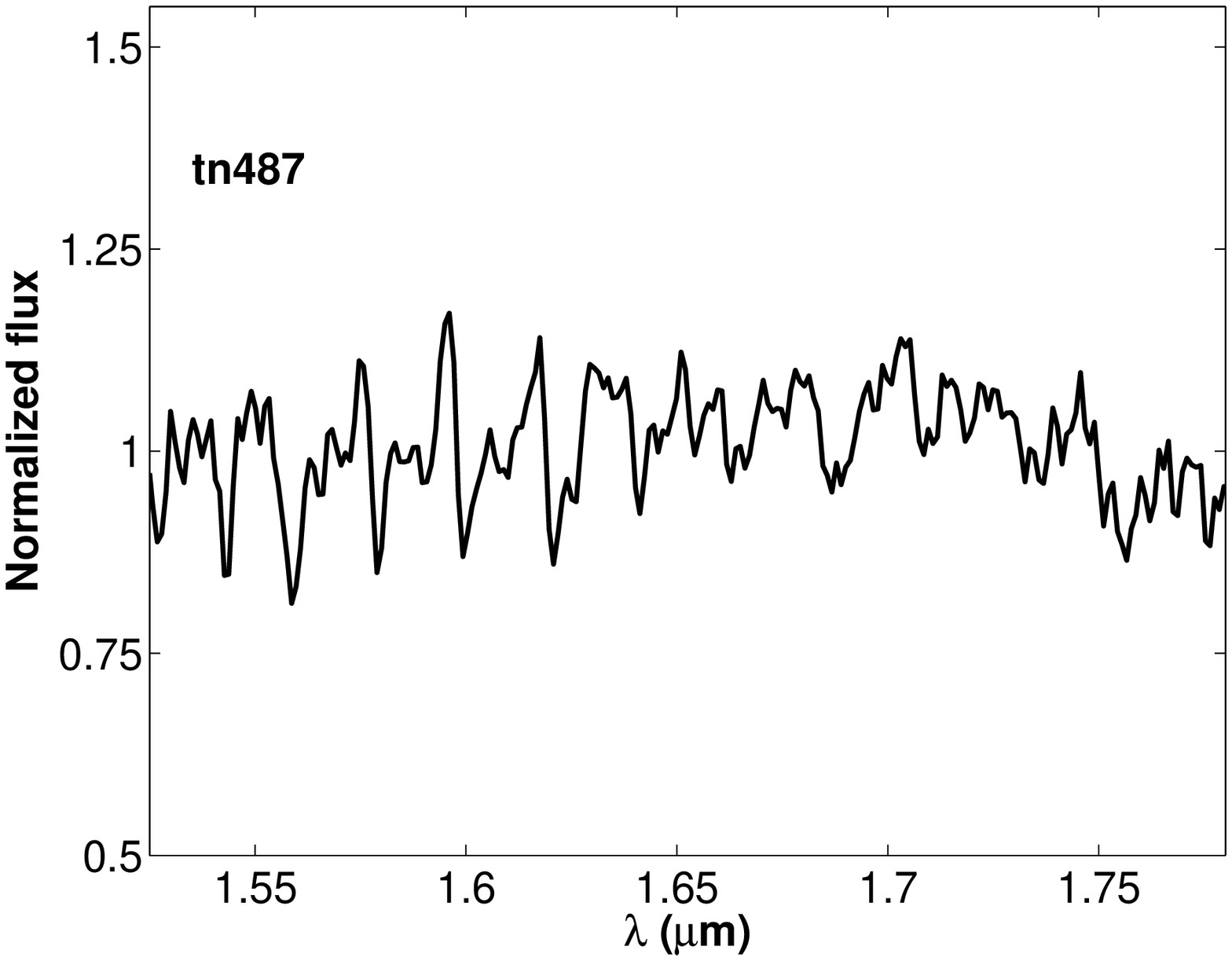}}\resizebox*{7cm}{4cm}{\includegraphics{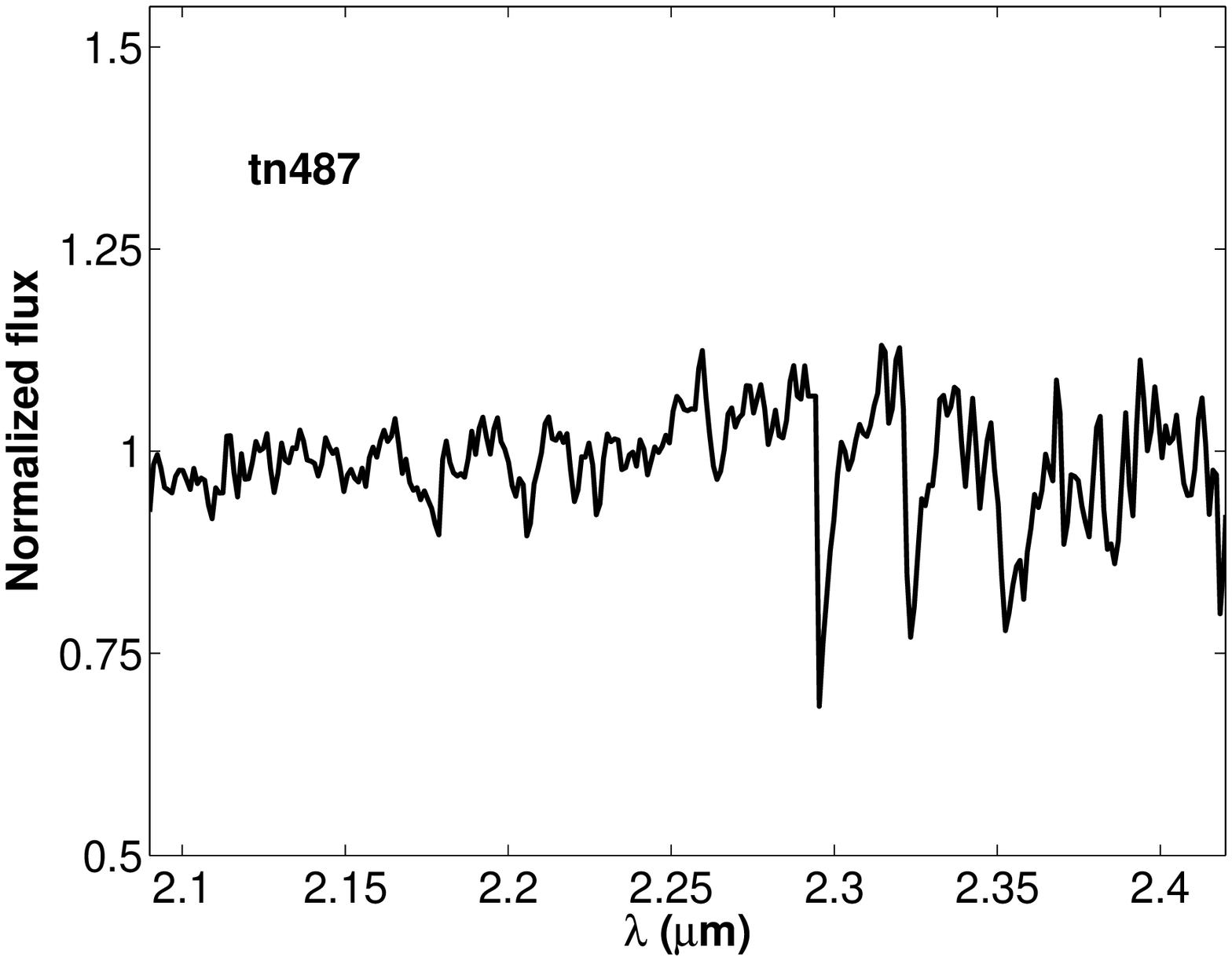}}
	\resizebox*{7cm}{4cm}{\includegraphics{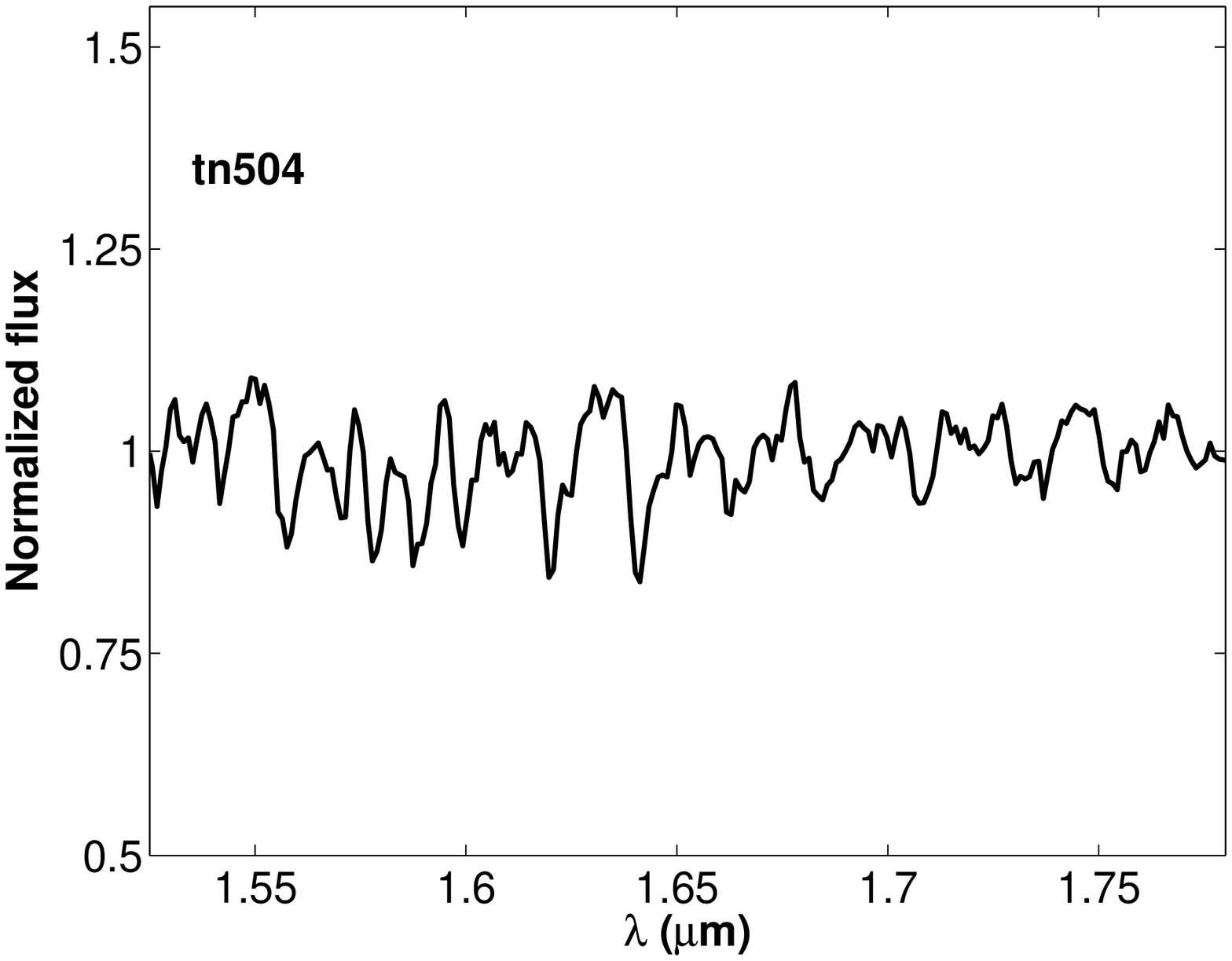}}\resizebox*{7cm}{4cm}{\includegraphics{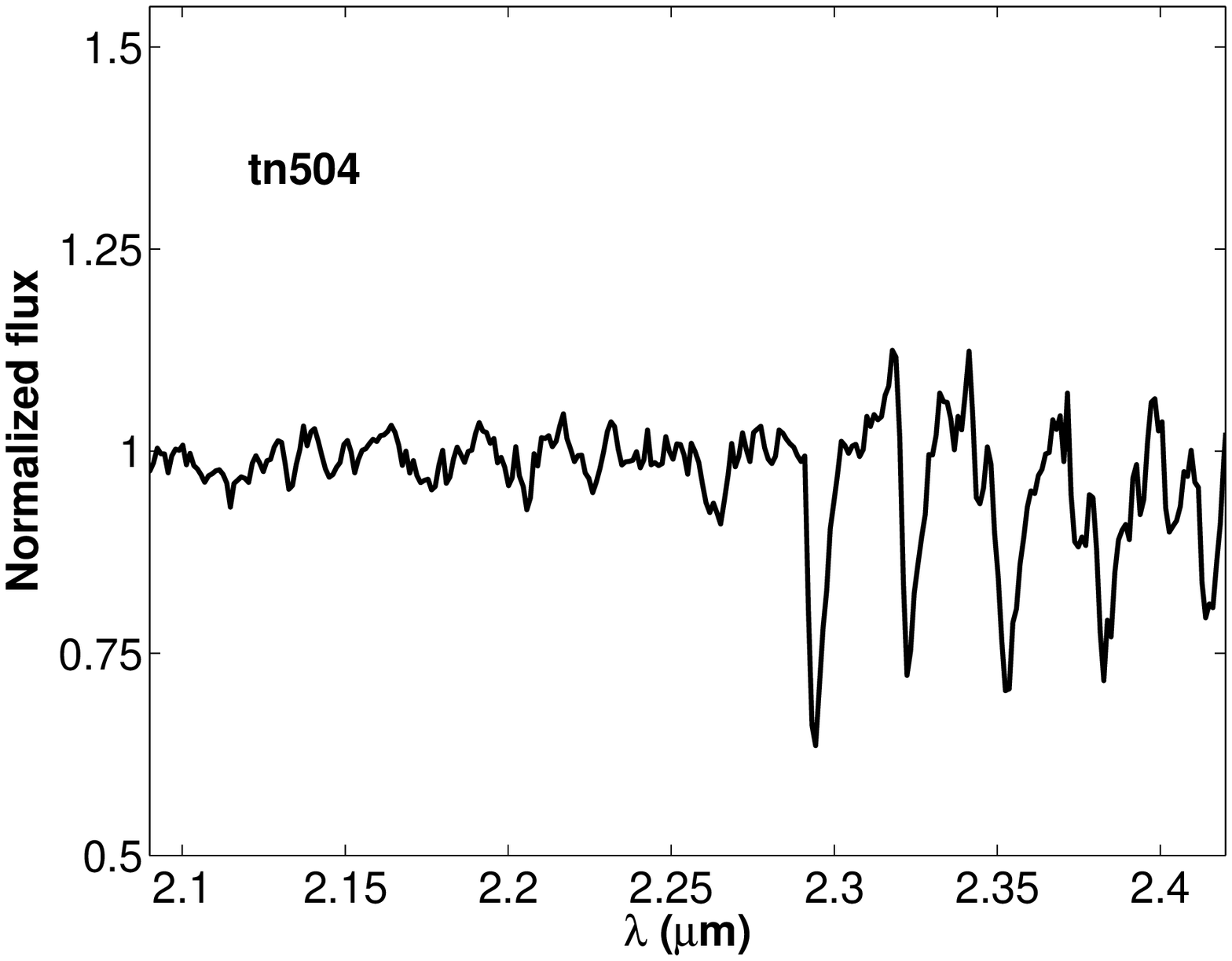}}
	\resizebox*{7cm}{4cm}{\includegraphics{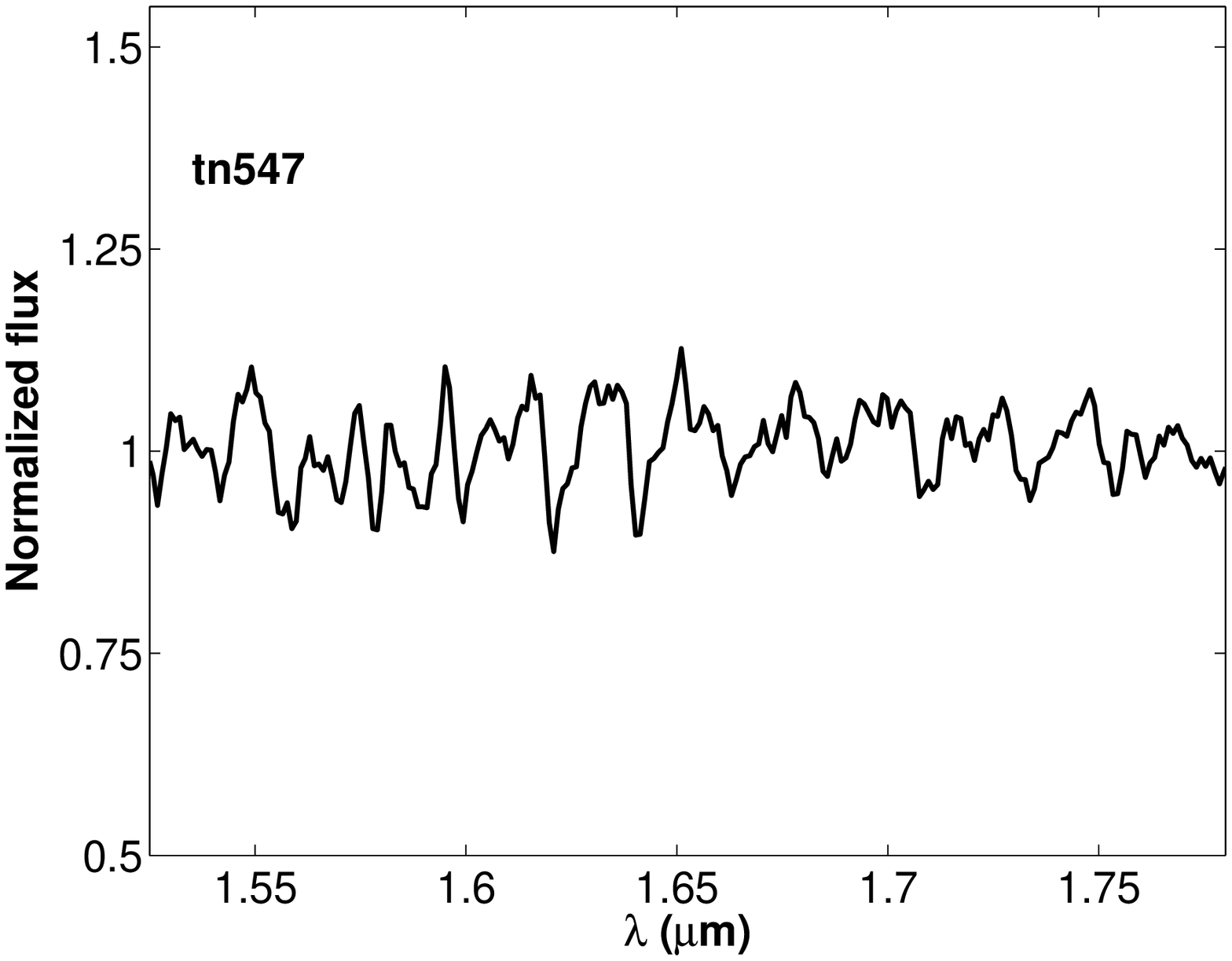}}\resizebox*{7cm}{4cm}{\includegraphics{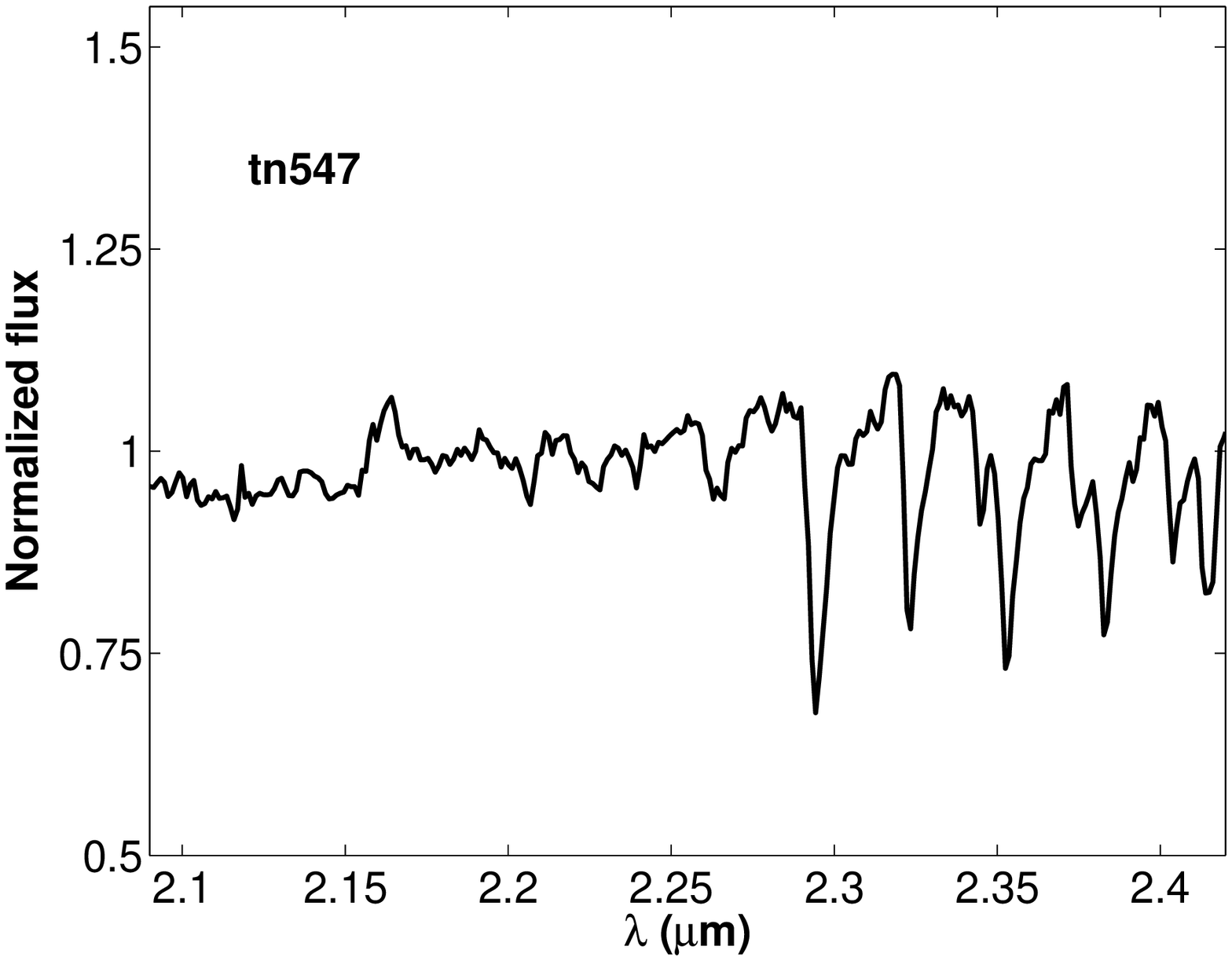}}
	\resizebox*{7cm}{4cm}{\includegraphics{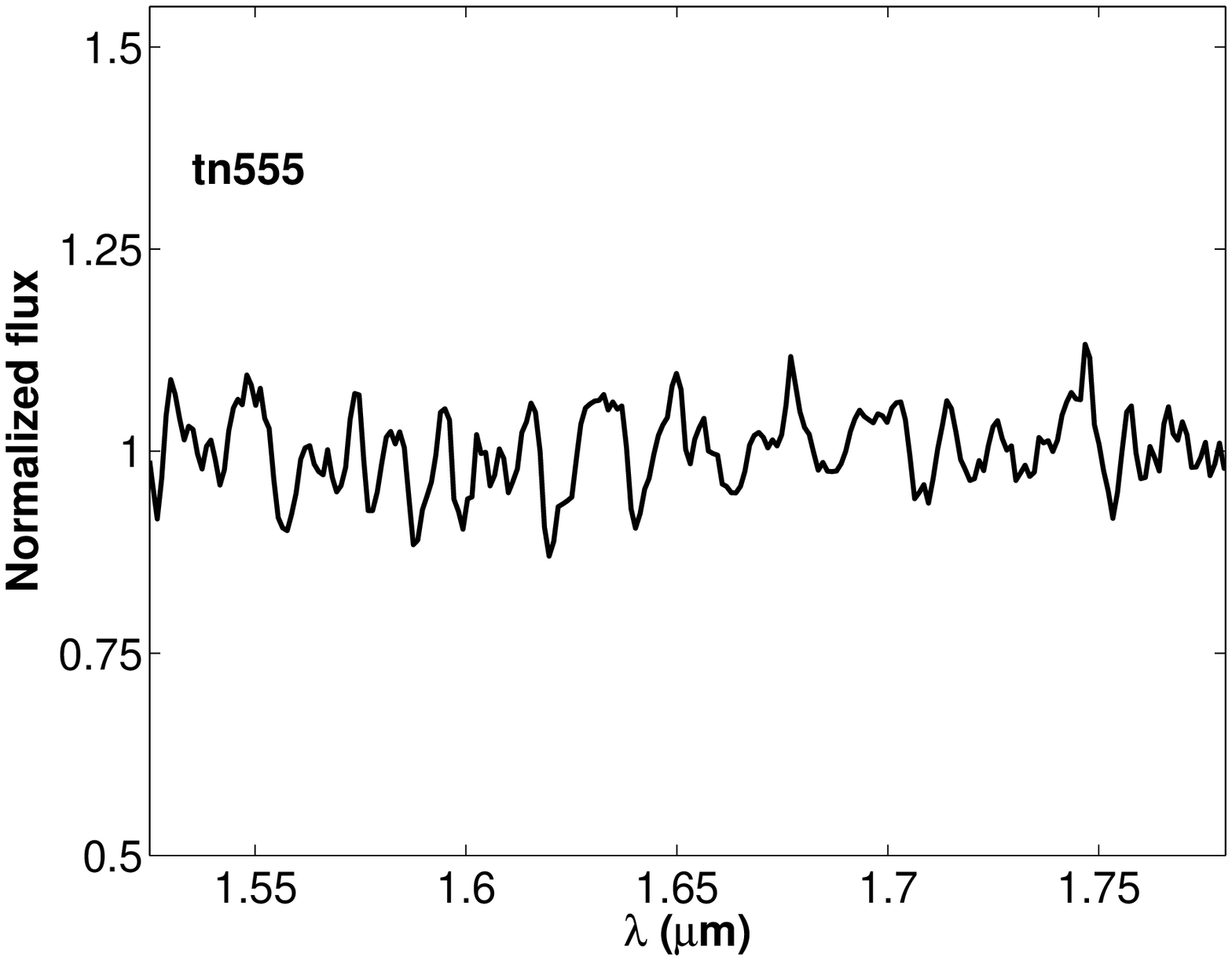}}\resizebox*{7cm}{4cm}{\includegraphics{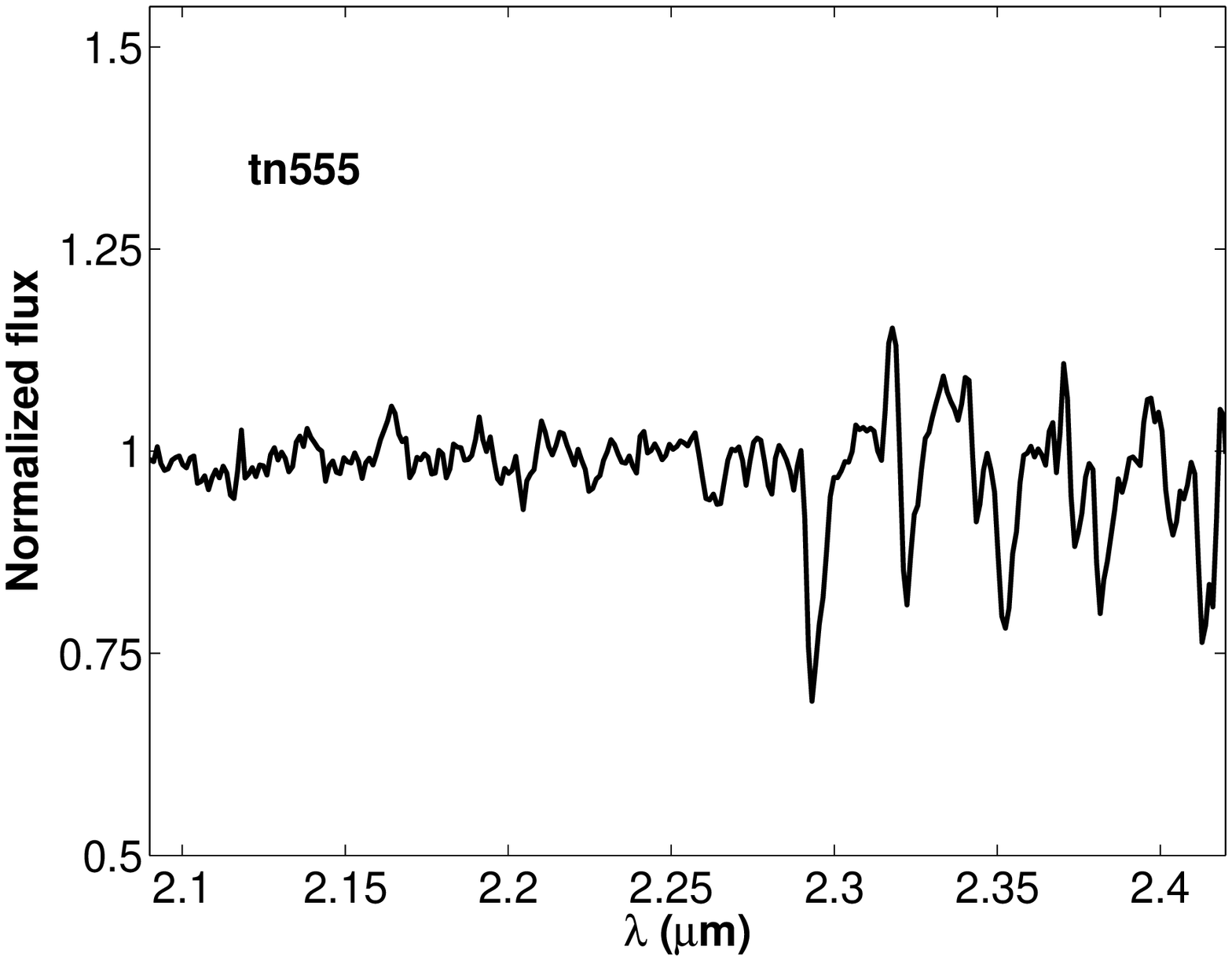}}
	\resizebox*{7cm}{4cm}{\includegraphics{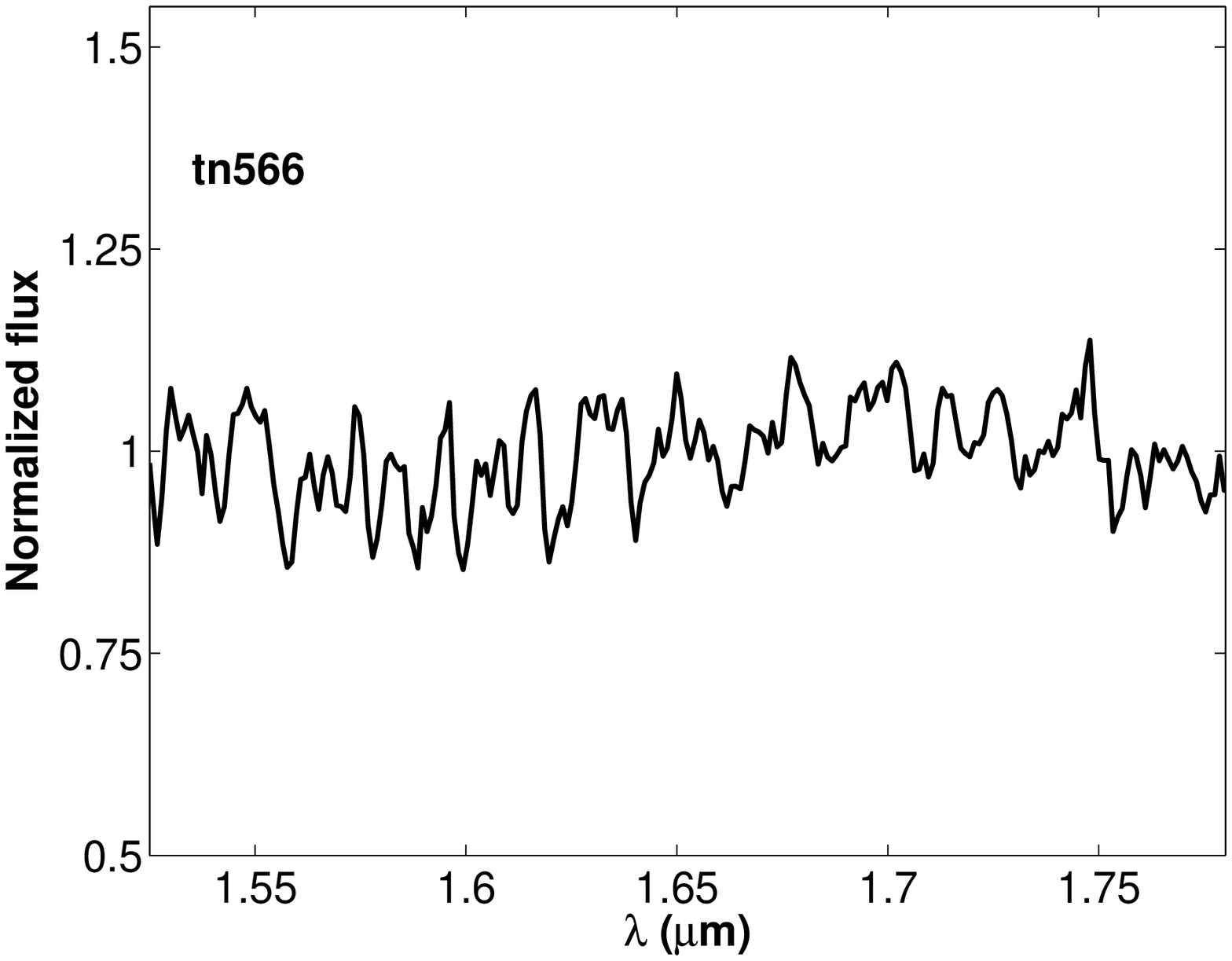}}\resizebox*{7cm}{4cm}{\includegraphics{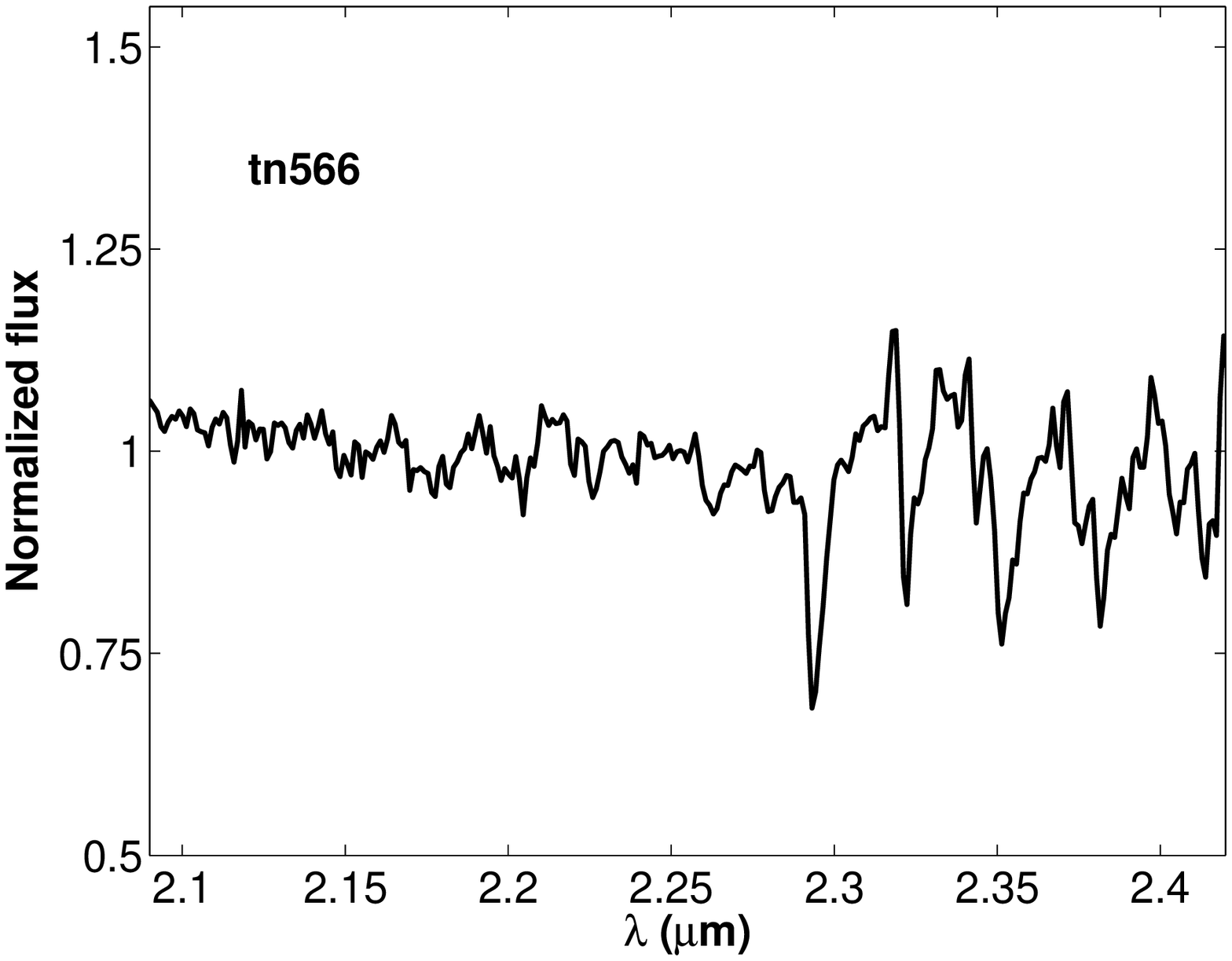}}
	\caption{Same as Fig. \ref{spec} but for l=27$^\circ$, b=0$^\circ$.}
	\end{figure*}

\newpage
\section{List of target stars}
	\begin{table*}[!h]
		\caption[]{Photometric characteristics of observed stars}
		\label{tabla1}
		\begin{center}
		\begin{footnotesize}
		\begin{tabular}{ccccccccc}
		\hline
		ID & $\alpha$ & $\delta$ & $K$ & $(J-K)$ & EW(Na) & EW(Ca)& EW(CO) & [Fe/H]\\ 
		\hline
		\hline
		&  &  &  &  l=07$^\circ$ & b=0$^\circ$ &  &  &  \\ 
		\hline
		tn016  &  18:01:57.9  &  -22:51:20.3   &  9.7  &  3.6  &  3.6  &  4.9  &  16.0  &  -0.15  \\
		tn017  &  18:01:58.1  &  -22:51:31.9   &  8.7  &  5.4  &  5.2  &  2.0  &  21.8  &  -0.56  \\
		tn020  &  18:01:59.7  &  -22:51:24.9   &  9.6  &  5.1  &  1.3  &  1.0  &  13.5  &  -1.00  \\
		tn022  &  18:02:04.7  &  -22:52:35.8   &  9.8  &  4.8  &  2.1  &  4.2  &  22.8  &  -0.43  \\
		tn023  &  18:02:04.8  &  -22:52:55.2   &  9.9  &  3.5  &  1.6  &  3.0  &  18.6  &  -0.71  \\
		tn045  &  18:01:11.4  &  -22:45:40.6   &  10.4  &  3.5  & 1.2  &  1.8  &  6.3  &  -1.15  \\
		tn046  &  18:01:13.1  &  -22:46:35.6   &  8.9  &  6.2  &  4.6  &  5.2  &  27.5  &  -0.10  \\
		tn054  &  18:01:18.3  &  -22:45:56.1   &  7.0  &  4.0  &  4.8  &  4.4  &  24.2  &  -0.24  \\
		tn064  &  18:01:20.4  &  -22:45:36.8   &  8.2  &  3.7  &  3.9  &  4.3  &  22.4  &  -0.23  \\
		tn068  &  18:01:22.4  &  -22:47:59.3   &  9.7  &  3.6  &  3.7  &  4.6  &  23.5  &  -0.19  \\
		tn069  &  18:01:22.9  &  -22:45:23.5   &  10.3  &  3.6  &  3.4  & 4.4  &  22.5  &  -0.24  \\
		tn070  &  18:01:23.0  &  -22:45:05.0   &  10.3  &  3.3  &  2.4  & 2.2  &  17.5  &  -0.66  \\
		tn576  &  18:02:03.8  &  -22:51:55.3   &  8.3  &  7.8  &  3.3  &  5.5  &  27.4  &  -0.06  \\
		\hline
		&  &  &  &  l=12$^\circ$ & b=0$^\circ$ &  &  &  \\ 
		\hline
		tn098  &  18:12:33.8  &  -18:30:33.0   &  8.0  &  3.3  &  3.1  &  4.7  &  26.8  &  -0.23  \\
		tn099  &  18:12:34.4  &  -18:30:54.6   &  9.5  &  3.0  &  1.3  &  2.8  &  16.1  &  -0.86  \\
		tn100  &  18:12:34.6  &  -18:30:14.2   &  7.5  &  3.3  &  2.9  &  5.1  &  19.5  &  -0.14  \\
		tn102  &  18:12:36.2  &  -18:30:49.4   &  8.3  &  5.5  &  4.2  &  5.3  &  28.7  &  -0.09  \\
		tn110  &  18:12:24.1  &  -18:28:32.8   &  7.3  &  4.3  &  1.2  &  1.9  &  18.5  &  -0.93  \\
		tn116  &  18:12:28.3  &  -18:28:40.5   &  9.4  &  5.1  &  1.2  &  3.2  &  17.2  &  -0.83  \\
		tn138  &  18:11:38.4  &  -18:25:42.5   &  9.5  &  2.6  &  2.2  &  3.9  &  17.2  &  -0.50  \\
		tn157  &  18:11:50.1  &  -18:24:22.7   &  9.8  &  2.9  &  13.1  &  10.4  &  28.1  &  -2.13  \\
		tn158  &  18:11:50.1  &  -18:24:32.7   &  6.9  &  5.4  &  --  &  1.2  &  23.0  &  --  \\
		tn160  &  18:11:52.0  &  -18:25:25.8   &  8.8  &  5.9  &  0.6  &  0.4  &  8.8  &  -1.31  \\
		tn185  &  18:11:45.3  &  -18:23:54.5   &  10.1  &  3.1  &  1.9  &  5.0  &  16.4  &  -0.37  \\
		tn186  &  18:11:45.4  &  -18:23:21.7   &  7.2  &  3.2  &  4.8  &  4.3  &  22.7  &  -0.24  \\
		tn191  &  18:11:48.4  &  -18:22:50.7   &  9.1  &  3.1  &  0.6  &  2.3  &  17.2  &  -1.11  \\
		tn192  &  18:11:48.5  &  -18:23:41.8   &  9.5  &  3.1  &  0.8  &  3.6  &  15.5  &  -0.90  \\
		tn581  &  18:12:35.9  &  -18:30:35.9   &  9.9  &  6.6  &  1.4  &  3.0  &  12.3  &  -0.86  \\
		\hline
		&  &  &  &  l=15$^\circ$ & b=0$^\circ$ &  &  &  \\ 
		\hline
		tn199  &  18:18:25.0  &  -15:54:02.5   &  10.0  &  3.1  &  0.5  &  2.0  &  9.2  &  -1.29  \\
		tn203  &  18:18:15.8  &  -15:52:25.2   &  9.2  &  5.1  &  2.9  &  5.2  &  29.8  &  -0.20  \\
		tn204  &  18:18:15.8  &  -15:51:50.5   &  10.1  &  4.6  &  5.0  &  4.1  &  24.1  &  -0.31  \\
		tn212  &  18:18:19.4  &  -15:51:26.0   &  8.4  &  5.2  &  5.2  &  5.1  &  23.0  &  -0.13  \\
		tn213  &  18:18:19.4  &  -15:50:45.3   &  9.6  &  3.6  &  4.9  &  4.7  &  23.1  &  -0.19  \\
		tn214  &  18:18:19.4  &  -15:49:57.6   &  6.9  &  2.2  &  2.3  &  3.4  &  30.4  &  -0.61  \\
		tn225  &  18:18:25.9  &  -15:50:11.8   &  9.1  &  3.1  &  1.2  &  2.5  &  14.8  &  -0.92  \\
		tn233  &  18:18:29.9  &  -15:50:41.8   &  9.5  &  4.0  &  5.0  &  5.3  &  23.0  &  -0.06  \\
		tn253  &  18:17:35.4  &  -15:44:54.8   &  9.7  &  2.0  &  0.1  &  0.6  &  1.1  &  -1.72  \\
		tn254  &  18:17:35.4  &  -15:45:50.8   &  9.6  &  3.2  &  3.8  &  4.5  &  19.7  &  -0.20  \\
		tn255  &  18:17:36.3  &  -15:45:32.5   &  8.7  &  5.9  &  3.4  &  3.1  &  24.7  &  -0.45  \\
		tn258  &  18:17:37.9  &  -15:46:27.7   &  8.6  &  3.1  &  2.6  &  4.0  &  17.9  &  -0.41  \\
		tn259  &  18:17:38.6  &  -15:47:13.3   &  9.8  &  3.8  &  2.4  &  3.3  &  16.5  &  -0.55  \\
		tn261  &  18:17:38.8  &  -15:47:40.3   &  5.8  &  4.7  &  4.0  &  5.3  &  25.7  &  -0.04  \\
		tn590  &  18:17:38.7  &  -15:48:05.7   &  9.5  &  5.6  &  2.9  &  4.3  &  25.5  &  -0.32  \\
		tn593  &  18:17:41.0  &  -15:47:34.0   &  9.9  &  6.9  &  2.8  &  3.9  &  18.4  &  -0.39  \\
		\hline
		\end{tabular}
		\end{footnotesize}
		\end{center}
	\end{table*}

	\addtocounter{table}{-1}
	\begin{table*}[!h]
		\caption[]{Photometric characteristics of observed stars (cont.)}
		\label{tabla2}
		\begin{center}
		\begin{footnotesize}
		\begin{tabular}{ccccccccc}
		
		\hline
		ID & $\alpha$ & $\delta$ & $K$ & $(J-K)$ & EW(Na) & EW(Ca)& EW(CO) & [Fe/H]\\ 
		\hline
		\hline
		&  &  &  &  l=20$^\circ$ & b=0$^\circ$ &  &  & \\ 
		\hline
		tn284  &  18:27:56.3  &  -11:25:31.8   &  9.6  &  2.9  &  5.3  &  4.8  &  12.2  &  -0.29  \\
		tn285  &  18:27:56.4  &  -11:26:50.0   &  9.1  &  2.3  &  2.4  &  2.7  &  16.5  &  -0.63  \\
		tn290  &  18:27:58.6  &  -11:26:11.4   &  6.7  &  2.6  &  3.3  &  4.3  &  20.0  &  -0.27  \\
		tn293  &  18:27:59.2  &  -11:27:17.1   &  5.4  &  3.3  &  5.2  &  5.7  &  29.2  &  -0.04  \\
		tn298  &  18:28:00.8  &  -11:28:57.4   &  7.7  &  3.7  &  2.7  &  5.5  &  25.3  &  -0.10  \\
		tn300  &  18:28:01.3  &  -11:27:25.4   &  6.1  &  3.1  &  3.2  &  3.7  &  25.5  &  -0.38  \\
		tn305  &  18:28:02.5  &  -11:25:55.9   &  7.1  &  2.8  &  3.2  &  4.7  &  23.0  &  -0.19  \\
		tn314  &  18:28:05.4  &  -11:25:34.4   &  8.8  &  3.1  &  2.8  &  5.0  &  23.7  &  -0.19  \\
		tn319  &  18:28:06.5  &  -11:27:10.7   &  6.2  &  3.1  &  3.2  &  4.4  &  27.8  &  -0.30  \\
		tn340  &  18:28:13.5  &  -11:27:49.6   &  9.9  &  2.5  &  4.9  &  3.5  &  8.9  &  -0.54  \\
		tn361  &  18:27:18.8  &  -11:20:39.0   &  8.2  &  2.4  &  5.4  &  4.5  &  19.5  &  -0.28  \\
		tn362  &  18:27:18.8  &  -11:19:53.5   &  6.6  &  2.6  &  5.8  &  4.1  &  21.2  &  -0.40  \\
		tn364  &  18:27:19.6  &  -11:21:46.7   &  9.8  &  4.4  &  4.4  &  9.5  &  18.0  &  1.32  \\
		tn365  &  18:27:19.6  &  -11:19:23.8   &  9.9  &  2.6  &  1.7  &  3.3  &  15.5  &  -0.70  \\
		tn367  &  18:27:19.7  &  -11:18:53.2   &  9.6  &  2.0  &  1.0  &  2.2  &  17.3  &  -0.97  \\
		tn368  &  18:27:19.7  &  -11:19:13.0   &  10.0  &  2.2  &  0.8  &  2.2  &  18.4  &  -1.02  \\
		tn374  &  18:27:21.7  &  -11:21:05.8   &  6.4  &  3.3  &  5.2  &  4.5  &  22.8  &  -0.26  \\
		tn377  &  18:27:22.7  &  -11:21:48.3   &  8.8  &  4.4  &  4.6  &  4.6  &  23.8  &  -0.18  \\
		tn381  &  18:27:24.3  &  -11:20:26.3   &  8.8  &  2.7  &  7.8  &  9.7  &  22.3  &  0.80  \\
		tn599  &  18:27:26.6  &  -11:21:59.2   &  9.7  &  3.6  &  4.7  &  3.7  &  20.8  &  -0.35  \\
		\hline
		&  &  &  &  l=26$^\circ$ & b=0$^\circ$ &  &  &  \\ 
		\hline
		tn417  &  18:39:18.5  &  -06:07:13.7   &  7.2  &  4.6  &  3.4  &  4.1  &  20.4  &  -0.29  \\
		tn418  &  18:39:18.6  &  -06:07:05.3   &  9.6  &  3.2  &  4.7  &  12.3  &  24.9  &  2.71  \\
		tn420  &  18:39:19.6  &  -06:08:06.8   &  7.5  &  4.8  &  4.6  &  4.7  &  26.5  &  -0.19  \\
		tn422  &  18:39:20.9  &  -06:08:16.4   &  9.9  &  3.0  &  1.5  &  1.9  &  11.0  &  -0.96  \\
		tn423  &  18:39:22.6  &  -06:07:06.8   &  8.1  &  3.2  &  2.3  &  4.6  &  24.7  &  -0.35  \\
		tn425  &  18:39:25.2  &  -06:09:12.3   &  7.1  &  4.4  &  2.6  &  4.9  &  28.1  &  -0.28  \\
		tn433  &  18:39:15.1  &  -06:05:19.1   &  5.6  &  3.1  &  4.4  &  4.2  &  28.3  &  -0.30  \\
		tn434  &  18:39:16.0  &  -06:05:03.2   &  6.1  &  2.9  &  2.4  &  3.8  &  27.4  &  -0.49  \\
		tn436  &  18:39:18.3  &  -06:05:42.4   &  6.7  &  3.5  &  3.1  &  5.1  &  24.3  &  -0.12  \\
		tn447  &  18:38:30.3  &  -06:02:32.0   &  9.7  &  5.5  &  5.7  &  4.6  &  24.0  &  -0.30  \\
		tn448  &  18:38:30.6  &  -06:02:24.8   &  8.9  &  3.2  &  3.8  &  3.2  &  19.7  &  -0.40  \\
		tn452  &  18:38:34.8  &  -06:01:24.3   &  9.9  &  5.0  &  5.9  &  3.3  &  20.2  &  -0.53  \\
		tn456  &  18:38:36.6  &  -06:02:16.4   &  8.2  &  3.5  &  5.6  &  3.0  &  19.4  &  -0.53  \\
		tn457  &  18:38:37.6  &  -06:03:15.4   &  10.0  &  3.2  &  6.5  &  5.9  &  18.4  &  -0.16  \\
		tn460  &  18:38:38.7  &  -06:00:57.8   &  10.1  &  3.5  &  1.6  &  4.5  &  16.6  &  -0.50  \\
		tn461  &  18:38:38.8  &  -06:00:21.5   &  9.3  &  3.9  &  0.2  &  3.0  &  16.2  &  -1.15  \\
		tn466  &  18:38:40.0  &  -06:00:35.6   &  9.6  &  2.7  &  1.3  &  2.4  &  16.1  &  -0.88  \\
		tn467  &  18:38:40.0  &  -06:00:08.6   &  9.0  &  3.9  &  1.8  &  3.6  &  17.7  &  -0.61  \\
		tn469  &  18:38:41.0  &  -06:03:00.8   &  6.7  &  2.5  &  3.9  &  4.3  &  20.9  &  -0.23  \\
		tn471  &  18:38:41.9  &  -06:02:16.9   &  9.1  &  3.6  &  2.1  &  4.1  &  20.3  &  -0.46  \\
		tn472  &  18:38:42.0  &  -06:01:25.1   &  8.0  &  2.9  &  2.3  &  4.1  &  24.6  &  -0.43  \\
		tn473  &  18:38:42.0  &  -06:02:44.6   &  9.8  &  2.9  &  --  &  2.0  &  13.4  &  --  \\
		tn605  &  18:39:16.6  &  -06:05:39.7   &  8.8  &  8.2  &  3.5  &  4.7  &  30.3  &  -0.26  \\
		tn606  &  18:38:37.4  &  -06:03:39.4   &  9.8  &  6.7  &  7.4  &  3.9  &  36.2  &  -1.02  \\
		\hline
		\end{tabular}
		\end{footnotesize}
		\end{center}
	\end{table*}
	
	\addtocounter{table}{-1}
	\begin{table*}[!h]
		\caption[]{Photometric characteristics of observed stars (cont.)}
		\label{tabla2}
		\begin{center}
		\begin{footnotesize}
		\begin{tabular}{ccccccccc}
		
		\hline
		ID & $\alpha$ & $\delta$ & $K$ & $(J-K)$ & EW(Na) & EW(Ca)& EW(CO) & [Fe/H]\\ 
		\hline
		\hline
		&  &  &  &  l=27$^\circ$ & b=0$^\circ$ &  &  &  \\ 
		\hline
		tn487  &  18:40:59.6  &  -05:15:15.7   &  5.1  &  3.3  &  4.7  &  1.5  &  28.4  &  -0.60  \\
		tn490  &  18:41:00.3  &  -05:14:24.3   &  9.2  &  4.4  &  2.8  &  5.5  &  18.4  &  -0.09  \\
		tn491  &  18:41:00.8  &  -05:15:27.8   &  8.6  &  3.1  &  3.9  &  3.5  &  22.0  &  -0.36  \\
		tn497  &  18:41:03.7  &  -05:14:40.6   &  8.1  &  4.3  &  3.9  &  1.7  &  20.7  &  -0.53  \\
		tn503  &  18:41:06.7  &  -05:15:35.3   &  8.9  &  3.8  &  2.0  &  5.6  &  17.1  &  -0.20  \\
		tn504  &  18:41:06.7  &  -05:15:8.5   &  5.1  &  2.6  &  4.3  &  5.5  &  29.1  &  -0.04  \\
		tn511  &  18:41:09.1  &  -05:15:45.8   &  8.9  &  2.2  &  2.6  &  0.9  &  24.5  &  -0.67  \\
		tn512  &  18:41:09.1  &  -05:14:24.8   &  10.4  &  2.9  &  4.5  &  4.8  &  27.7  &  -0.17  \\
		tn515  &  18:41:11.1  &  -05:14:59.8   &  10.3  &  4.4  &  3.7  &  1.5  &  12.1  &  -0.64  \\
		tn517  &  18:41:11.4  &  -05:12:21.6   &  8.7  &  4.6  &  6.9  &  3.7  &  20.4  &  -0.68  \\
		tn518  &  18:41:12.0  &  -05:12:1.3   &  8.1  &  2.4  &  6.0  &  2.9  &  20.1  &  -0.60  \\
		tn519  &  18:41:12.1  &  -05:13:5.8   &  6.2  &  2.7  &  4.2  &  4.2  &  29.4  &  -0.31  \\
		tn520  &  18:41:13.0  &  -05:14:46.7   &  9.5  &  2.3  &  2.5  &  3.9  &  22.8  &  -0.42  \\
		tn524  &  18:41:14.7  &  -05:12:56.3   &  7.5  &  3.0  &  7.7  &  2.8  &  21.2  &  -1.02  \\
		tn525  &  18:41:14.8  &  -05:13:7.8   &  4.6  &  2.6  &  6.8  &  4.2  &  15.4  &  -0.61  \\
		tn547  &  18:40:22.4  &  -05:09:33.9   &  9.1  &  4.2  &  3.3  &  4.3  &  22.4  &  -0.26  \\
		tn552  &  18:40:24.1  &  -05:08:15.9   &  6.1  &  2.0  &  2.3  &  3.5  &  20.7  &  -0.51  \\
		tn555  &  18:40:26.0  &  -05:08:47.4   &  7.9  &  5.5  &  2.9  &  4.7  &  20.2  &  -0.22  \\
		tn557  &  18:40:28.7  &  -05:08:41.2   &  8.9  &  2.5  &  3.1  &  2.7  &  20.5  &  -0.50  \\
		tn559  &  18:40:29.5  &  -05:07:11.0   &  8.9  &  2.5  &  3.3  &  3.7  &  21.9  &  -0.35  \\
		tn566  &  18:40:32.2  &  -05:09:42.8   &  8.3  &  4.7  &  3.0  &  3.7  &  22.4  &  -0.39  \\
		tn617  &  18:41:09.1  &  -05:12:57.2   &  9.7  &  2.1  &  4.4  &  3.7  &  25.2  &  -0.34  \\
		tn622  &  18:40:26.2  &  -05:08:27.0   &  9.3  &  6.6  &  6.7  &  9.5  &  18.6  &  1.00  \\
		tn623  &  18:40:29.5  &  -05:07:5.4   &  10.0  &  5.1  &  3.6  &  5.7  &  20.5  &  0.04  \\
		\hline
		\end{tabular}
		\end{footnotesize}
		\end{center}
	\end{table*}

\end{appendix}
\end{onecolumn}
\end{document}